\documentclass[twocolumn]{aastex631}

\usepackage{amsmath}	
\usepackage{comment}
\usepackage{graphicx}
\usepackage{bigdelim,multirow}
\usepackage{paralist}
\usepackage{textgreek}
\usepackage{xargs}
\usepackage{xspace}
\usepackage{makecell}
\newcommand{\subref}[2]{\hyperref[#1]{\ref{#1}{#2}}}
\newcommand{\subreflett}[2]{\hyperref[#1]{#2}}
\usepackage{hyperref}
\hypersetup{
    colorlinks=true,
    linkcolor=blue,
    citecolor=blue,
    filecolor=magenta,      
    urlcolor=blue,
    }

\graphicspath{{./}{figures/}}
\usepackage{etoolbox}
\makeatletter
\newcommand\sendemail[3]{
\edef\@tempa{mailto:#1?subject=#2 }%
\edef\@tempb{\expandafter\html@spaces\@tempa\@empty}%
\href{\@tempb}{#3}}

\catcode\%=11
\def\html@spaces#1 #2{#1
\catcode\%=14
\makeatother

\newcommand{\todo}[1]{\textcolor{magenta}{[#1]}}
\newcommand{\jantodo}[1]{\textcolor{green}{Jan: #1}}
\newcommand{\alextodo}[1]{\textcolor{green}{Alex: #1}}
\newcommand{\stefanotodo}[1]{\textcolor{green}{Stefano: #1}}
\newcommand{\fdetodo}[1]{\textcolor{green}{FDE: #1}}

\newcommand{\citationneeded}{\textcolor{ForestGreen}{$^{\rm citation\;needed}$}}
\let\oldtextsigma\textsigma
\renewcommand{\textsigma}{\oldtextsigma\xspace}
\let\oldAA\AA
\renewcommand{\AA}{\text{\oldAA}\xspace}
\let\oldtextdegree\textdegree
\renewcommand{\textdegree}{\oldtextdegree\xspace}

\newcommand{\kms}{\ensuremath{\mathrm{km\,s^{-1}}}\xspace}
\newcommand{\Msun}{\ensuremath{{\rm M}_\odot}\xspace}
\newcommand{\Zsun}{\ensuremath{{\rm Z}_\odot}\xspace}
\newcommand{\yr}{\ensuremath{{\rm yr}}\xspace}
\newcommand{\Myr}{\ensuremath{{\rm Myr}}\xspace}
\newcommand{\Gyr}{\ensuremath{{\rm Gyr}}\xspace}
\newcommand{\peryr}{\ensuremath{{\rm yr^{-1}}}\xspace}
\newcommand{\Lsun}{\hbox{\,${\rm L}_\odot$}}
\newcommand{\mum}{\text{\textmu m}\xspace}
\newcommand{\kpc}{\text{kpc}\xspace}
\newcommand{\ZH}{\text{[Z/H]}\xspace}

\newcommandx{\lambdar}[2][1=R,2=]{\ensuremath{\lambda_{\rm {#1}}{#2}}\xspace}
\newcommand{\eps}{\ensuremath{\epsilon}\xspace}
\newcommand{\mstar}{\ensuremath{M_\star}\xspace}
\newcommand{\mdyn}{\ensuremath{M_\mathrm{dyn}}\xspace}
\newcommand{\re}{\ensuremath{R_\mathrm{e}}\xspace}
\newcommand{\vstar}{\ensuremath{v_\star}\xspace}
\newcommand{\vnai}{\ensuremath{v_{\NaI}}\xspace}
\newcommand{\sigmastar}{\ensuremath{\sigma_\star}\xspace}
\newcommand{\sigmaestar}{\ensuremath{\sigma_{\star,\mathrm{e}}}\xspace}
\newcommand{\vperc}[1]{\ensuremath{v_{#1}}\xspace}

\newcommand{\nelec}{\ensuremath{n_\mathrm{e}}\xspace}

\newcommandx{\fluxdcgs}[1][1=-20]{$\times 10^{[#1]}$~erg~s$^{-1}$~cm$^{-2}$~\AA$^{-1}$\xspace}
\newcommandx{\fluxcgs}[2][1=-20,2=\ensuremath{\times}]{${#2}10^{#1}$~erg~s$^{-1}$~cm$^{-2}$\xspace}
\newcommandx{\powercgs}[1][1=44]{$\times 10^{#1}$~erg~s$^{-1}$\xspace}
\newcommand{\Av}{\ensuremath{A_V}\xspace}

\newcommand{\jwst}{\textit{JWST}\xspace}
\newcommand{\hst}{\textit{HST}\xspace}

\newcommand{\ppxf}{{\sc ppxf}\xspace}
\newcommand{\fsps}{{\sc fsps}\xspace}
\newcommand{\prospector}{{\sc prospector}\xspace}
\newcommand{\bagpipes}{{\sc bagpipes}\xspace}
\newcommand{\beagle}{{\sc beagle}\xspace}
\newcommand{\emcee}{{\sc emcee}\xspace}
\newcommand{\cloudy}{{\sc cloudy}\xspace}
\newcommand{\fitsmap}{{\sc fitsmap}\xspace}
\newcommand{\eazy}{{\sc eazy}\xspace}
\newcommand{\empt}{{\sc eMPT}\xspace}
\newcommandx{\mappings}[1][1=]{{\sc mappings{#1}}\xspace}

\newcommand{\mediumjwstgn}{medium/\jwst~GN\xspace}
\newcommand{\mediumhstgn}{medium/\hst~GN\xspace}
\newcommand{\mediumhstgs}{medium/\hst~GS\xspace}
\newcommand{\mediumjwstgs}{medium/\jwst~GS\xspace}
\newcommand{\mediumjwstgsb}{medium/\jwst~GS~1180\xspace}
\newcommand{\ultradeep}{ultradeep~GS\xspace}

\newcommand{\Lyalpha}{\text{Ly\textalpha}\xspace}
\newcommand{\Halpha}{\text{H\textalpha}\xspace}
\newcommand{\Hbeta}{\text{H\textbeta}\xspace}
\newcommand{\Hgamma}{\text{H\textgamma}\xspace}
\newcommand{\Hdelta}{\text{H\textdelta}\xspace}
\newcommand{\Hepsilon}{\text{H\textepsilon}\xspace}
\newcommand{\Hzeta}{\text{H\textdzeta}\xspace}
\newcommand{\Paalpha}{\text{Pa\textalpha}\xspace}
\newcommand{\Pabeta}{\text{Pa\textbeta}\xspace}
\newcommand{\Pagamma}{\text{Pa\textgamma}\xspace}
\newcommand{\Padelta}{\text{Pa\textdelta}\xspace}

\newcommandx{\permittedEL}[6][1=O,2=III,3=,4=,5=,6=]{\text{{#1}\,{\sc {#2}}{#3}{#4}{#5}{#6}}\xspace}
\newcommandx{\semiforbiddenEL}[6][1=O,2=III,3=,4=,5=,6=]{\text{{#1}\,{\sc{#2}}]{#3}{#4}{#5}{#6}}\xspace}
\newcommandx{\forbiddenEL}[6][1=O,2=III,3=,4=,5=,6=]{\text{[{#1}\,{\sc{#2}}]{#3}{#4}{#5}{#6}}\xspace}

\newcommand{\EW}[1]{\text{EW(#1)}\xspace}

\newcommand{\HI}{\permittedEL[H][i]}
\newcommand{\HII}{\permittedEL[H][ii]}

\newcommand{\NV}{\permittedEL[N][v]}
\newcommandx{\NVL}[1][1=1243]{\permittedEL[N][v][\textlambda][#1]}
\newcommandx{\NVall}{\permittedEL[N][v][\textlambda][\textlambda][1239,][1243]}

\newcommandx{\CIIL}[1][1=232x]{\semiforbiddenEL[C][ii][\textlambda][#1]}
\newcommandx{\CIIall}{\semiforbiddenEL[C][ii][\textlambda][\textlambda][2323.5--][2328.1]}

\newcommand{\NIV}{\semiforbiddenEL[N][iv]}
\newcommandx{\NIVL}[1][1=1486]{\semiforbiddenEL[N][iv][\textlambda][#1]}

\newcommand{\CIV}{\permittedEL[C][iv]}
\newcommandx{\CIVL}[1][1=1550]{\permittedEL[C][iv][\textlambda][#1]}
\newcommand{\CIVall}{\permittedEL[C][iv][\textlambda][\textlambda][1549,][1551]}

\newcommand{\HeII}{\permittedEL[He][ii]}
\newcommandx{\HeIIL}[1][1=1640]{\permittedEL[He][ii][\textlambda][#1]}

\newcommand{\semiOIII}{\semiforbiddenEL[O][iii]}
\newcommandx{\semiOIIIL}[1][1=1666]{\semiforbiddenEL[O][iii][\textlambda][#1]}
\newcommand{\semiOIIIall}{\semiforbiddenEL[O][iii][\textlambda][\textlambda][1661,][1666]}

\newcommand{\NIII}{\semiforbiddenEL[N][iii]}
\newcommandx{\NIIIL}[1][1=1750]{\semiforbiddenEL[N][iii][\textlambda][#1]}
\newcommand{\NIIIall}{\semiforbiddenEL[N][iii][\textlambda][\textlambda][1747--][1754]}

\newcommandx{\CIII}{\semiforbiddenEL[C][iii]}
\newcommandx{\CIIIL}[1][1=1909]{\semiforbiddenEL[C][iii][\textlambda][#1]}
\newcommand{\CIIIall}{\semiforbiddenEL[C][iii][\textlambda][\textlambda][1907,][1909]}

\newcommand{\NeIV}{\forbiddenEL[Ne][iv]}
\newcommandx{\NeIVL}[1][1=2424]{\forbiddenEL[Ne][iv][\textlambda][#1]}
\newcommand{\NeIVall}{\forbiddenEL[Ne][iv][\textlambda][\textlambda][2422,][2424]}

\newcommand{\MgII}{\permittedEL[Mg][ii]}
\newcommandx{\MgIIL}[1][1=2803]{\permittedEL[Mg][ii][\textlambda][#1]}
\newcommand{\MgIIall}{\permittedEL[Mg][ii][\textlambda][\textlambda][2796,][2803]}

\newcommand{\NeV}{\forbiddenEL[Ne][v]}
\newcommandx{\NeVL}[1][1=3426]{\forbiddenEL[Ne][v][\textlambda][#1]}
\newcommand{\NeVall}{\forbiddenEL[Ne][v][\textlambda][\textlambda][3346,][3426]}

\newcommand{\OII}{\forbiddenEL[O][ii]}
\newcommandx{\OIIL}[1][1=3727]{\forbiddenEL[O][ii][\textlambda][#1]}
\newcommand{\OIIall}{\forbiddenEL[O][ii][\textlambda][\textlambda][3726,][3729]}

\newcommand{\NeIII}{\forbiddenEL[Ne][iii]}
\newcommandx{\NeIIIL}[1][1=3869]{\forbiddenEL[Ne][iii][\textlambda][#1]}
\newcommand{\NeIIIall}{\forbiddenEL[Ne][iii][\textlambda][\textlambda][3869,][3968]}

\newcommand{\OIII}{\forbiddenEL[O][iii]}
\newcommandx{\OIIIL}[1][1=5007]{\forbiddenEL[O][iii][\textlambda][#1]}
\newcommand{\OIIIall}{\forbiddenEL[O][iii][\textlambda][\textlambda][4959,][5007]}

\newcommand{\NaI}{\permitted[Na][i]}
\newcommandx{\NaIL}[1][1=5890]{\permittedEL[Na][i][\textlambda][#1]}
\newcommand{\NaIall}{\permittedEL[Na][i][\textlambda][\textlambda][5890,][5896]}

\newcommand{\OI}{\forbiddenEL[O][i]}
\newcommandx{\OIL}[1][1=6300]{\forbiddenEL[O][i][\textlambda][#1]}
\newcommand{\OIall}{\forbiddenEL[O][i][\textlambda][\textlambda][6300,][6363]}

\newcommand{\HeI}{\permittedEL[He][i]}
\newcommandx{\HeIL}[1][1=10830]{\permittedEL[He][i][\textlambda][#1]}

\newcommand{\NII}{\forbiddenEL[N][ii]}
\newcommandx{\NIIL}[1][1=6584]{\forbiddenEL[N][ii][\textlambda][#1]}
\newcommand{\NIIall}{\forbiddenEL[N][ii][\textlambda][\textlambda][6549,][6584]}

\newcommand{\SII}{\forbiddenEL[S][ii]}
\newcommand{\SIIL}[1][1=6717]{\forbiddenEL[S][ii][\textlambda][#1]}
\newcommand{\SIIall}{\forbiddenEL[S][ii][\textlambda][\textlambda][6717,][6730]}

\newcommandx{\OIIAuL}[1][1=7325]{\forbiddenEL[O][ii][\textlambda][#1]}
\newcommand{\OIIAuall}{\forbiddenEL[O][ii][\textlambda][\textlambda][7319--][7332]}

\newcommand{\SIII}{\forbiddenEL[S][iii]}
\newcommandx{\SIIIL}[1][1=9532]{\forbiddenEL[S][iii][\textlambda][#1]}
\newcommand{\SIIIall}{\forbiddenEL[S][iii][\textlambda][\textlambda][9069,][9532]}

\newcommandx{\SIIAuL}[1][1=10290]{\forbiddenEL[S][ii][\textlambda][#1]}
\newcommand{\SIIAuall}{\forbiddenEL[S][ii][\textlambda][\textlambda][10290--][10373]}

\newcommand{\hda}{\ensuremath{\mathrm{H\text{\textdelta}_A}}\xspace}
\newcommand{\hga}{\ensuremath{\mathrm{H\text{\textgamma}_A}}\xspace}
\defcitealias{bunker+2023b}{B23}

\submitjournal{ApJS}

\begin{document}

\title{JADES Data Release 3 -- NIRSpec/MSA spectroscopy for 4,000 galaxies in the GOODS fields}

\author[0000-0003-3458-2275]{Francesco D'Eugenio}
\altaffiliation{These authors contributed equally to this work}
\affiliation{Kavli Institute for Cosmology, University of Cambridge, Madingley Road, Cambridge CB3 0HA, UK} 
\affiliation{Cavendish Laboratory, University of Cambridge, 19 JJ Thomson Avenue, Cambridge CB3 0HE, UK}
\correspondingauthor{Francesco D'Eugenio}
\email{francesco.deugenio@gmail.com}

\author[0000-0002-0450-7306]{Alex J.\ Cameron}
\altaffiliation{These authors contributed equally to this work}
\affiliation{Department of Physics, University of Oxford, Denys Wilkinson Building, Keble Road, Oxford OX1 3RH, UK}

\author[0000-0001-6010-6809]{Jan Scholtz}
\altaffiliation{These authors contributed equally to this work}
\affiliation{Kavli Institute for Cosmology, University of Cambridge, Madingley Road, Cambridge CB3 0HA, UK} 
\affiliation{Cavendish Laboratory, University of Cambridge, 19 JJ Thomson Avenue, Cambridge CB3 0HE, UK}

\author[0000-0002-6719-380X]{Stefano Carniani}
\altaffiliation{These authors contributed equally to this work}
\affiliation{Scuola Normale Superiore, Piazza dei Cavalieri 7, I-56126 Pisa, Italy}

\author[0000-0002-4201-7367]{Chris J. Willott}
\altaffiliation{These authors contributed equally to this work}
\affil{NRC Herzberg, 5071 West Saanich Rd, Victoria, BC V9E 2E7, Canada}

\author[0000-0002-9551-0534]{Emma Curtis-Lake}
\affiliation{Centre for Astrophysics Research, Department of Physics, Astronomy and Mathematics, University of Hertfordshire, Hatfield AL10 9AB, UK}

\author[0000-0002-8651-9879]{Andrew J. Bunker}
\affiliation{Department of Physics, University of Oxford, Denys Wilkinson Building, Keble Road, Oxford OX1 3RH, UK}

\author[0000-0002-7392-7814]{Eleonora Parlanti}
\affiliation{Scuola Normale Superiore, Piazza dei Cavalieri 7, I-56126 Pisa, Italy}

\author[0000-0002-4985-3819]{Roberto Maiolino}
\affiliation{Kavli Institute for Cosmology, University of Cambridge, Madingley Road, Cambridge CB3 0HA, UK} 
\affiliation{Cavendish Laboratory, University of Cambridge, 19 JJ Thomson Avenue, Cambridge CB3 0HE, UK}
\affiliation{Department of Physics and Astronomy, University College London,
Gower Street, London WC1E 6BT, UK}

\author[0000-0001-9262-9997]{Christopher N. A. Willmer}
\affiliation{Steward Observatory, University of Arizona, 933 North Cherry Avenue, Tucson, AZ 85721, USA}

\author[0000-0002-6780-2441]{Peter Jakobsen}
\affiliation{Cosmic Dawn Center (DAWN), Copenhagen, Denmark 2. Niels Bohr Institute, University of Copenhagen, Jagtvej 128, DK-2200, Copenhagen, Denmark}

\author[0000-0002-4271-0364]{Brant E. Robertson}
\affiliation{Department of Astronomy and Astrophysics, University of California, Santa Cruz, 1156 High Street, Santa Cruz, CA 95064, USA}

\author[0000-0002-9280-7594]{Benjamin D.\ Johnson}
\affiliation{Center for Astrophysics $|$ Harvard \& Smithsonian, 60 Garden St., Cambridge MA 02138 USA}

\author[0000-0002-8224-4505]{Sandro Tacchella}
\affiliation{Kavli Institute for Cosmology, University of Cambridge, Madingley Road, Cambridge CB3 0HA, UK} 
\affiliation{Cavendish Laboratory, University of Cambridge, 19 JJ Thomson Avenue, Cambridge CB3 0HE, UK}

\author[0000-0002-1617-8917]{Phillip A.\ Cargile}
\affiliation{Center for Astrophysics $|$ Harvard \& Smithsonian, 60 Garden St., Cambridge MA 02138 USA}

\author[0000-0002-7028-5588]{Tim Rawle}
\affiliation{European Space Agency (ESA), European Space Astronomy Centre (ESAC), Camino Bajo del Castillo s/n, 28692 Villafranca del Castillo, Madrid, Spain}

\author[0000-0001-7997-1640]{Santiago Arribas}
\affiliation{Centro de Astrobiolog\'ia (CAB), CSIC–INTA, Cra. de Ajalvir Km.~4, 28850- Torrej\'on de Ardoz, Madrid, Spain}

\author[0000-0002-7636-0534]{Jacopo Chevallard}
\affiliation{Department of Physics, University of Oxford, Denys Wilkinson Building, Keble Road, Oxford OX1 3RH, UK}

\author[0000-0002-2678-2560]{Mirko Curti}
\affiliation{European Southern Observatory, Karl-Schwarzschild-Strasse 2, 85748 Garching, Germany}

\author[0000-0003-1344-9475]{Eiichi Egami}
\affiliation{Steward Observatory, University of Arizona, 933 North Cherry Avenue, Tucson, AZ 85721, USA}

\author[0000-0002-2929-3121]{Daniel J.\ Eisenstein}
\affiliation{Center for Astrophysics $|$ Harvard \& Smithsonian, 60 Garden St., Cambridge MA 02138 USA}

\author[0000-0002-5320-2568]{Nimisha Kumari}
\affiliation{AURA for European Space Agency, Space Telescope Science Institute, 3700 San Martin Drive. Baltimore, MD, 21210}

\author[0000-0002-3642-2446]{Tobias J.\ Looser}
\affiliation{Kavli Institute for Cosmology, University of Cambridge, Madingley Road, Cambridge CB3 0HA, UK} 
\affiliation{Cavendish Laboratory, University of Cambridge, 19 JJ Thomson Avenue, Cambridge CB3 0HE, UK}

\author[0000-0002-7893-6170]{Marcia J. Rieke}
\affiliation{Steward Observatory, University of Arizona, 933 North Cherry Avenue, Tucson, AZ 85721, USA}

\author[0000-0001-5171-3930]{Bruno Rodríguez Del Pino}
\affiliation{Centro de Astrobiolog\'ia (CAB), CSIC–INTA, Cra. de Ajalvir Km.~4, 28850- Torrej\'on de Ardoz, Madrid, Spain}

\author[0000-0001-5333-9970]{Aayush Saxena}
\affiliation{Department of Physics, University of Oxford, Denys Wilkinson Building, Keble Road, Oxford OX1 3RH, UK}
\affiliation{Department of Physics and Astronomy, University College London, Gower Street, London WC1E 6BT, UK}

\author[0000-0003-4891-0794]{Hannah \"Ubler}
\affiliation{Kavli Institute for Cosmology, University of Cambridge, Madingley Road, Cambridge CB3 0HA, UK} 
\affiliation{Cavendish Laboratory, University of Cambridge, 19 JJ Thomson Avenue, Cambridge CB3 0HE, UK}

\author[0000-0001-8349-3055]{Giacomo Venturi}
\affiliation{Scuola Normale Superiore, Piazza dei Cavalieri 7, I-56126 Pisa, Italy}

\author[0000-0002-7595-121X]{Joris Witstok}
\affiliation{Kavli Institute for Cosmology, University of Cambridge, Madingley Road, Cambridge CB3 0HA, UK} 
\affiliation{Cavendish Laboratory, University of Cambridge, 19 JJ Thomson Avenue, Cambridge CB3 0HE, UK}

\author[0000-0003-0215-1104]{William M.\ Baker}
\affiliation{Kavli Institute for Cosmology, University of Cambridge, Madingley Road, Cambridge CB3 0HA, UK} 
\affiliation{Cavendish Laboratory, University of Cambridge, 19 JJ Thomson Avenue, Cambridge CB3 0HE, UK}

\author[0000-0003-0883-2226]{Rachana Bhatawdekar}
\affiliation{European Space Agency (ESA), European Space Astronomy Centre (ESAC), Camino Bajo del Castillo s/n, 28692 Villanueva de la Cañada, Madrid, Spain}

\author[0000-0001-8470-7094]{Nina Bonaventura}
\affiliation{Steward Observatory University of Arizona 933 N. Cherry Avenue Tucson AZ 85721, USA}

\author[0000-0003-4109-304X]{Kristan Boyett}
\affiliation{School of Physics, University of Melbourne, Parkville 3010, VIC, Australia}
\affiliation{ARC Centre of Excellence for All Sky Astrophysics in 3 Dimensions (ASTRO 3D), Australia}

\author[0000-0003-3458-2275]{Stephane Charlot}
\affiliation{Sorbonne Universit\'e, CNRS, UMR 7095, Institut d'Astrophysique de Paris, 98 bis bd Arago, 75014 Paris, France}

\author[0000-0002-9708-9958]{A.\ Lola Danhaive}
\affiliation{Kavli Institute for Cosmology, University of Cambridge, Madingley Road, Cambridge CB3 0HA, UK} 
\affiliation{Cavendish Laboratory, University of Cambridge, 19 JJ Thomson Avenue, Cambridge CB3 0HE, UK}

\author[0000-0001-9262-9997]{Kevin N. Hainline}
\affiliation{Steward Observatory, University of Arizona, 933 North Cherry Avenue, Tucson, AZ 85721, USA}

\author[0000-0002-8543-761X]{Ryan Hausen}
\affiliation{Department of Physics and Astronomy, The Johns Hopkins University, 3400 N. Charles St., Baltimore, MD 21218}

\author[0000-0003-4337-6211]{Jakob M.\ Helton}
\affiliation{Steward Observatory, University of Arizona, 933 North Cherry Avenue, Tucson, AZ 85721, USA}

\author[0000-0002-1660-9502]{Xihan Ji}
\affiliation{Kavli Institute for Cosmology, University of Cambridge, Madingley Road, Cambridge CB3 0HA, UK} 
\affiliation{Cavendish Laboratory, University of Cambridge, 19 JJ Thomson Avenue, Cambridge CB3 0HE, UK}

\author[0000-0001-7673-2257]{Zhiyuan Ji}
\affiliation{Steward Observatory, University of Arizona, 933 North Cherry Avenue, Tucson, AZ 85721, USA}

\author[0000-0002-0267-9024]{Gareth C.\ Jones}
\affiliation{Department of Physics, University of Oxford, Denys Wilkinson Building, Keble Road, Oxford OX1 3RH, UK}

\author[0000-0002-0267-9024]{Ignas Joud\v{z}balis}
\affiliation{Kavli Institute for Cosmology, University of Cambridge, Madingley Road, Cambridge CB3 0HA, UK} 
\affiliation{Cavendish Laboratory, University of Cambridge, 19 JJ Thomson Avenue, Cambridge CB3 0HE, UK}

\author[0000-0003-0695-4414]{Michael V.\ Maseda}
\affiliation{Department of Astronomy, University of Wisconsin-Madison, 475 N. Charter St., Madison, WI 53706 USA}

\author[0000-0003-4528-5639]{Pablo G. P\'erez-Gonz\'alez}
\affiliation{Centro de Astrobiolog\'ia (CAB), CSIC–INTA, Cra. de Ajalvir Km.~4, 28850- Torrej\'on de Ardoz, Madrid, Spain}

\author[0000-0002-0362-5941]{Michele Perna}
\affiliation{Centro de Astrobiolog\'ia (CAB), CSIC–INTA, Cra. de Ajalvir Km.~4, 28850- Torrej\'on de Ardoz, Madrid, Spain}

\author[0000-0001-8630-2031]{Dávid Puskás}
\affiliation{Kavli Institute for Cosmology, University of Cambridge, Madingley Road, Cambridge CB3 0HA, UK} 
\affiliation{Cavendish Laboratory, University of Cambridge, 19 JJ Thomson Avenue, Cambridge CB3 0HE, UK}

\author[0000-0003-4702-7561]{Irene Shivaei}
\affiliation{Centro de Astrobiolog\'ia (CAB), CSIC–INTA, Cra. de Ajalvir Km.~4, 28850- Torrej\'on de Ardoz, Madrid, Spain}

\author[0000-0000-0000-0000]{Maddie S.\ Silcock}
\affiliation{Centre for Astrophysics Research, Department of Physics, Astronomy and Mathematics, University of Hertfordshire, Hatfield AL10 9AB, UK}

\author[0000-0003-4770-7516]{Charlotte Simmonds}
\affiliation{Kavli Institute for Cosmology, University of Cambridge, Madingley Road, Cambridge CB3 0HA, UK} 
\affiliation{Cavendish Laboratory, University of Cambridge, 19 JJ Thomson Avenue, Cambridge CB3 0HE, UK}

\author[0000-0001-8034-7802]{Renske Smit}
\affiliation{Astrophysics Research Institute, Liverpool John Moores University, 146 Brownlow Hill, Liverpool L3 5RF, UK}

\author[0000-0002-4622-6617]{Fengwu Sun}
\affiliation{Steward Observatory, University of Arizona, 933 North Cherry Avenue, Tucson, AZ 85721, USA}

\author[0000-0001-6917-4656]{Natalia C.\ Villanueva}
\affiliation{Kavli Institute for Cosmology, University of Cambridge, Madingley Road, Cambridge CB3 0HA, UK} 
\affiliation{Cavendish Laboratory, University of Cambridge, 19 JJ Thomson Avenue, Cambridge CB3 0HE, UK}

\author[0000-0003-2919-7495]{Christina C.\ Williams}
\affiliation{NSF’s National Optical-Infrared Astronomy Research Laboratory, 950 North Cherry Avenue, Tucson, AZ 85719, USA}
\affiliation{Steward Observatory, University of Arizona, 933 North Cherry Avenue, Tucson, AZ 85721, USA}

\author[0000-0003-3307-7525]{Yongda Zhu}
\affiliation{Steward Observatory, University of Arizona, 933 North Cherry Avenue, Tucson, AZ 85721, USA}

\begin{abstract}
We present the third data release of JADES, the \jwst Advanced Deep Extragalactic Survey, providing both imaging and spectroscopy in the two GOODS fields.
Spectroscopy consists of medium-depth and deep NIRSpec/MSA spectra of 4,000 targets, covering the spectral range $0.6\text{--}5.3$~\mum and observed with both the low-dispersion prism ($R=30\text{--}300$) and all three medium-resolution gratings ($R=500\text{--}1,500$).
We describe the observations, data reduction, sample selection, and target allocation. We measured 2,375 redshifts (2,053 from multiple emission lines); our targets span the range from $z=0.5$ up to $z=13$, including 404 at $z>5$.
The data release includes 2-d and 1-d fully reduced spectra, with slit-loss corrections and background subtraction optimized for point sources. We also provide redshifts and $S/N>5$ emission-line flux catalogs for the prism and grating spectra, and concise guidelines on how to use these data products.
Alongside spectroscopy, we are also publishing fully calibrated NIRCam imaging, which enables studying the JADES sample with the combined power of imaging and spectroscopy. Together, these data provide the largest statistical sample to date to characterize the properties of galaxy populations in the first billion years after the Big Bang.
\end{abstract}

\submitjournal{ApJS}
\keywords{}

\section{Introduction}\label{s.intro}
The long-awaited launch of \jwst has revolutionized our ability to observe the early universe.
Already in the first two years of operations, \jwst enabled an amazing number of discoveries and studies.
Many of these breakthroughs have been made possible by the unprecedented sensitivity and wavelength coverage of the NIRSpec instrument \citep{jakobsen+2022}.
These include the spectroscopic confirmation of galaxies beyond redshift $z=10$ \citep{curtis-lake+2023,arrabal-haro+2023,wang+2023} -- including through emission lines \citep{bunker+2023a};
the discovery of substantial neutral-gas absorption in galaxies at $z=9\text{--}11$ \citep[e.g.,][]{heintz+2023b,umeda+2023};
the first studies of metallicity and chemical abundances using well-known optical lines \citep[e.g.,][]{curti+2023a,nakajima+2023};
the discovery of massive, quiescent and old galaxies at $z=3\text{--}5$ \citep[e.g.,][]{carnall+2023b,glazebrook+2023};
the first `mini-quenched' galaxies \citep{looser+2024,strait+2023};
neutral-phase outflows in massive quiescent galaxies \citep{belli+2023,deugenio+2023c,davies+2024};
the discovery of bright, metal-poor active galactic nuclei \citep[e.g.,][]{kocevski+2023,uebler+2023};
the most distant active galactic nuclei \citep[AGN;][]{maiolino+2023a,goulding+2023};
and even tentative evidence of the first generation of stars \citep{maiolino+2023b}.

However, for spectroscopy, sample sizes are still small, of the order of tens to one hundred objects \citep{looser+2023,curti+2023b,nakajima+2023}.
The availability of large samples is key to characterizing the properties of galaxy populations, studying their cosmic evolution, and disentangling the intricate pattern of cause and effect between the observed properties of galaxies.
Studies of galaxies at redshifts $z<1$ rely on thousands or even hundreds of thousands of spectroscopic targets \citep[e.g.,][]{abazajian+2009,driver+2018,desi+2016a}.
By studying several physical properties at a time \citep[e.g.,][]{kauffmann+2003a,kauffmann+2003b,peng+2010,graves+faber2010}, or by using machine-learning techniques \citep[e.g.,][]{bluck+2022,baker+2022,barsanti+2023,walmsley+2023}, these studies have made tremendous progress in understanding the links between many galaxy properties like morphology, star-formation rate, age, gas fraction, star-formation efficiency, supermassive black-hole mass, local and global environment, and metallicity.
In the last decades, large spectroscopic surveys in the near-infrared have enabled the study of hundreds of galaxies at redshifts $z=1\text{--}4$ \citep[e.g.,][]{wisnioski+2015,stott+2016,kriek+2015}.
In the near future, new surveys will observe even more galaxies at redshifts $z \lesssim 4$ \citep{dalton+2012,tamura+2016,dejong+2019,maiolino+2020}.
However -- at least for the next decade -- nothing other than \jwst and, in particular, NIRSpec will be able to obtain deep rest-frame optical spectroscopy for large samples of galaxies at redshifts 3--10, the crucial early phases of galaxy formation.

The NIRSpec Micro Shutter Assembly \citep[MSA,][]{ferruit+2022} was designed to observe more than one hundred targets simultaneously, and is now the highest-multiplicity slit-based spectrograph in the near-infrared.
In the NIRSpec/MSA, this high multiplicity marries an unprecedented combination of large collecting area, low background, and long wavelength coverage, which all together enable us, for the first time, to efficiently observe large samples of galaxies at redshifts $z>4$, covering their strongest rest-frame optical features \citetext{e.g., \citealp{treu+2022}, \citealp{bezanson+2022}, \citealp{fujimoto+2023}, \citealp{oesch+2023}, \citealp{bunker+2023b}, hereafter: \citetalias{bunker+2023b}}.

One of the goals of JADES, the \jwst Advanced Deep Extragalactic Survey \citep{eisenstein_overview_2023}, is to observe a statistical sample of galaxies at redshifts $z>3$, thus enabling spectroscopic studies to move from the `discovery' stage to a more quantitative understanding of galaxy populations.
To enable this progress, JADES -- a collaboration between the \jwst NIRCam and NIRSpec GTO teams -- was designed to fully exploit the synergy between photometry and spectroscopy.
The JADES strategy divides the survey time between two tiers: medium-depth and deep observations \citep[for the least deep and widest tier of the NIRSpec GTO see][]{maseda+2024}.
All observations are in the two GOODS fields \citep[Great Observatories Origins Deep Survey;][]{giavalisco_goods_2004}, but the medium tier is divided between the GOODS South and North fields (hereafter, GOODS-S and GOODS-N), whereas the deep tier is in GOODS-S only.

In this article, we present NIRSpec observations of the deep tier from the JADES Origins Field \citep{eisenstein+2023b}, as well as
medium-depth spectra from both GOODS-S and GOODS-N.
We provide fully reduced and calibrated 1-d and 2-d spectra, as well as measurements of redshift and emission-line fluxes\footnote{Available on the JADES website \url{https://jades-survey.github.io/scientists/data.html}.}. The NIRSpec data presented here cover all JADES observations up to October~2023; spectroscopic data collected from November~2023 will be the subject of a future data release.  
In support of this spectroscopy, we also present previously unpublished NIRCam imaging in GOODS-N, including photometric catalogs and photometric redshifts.
After presenting the new NIRCam data (Section~\ref{sec:nircam}), we move to spectroscopy with a summary of the NIRSpec observations and sample selection
(Sections~\ref{sec:observations} and~\ref{sec:sample}) and of the data reduction (Section~\ref{s.datared}).
We then outline the measurements of spectroscopic redshifts and line fluxes (Sections~\ref{s.visinsp}--\ref{s.r1000}).
In Sections~\ref{s.quality} and~\ref{s.limitations} we present an assessment of the data products and guidelines for their use, and in Section~\ref{s.highl} we showcase exciting highlights from the current data. We conclude with a short summary and brief outlook (Section~\ref{s.conclusions}).

Note that the current data release employs the same data reduction as the previous NIRSpec data release \citepalias[DR1;][]{bunker+2023b}, with the only difference being a different algorithm for the measurement of emission-line fluxes.
This is the third JADES data release (hereafter: DR3), but only the second data release for spectroscopy; JADES DR2 included only NIRCam imaging.
Throughout this work, we use the AB~magnitude system \citep{oke+gunn1983}.

\section{Release of NIRCam Imaging in GOODS-N}\label{sec:nircam}

GOODS-N is a very important deep field, as it includes the iconic Hubble Deep Field \citep[HDF;][]{williams+1996} and the many other programs that followed in and around it.  Of particular note are the substantial HST optical and near-infrared imaging from GOODS \citep{giavalisco_goods_2004} and CANDELS \citep{grogin+2011,koekemoer+2011}, very deep X-ray imaging from Chandra \citep{alexander+2003,xue+2016}, sub/millimeter observations \citep{chapin+2009,mullaney+2012,magnelli+2013,cowie+2017}, and broad-band radio coverage \citep{morrison+2010,murphy+2017}.  This region also has extensive ground-based spectroscopy \citep[e.g.,][]{wirth+2004,treu+2005,newman+2013,kriek+2015}, HST grism spectroscopy \citep{momcheva+2016_3dhst_specz}, and hosts other JWST NIRSpec programs such as AURORA \citep[Program ID, PID~1914;][]{shapley+2021}.

As a part of JADES DR3, we are including images and catalogs from the NIRCam imaging in GOODS-N, observed as the Medium Prime part of JWST program 1181 (PI: Eisenstein).  These data were observed in February 2023 and include 7 overlapping medium-deep pointings, each with 8-9 separate filters.  Four of the pointings are mildly deeper than the other three; details are in \citet{eisenstein_overview_2023}.
The imaging data are supplemented with F182M, F210M, and shallow F444W imaging from the FRESCO survey \citep{oesch+2023_fresco}.

\begin{figure*}
  \includegraphics[width=\textwidth]{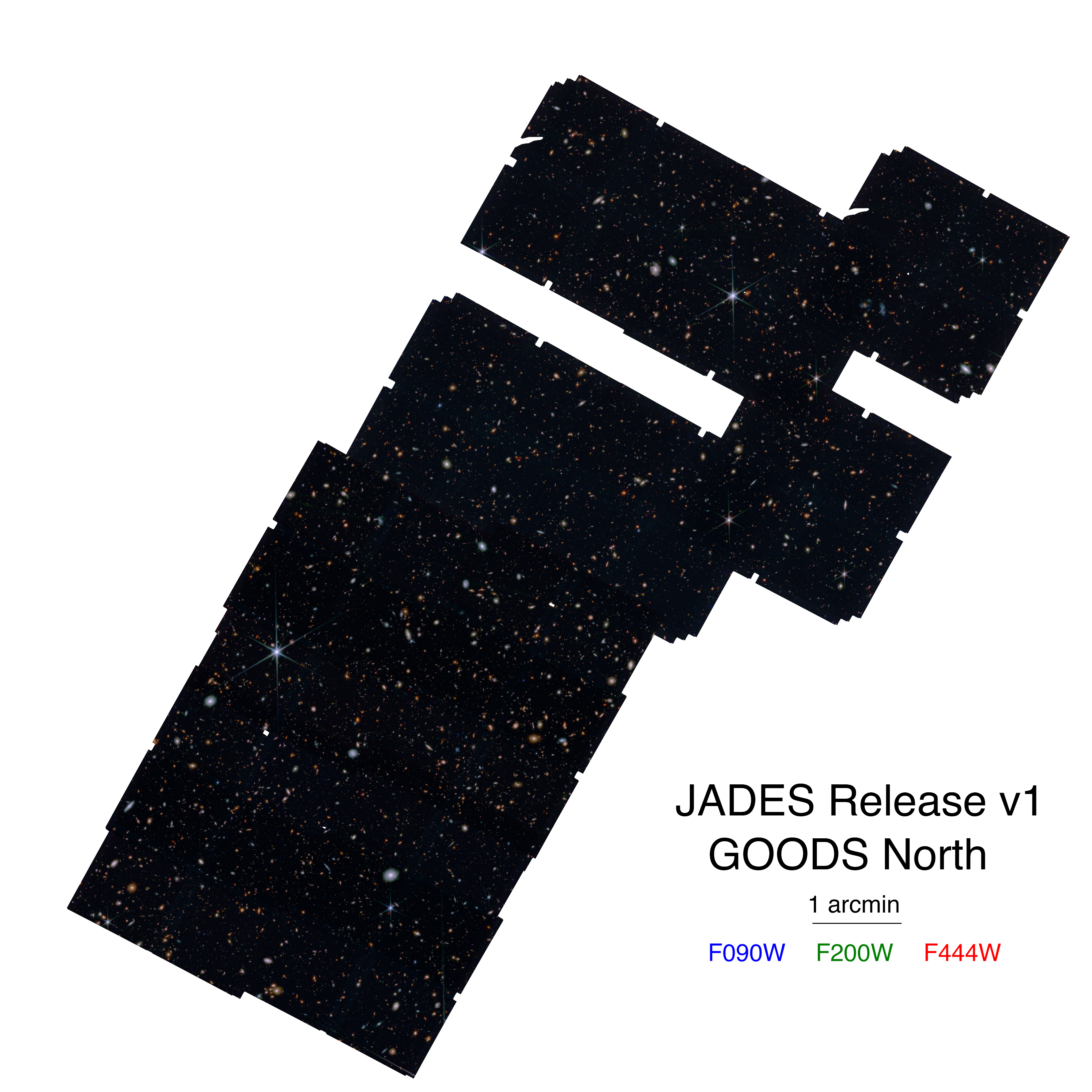}
  \caption{Mosaic of JADES JWST/NIRCam data in the GOODS-N field acquired in February 2023. The F444W, F200W, and F090W mosaics are shown as the red, green, and blue channels in this multicolor image. The scale bar indicates 1 arcmin.
  }\label{f.goodsn}
\end{figure*}

\begin{figure*}
  \includegraphics[width=\textwidth]{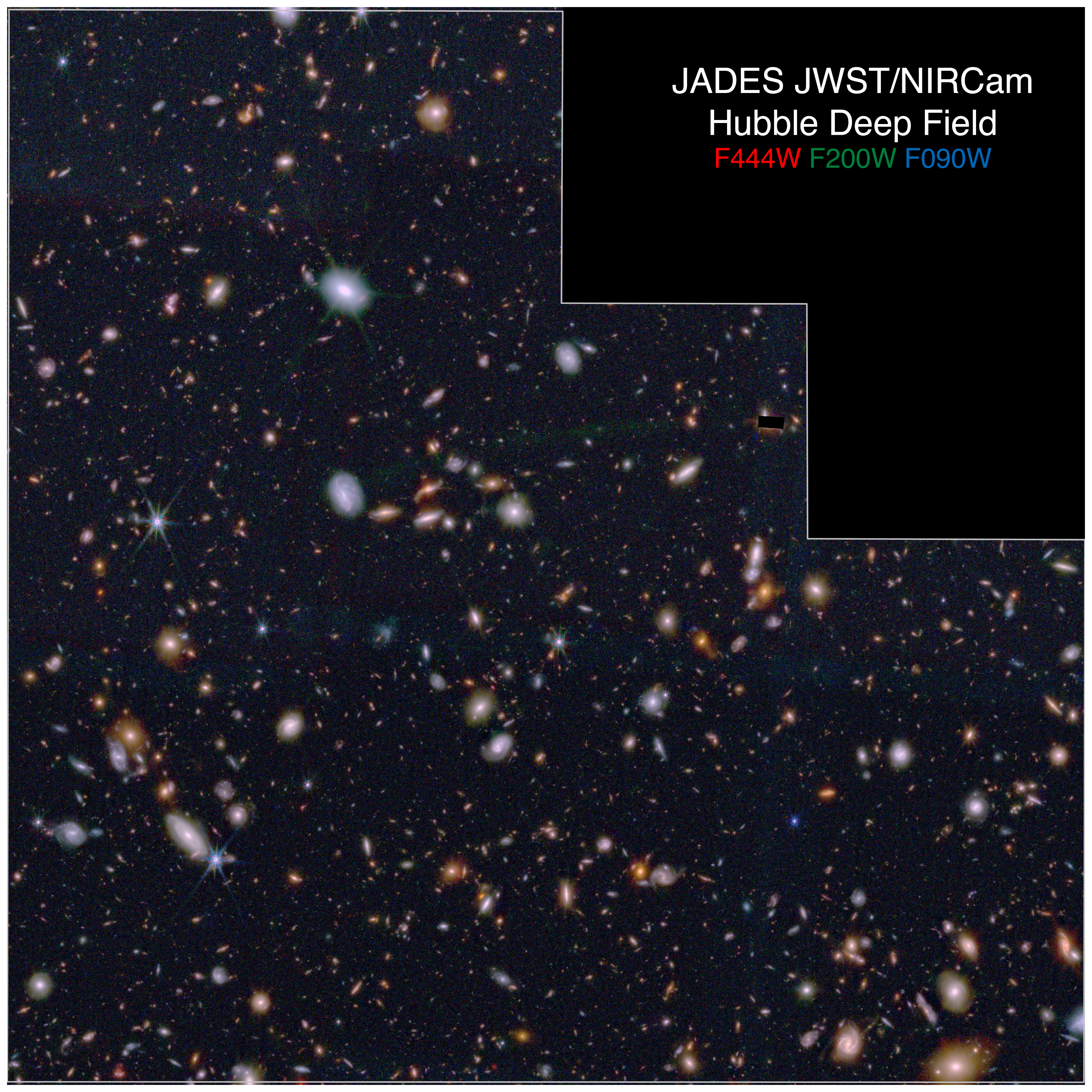}
  \caption{Mosaic of JADES JWST/NIRCam data covering the Hubble Deep Field region. Shown are the F444W, F200W, and F090W mosaics as the red-green-blue channels of this multicolor image. The iconic Hubble Deep Field \citep{williams+1996} footprint is shown as a silver line. We note the linear diffuse green feature in the center of the image results from local noise in the F200W image and is not an astrophysical object.
  }\label{f.hdf}
\end{figure*}

The reduction of these images follows closely the processing used for JADES DR1 \citep{rieke_jades_2023} and DR2 \citep{eisenstein+2023b}. For the photometric catalog release, we follow the methods described in \citet{rieke_jades_2023}, \citet{eisenstein+2023b}, and \citet{robertson+2023b}. These methods were engineered on the deeper JADES GOODS-S imaging. To remove some spurious extended low surface brightness sources, we conservatively remove objects with an average surface brightness of $SB<0.045$ nJy per pixel within the detection segmentation. Beyond this revision, the techniques follow exactly \citet{robertson+2023b} and will be described in more detail in Robertson et al.\ (in prep.).
As before, the JADES GOODS-N release includes mosaic imaging in all bands, object detection and photometric catalogs from the JADES reduction pipelines, and photometric redshifts determined using \eazy \citep{brammer+2008}.  An interactive \fitsmap \citep{hausen+robertson2022} website displaying the images, catalogs, and NIRSpec slit overlays and extracted spectra is available via \url{https://jades-survey.github.io/viewer/}. Figure \ref{f.goodsn} shows the extent of the JADES NIRCam imaging in the release, visualized as a red-green-blue false-color image using F444W, F200W, and F090W, respectively. To provide a sense of the depth of the image, which reaches $\sim 30$~mag in some JADES filters, we show in Figure \ref{f.hdf} the JADES NIRCam F444W/F200W/F090W image of the Hubble Deep Field footprint. This image highlights the advance of JWST in sensitivity and resolution, providing a deep near-infrared view of the iconic Hubble Deep Field.

\begin{figure*}
  \begin{center}
  \includegraphics[width=0.9\textwidth]{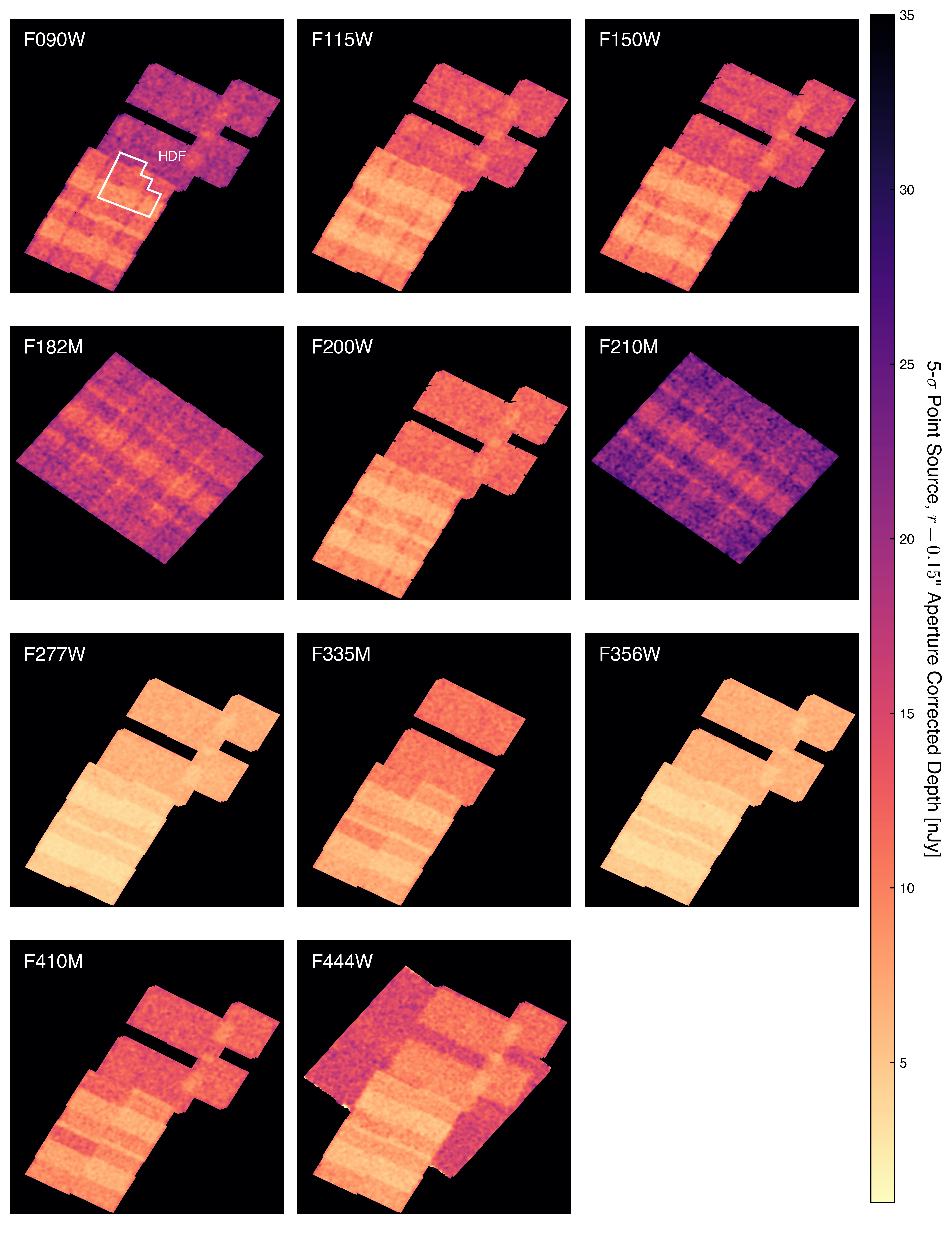}
  \end{center}
  \caption{Depth maps of the JWST/NIRCam imaging in the GOODS-N field. Shown are the local, aperture-corrected 5-$\sigma$ point source depths measured in circular apertures with radii $r=0.3$~arcsec. The F090W, F115W, F150W, F200W, F277W, F335M, F356W, and F410M images use data from the JADES Program. The F182M and F210M data were acquired by the FRESCO survey \citep{oesch+2023_fresco}. The deep F444W data are from the JADES Program, and the shallow/wide F444W tier is from FRESCO. The aperture correction is performed using the model point spread functions from \citet{ji+2023}. The color bar indicates the local depth in each filter, as measured in nJy. For reference, the outline of the Hubble Deep Field \citep{williams+1996} is shown in white on the F090W depth map.
  }\label{f.ncdepth}
\end{figure*}

In total, this JADES release covers 56 arcmin$^2$ of NIRCam imaging in GOODS-N, detecting 85709 distinct objects. A summary of the area, median depths, and median exposure times in each filter are provided in Table \ref{tab:depth}. Maps of the local 5-$\sigma$ point-source, aperture-corrected depths measured in circular apertures of radius $r=0.15$~arcsec are shown for each filter in Figure \ref{f.ncdepth}, along with the location of the HDF footprint for reference.

In releasing the JADES NIRSpec spectroscopic and NIRCam imaging data jointly, we note the unique synergy between these JWST datasets. Beyond the scientific synergies, the NIRCam data complement the NIRSpec data by providing targets, enabling consistency checks on the flux calibration of NIRSpec spectral modes, improved slit-loss corrections by providing information on source morphologies, and providing important checks on possible slit contamination by faint sources proximate to the primary NIRSpec targets. The design and execution of the NIRSpec spectroscopic surveys benefit from deep NIRCam imaging, and co-spatial NIRCam imaging data enhance the science return of complex NIRSpec MSA campaigns.

\begin{deluxetable*}{ccccc}
\tabletypesize{\footnotesize}
\tablecaption{GOODS-N JWST/NIRCam Depths\label{tab:depth}}
\tablehead{
  \colhead{Band} & \colhead{Area} & \colhead{Median Exposure Time} & \colhead{Median Depth$^a$} & \colhead{Median Depth}\\
  \colhead{} & \colhead{[arcmin$^{2}$]} & \colhead{[s]} & \colhead{[nJy]} & \colhead{[AB]}
}
\startdata
JADES F090W & 56.4 & 11338 & 9.8 & 29.92\\
JADES F115W & 56.4 & 22676 &6.9 & 29.30 \\
JADES F150W & 56.4 & 11338 &7.6 & 29.19\\
FRESCO F182M$^{b}$ & 63.0 & 12211 & 11.3 & 28.77\\
JADES F200W & 56.4 & 11338 & 6.4 & 29.38\\
FRESCO F210M$^{b}$ &  60.6 & 10558 & 14.4 & 28.50\\
JADES F277W & 55.5 & 11338 & 3.8 & 29.95\\
JADES F335M & 47.1 & 8503 & 6.0 & 29.45\\
JADES F356W & 55.5 & 11338 & 3.7 & 29.97\\
JADES F410M & 55.5 & 8503 & 7.7 & 29.19\\
JADES+FRESCO$^{b}$ F444W & 83.0 & 10393 & 6.4 & 29.38\\
\enddata
\tablecomments{$^a$ Median $r=0.15$~arcsec aperture corrected $5\sigma$ point-source depth. $^b$ This filter uses data from the FRESCO Program \citep{oesch+2023_fresco}.}
\end{deluxetable*}

\section{NIRSpec/MSA observations}
\label{sec:observations}

\begin{table*}
\tiny
\begin{center}
\caption{Summary of JADES NIRSpec/MSA observations. Under each disperser we report the (minimum/mean/maximum) exposure times; the minimum exposure time can be 0, due to disobedient shutters (for PRISM) and for protecting high-priority targets from overlap (for the gratings).}
\label{tab:obssummary}
\begin{tabular}{lccclcccccc}
  \hline
PID          &  Field  & Depth  & Selection & Tier name                     & PRISM & G140M & G235M & G395M & Targets & Release \\ 
             &         &        &           &                               &   [h] &   [h] &   [h] &   [h] &         &         \\ 
\hline
1210     & GOODS-S &   Deep & HST/JWST &        \verb|goods-s-deephst| & ( 7.8/16.5/28.0) & ( 2.3/ 4.1/ 7.0) & ( 2.3/ 4.1/ 7.0) & ( 2.3/ 4.1/ 7.0) & 253 & \citetalias{bunker+2023b} \\
\hline
1180$^a$ & GOODS-S & Medium &      HST &      \verb|goods-s-mediumhst| & ( 0.9/ 1.0/ 4.3) & ( 0.9/ 1.0/ 4.3) & ( 0.9/ 1.0/ 4.3) & ( 0.9/ 1.0/ 4.3) & 1342 & This work \\
1180     & GOODS-S & Medium &     JWST & \verb|goods-s-mediumjwst1180| & ( 0.3/ 2.1/ 5.2) & ( 0.9/ 1.8/ 4.3) & ( 0.9/ 1.8/ 4.3) & ( 0.9/ 1.8/ 4.3) & 533 & " \\
1181     & GOODS-N & Medium &      HST &      \verb|goods-n-mediumhst| & ( 0.6/ 2.0/ 6.9) & ( 0.9/ 1.0/ 3.5) & ( 0.9/ 1.0/ 3.5) & ( 0.9/ 1.0/ 3.5) & 853 & " \\
1181     & GOODS-N & Medium &     JWST &     \verb|goods-n-mediumjwst| & ( 0.3/ 1.6/ 5.2) & ( 0.9/ 1.7/ 5.2) & ( 0.9/ 1.7/ 5.2) & ( 0.9/ 1.7/ 5.2) & 709 & " \\
1286     & GOODS-S & Medium &     JWST &     \verb|goods-s-mediumjwst| & ( 0.5/ 2.1/ 2.2) & ( 0.7/ 2.1/ 2.2) & ( 0.9/ 2.4/ 2.6) & ( 0.9/ 2.4/ 2.6) & 169 & " \\
3215     & GOODS-S &   Deep &      HST &      \verb|goods-s-ultradeep| & ( 2.8/32.4/61.6) & ( 2.8/ 7.7/11.2) &        ---       & (11.2/23.0/33.6) & 228 & " \\
\hline
1286$^b$ & GOODS-S & Medium &     JWST &      ---                      &                  &                  &                  &                  &     & 2025 \\
1287     & GOODS-S & Deep   &     JWST &      ---                      &                  &                  &                  &                  &     & " \\
  \hline
\end{tabular}
\end{center}
$^a$ Two-thirds of these observations were affected by `shorts'. See~\ref{s.1180} and Appendix~\ref{a.shorts} for more details.
$^b$ This data release includes only 1 of 8 observations from PID~1286; the remaining 7 observations were obtained in Decmeber~2023 and will be part of a future data release.
\end{table*}

All observations used NIRSpec in Multi-Object Spectroscopy mode, with the NIRSpec/MSA \citep{ferruit+2022}. The MSA configurations were planned as described in Section~\ref{sec:sample} using the strategy detailed in \cite{eisenstein_overview_2023}.
For each visit, a set of target acquisition (TA) objects (stars and compact galaxies) were identified on the same images as those used for measuring the positions of the science targets. All TA objects were visually inspected to ensure they were compact, symmetric, and did not have color gradients or nearby sources. All TAs used the NIRSpec CLEAR filter and longest readout time (mode NRSRAPIDD6) because the GOODS fields only contain faint TA objects. Further details on the TA setups are provided in \cite{eisenstein_overview_2023}. All 35 TAs performed for JADES so far have been successful.

There were some technical issues with JADES NIRSpec observations that resulted in visits being skipped or having only partial data collection. The two sources of these issues are guide-star acquisition or re-acquisition failures from the Fine Guidance Sensor and `shorts' \citep[electrical short circuits with the NIRSpec MSA;][]{rawle+2022}. For cases where observations of a MSA configuration were partially successful, our strategy is to return with the same MSA configuration one year later to complete the data acquisition on the same targets. For cases where no data were obtained for a MSA configuration, we have replanned at either the same or at a different orientation. 

Table \ref{tab:obssummary} gives a summary of all JADES NIRSpec observations, including the typical integration time (we provide the minimum, mean and maximum integration time for each spectral configuration).
In the following subsections, we provide an outline of the observations that are the subject of this data release. The DR3 electronic files provide more details such as observation date and actual integration times per target.
We generally describe the observations with a label structured as `depth/selection', where depth is either `Medium', `Deep', or `Ultradeep', and selection is either `HST' or `JWST', depending on how the majority of targets was selected; these labels are then `translated' into the \verb|TIER| column in the published tables, and are part of the file names for the published spectra.

\subsection{1210: GOODS-S Deep/HST (+JWST)}\label{s.1210}

Observations from the first deep tier of JADES were already presented and extensively described in \citetalias{bunker+2023b}. We refer the reader to that article for all information concerning the observations in program~1210.

\subsection{1180: GOODS-S Medium/HST (+JWST)}\label{s.1180}

The intent of this program was to observe galaxies mainly selected from HST imaging, before any JWST NIRCam images were available. Six observations were planned, each consisting of two pointings offset by $\approx 7$~arcsec to dither over the short-wave detector gap in the parallel NIRCam observations. Unfortunately, these observations in October 2022 were heavily impacted by MSA shorts \citetext{see also \citealp{eisenstein_overview_2023} for the specific case of shorts in program 1180}. As a result, only four of the 12 target sets were successfully observed in 2022, with three more completed in October 2023.
Even though these failed observations were compensated with replanned observations (as described below), they can still be used to measure redshifts for sufficiently bright galaxies.
For this reason, we provide reduced data also for the shorts-affected observations.
Examples of shorts-contaminated data are provided in Appendix~\ref{a.shorts}; a `data-reduction flag' signals spectra affected by data-reduction problems; this includes all spectra from shorts-affected observations, including when the data quality was not severely affected (\verb|DR_flag|).
The time for the five remaining pointings was replanned as two observations targeted from NIRCam imaging, with target selection criteria closely matching that of GOODS-S Medium/JWST program 1286. These are referred to as 1180 GOODS-S Medium/JWST.

The seven completed GOODS-S Medium/HST pointings were observed in a single 3-point nodding pattern with each of the PRISM, G140M, G235M and G395M dispersers yielding total integration times of 3.8, 3.1, 3.1 and 3.1 ksec, respectively. The highest priority targets were observed in more than one pointing when possible, so some targets have longer integration times. 

The two replanned observations in 1180 are known as 1180 GOODS-S Medium/JWST. Due to the amount of time available, observation 134 has two sub-pointings whereas observation 135 has three. Each sub-pointing has the same dispersers and integration times as in Medium/HST above. The pointings were no longer constrained to be offset by 7~arcsec, and we adopted a strategy to maximize the exposure times of different objects.
These observations were executed in January 2023.

\subsection{1181: GOODS-N Medium/HST}

This program consists of four observations planned similarly to the six of GOODS-S Medium/HST, with each observation comprising two pointings offset by $\approx 7$~arcsec. However, for GOODS-N, the total time used with the PRISM per pointing was increased to 6.3 ksec, using two sets of the 3-point nodding pattern. All but one observations were successfully completed in February 2023.

\subsection{1181: GOODS-N Medium/JWST}

This program was planned using JWST NIRCam imaging for target selection. Four observations were planned, each with three sub-pointings offset by $\approx 1$~arcsec. Three of the observations were completed between April and May 2023. The fourth was delayed until late May by a guide-star acquisition failure and then partially impacted by MSA shorts. It is scheduled to be completed in May 2024. Each sub-pointing uses the PRISM, G140M, G235M, G395M and G395H dispersers with total integration times of 3.1 ksec each. The MSA configurations are designed to maximize commonality of the targets in each of the three sub-pointings to yield total integration times of 9.3 ksec per disperser.

\subsection{1286: GOODS-S Medium/JWST}

Program ID 1286 is the main GOODS-S Medium/JWST program. These observations were planned the same way as those of GOODS-N Medium/JWST above, each with three sub-pointings separated by $\approx 1$~arcsec. Only one of the eight observations was executed during Cycle 1 in January 2023. The remaining seven were observed in October and December 2023. Only the visit from January 2023 is included in this data release. For this visit the integration times per sub-pointing in some dispersers were reduced from 3.1 to 2.7 ksec to fit within the available allocation.

\subsection{3215: GOODS-S Ultra-Deep/JWST}\label{s.3215}

Program ID 3215 is a Cycle 2 GO program that builds on the parallel NIRSpec and NIRCam JADES observations in program 1210 \citep{eisenstein+2023b}. The NIRSpec MSA observation consists of five sub-pointings within 2~arcsec, close to the three sub-pointings of Deep/HST. For each sub-pointing the total integration time is 
33 ksec in each of PRISM and G395M and 8.3 ksec in G140M. Four sub-pointings were successfully executed in October 2023. The fifth suffered from a bright MSA short. All of the grating and 25\% of the PRISM exposures in visit 5 were unusable. These are scheduled to be re-observed in October 2024.

\begin{figure*}
  \includegraphics[width=\textwidth]{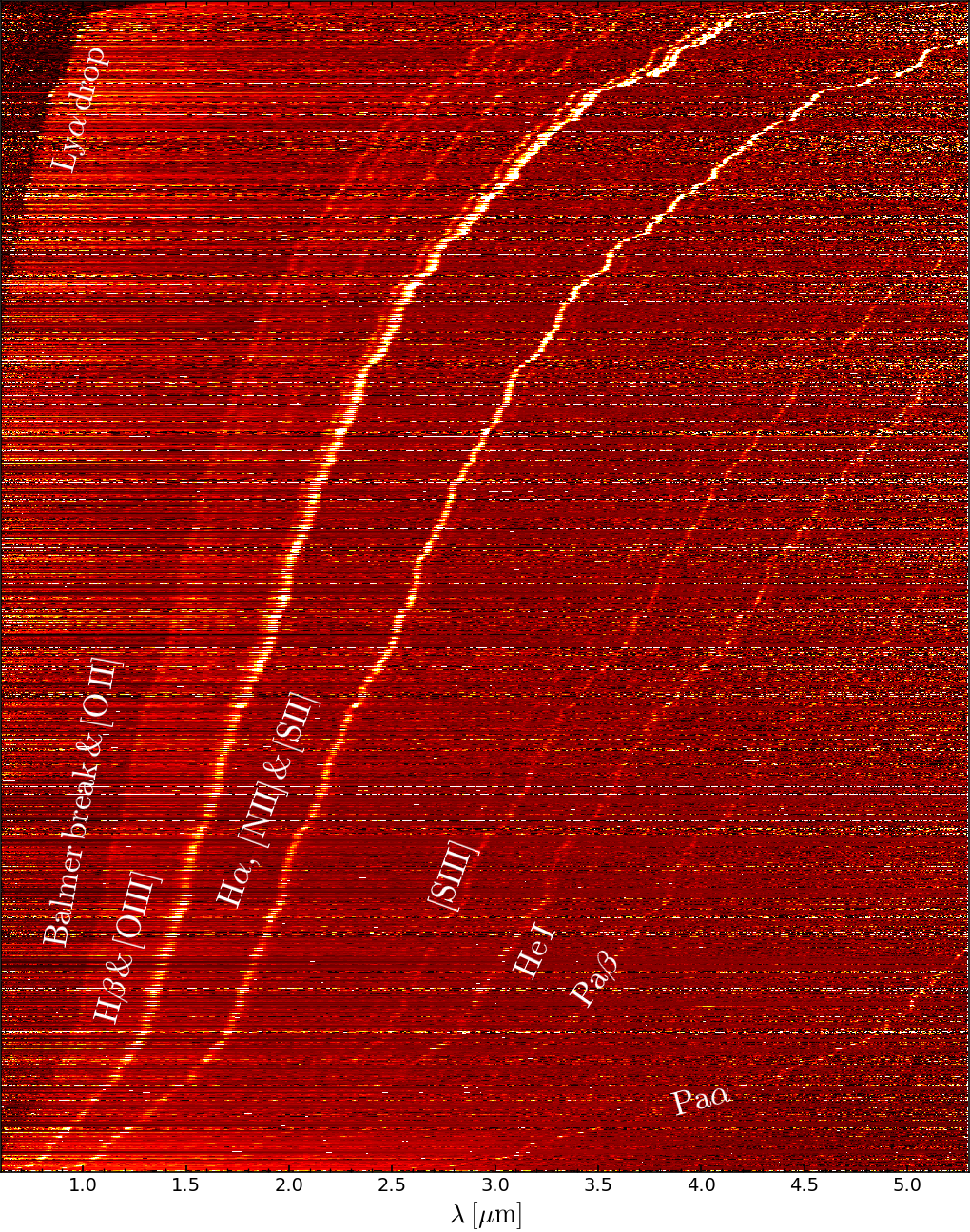}
  \caption{A subset of the observed spectra with multiple emission lines, sorted from bottom to top by increasing redshift. A number of continuum and line features is apparent, illustrating the simultaneous coverage of the rest-frame UV and optical ranges before cosmic noon.}\label{f.andy}
\end{figure*}

A subset of the observed spectra is shown in Figure~\ref{f.andy}, where the targets are in order of increasing redshift from the bottom to the top row. The varying noise level reflects the combination of exposures of different depth.

\section{Sample selection}\label{sec:sample}

As described in our first JADES NIRSpec data release paper \citepalias{bunker+2023b}, we employ a priority class system to most efficiently use the micro shutter array of NIRSpec (see \citealt{ferruit+2022}). The highest priority classes are reserved for objects that are few in number (i.e. having low sky density), typically targeting very high redshift candidates. We use these galaxies to optimize each pointing of NIRSpec (see Section~\ref{s.pointing}). Lower-priority classes contain many more galaxies, only a fraction of which actually get placed on shutters. With this, we aim to achieve a statistical sample, and JADES aims to span galaxy evolution from Cosmic Noon to within the Epoch of Reionization.
As outlined in \citet{eisenstein+2023a}, the JADES spectroscopy has a tiered `wedding cake' survey design, where a smaller number of deep pointings (with long exposure times) are supplemented with medium-depth pointings covering a larger area.
As JADES spans a range of science cases, there is not a single selection function for the spectroscopic sample. Instead, each tier has its own prioritization scheme (Tables~\ref{tab:priorities_Medium_HST}--\ref{tab:priorities_3215}). However, each of these schemes is structured in broadly the same way, with the exact criteria changing to reflect the differing input catalogs and differing sensitivity of the observations in each tier.

\citetalias{bunker+2023b} presented deep spectroscopic observations around the Hubble Ultra Deep Field, including details of the prioritization scheme used for target selection.
The data release presented in this paper predominantly introduces the medium tiers of JADES, and the new deep pointing from Program ID~3215.

Some spectroscopic observations were obtained before having JWST/NIRCam imaging. For these, we had to rely on targets selected from existing imaging, predominantly HST, augmented by data from other facilities - hereafter, we refer to observations planned in this way as `Medium/HST' (Section~\ref{sub:targets_medium_hst}).

Later observations benefited from JWST/NIRCam imaging, and targets were selected from these new datasets where possible. This is called Medium/JWST (Section~\ref{sub:targets_medium_jwst}). However, we note that in some Medium/JWST observations, the MSA footprint extended beyond the NIRCam coverage. These areas of the MSA had to be filled by HST-based catalogs with selection criteria detailed in Column~4 of Table~\ref{tab:priorities_Medium_JWST}, which was designed to mimic the Medium/JWST criteria as best as possible.

We note that Medium/JWST observations were typically deeper than Medium/HST, so although the overall aims of the two tiers are similar, the exact magnitude cuts are different. This is discussed in more detail below.

\subsection{Deep/HST}

The sample selection for program~1210 is described in \citetalias{bunker+2023b}.

\subsection{Medium/HST}\label{sub:targets_medium_hst}

Our observations span the two well studied fields GOODS-S and GOODS-N, meaning that our HST-based input target catalogs for our initial spectroscopic observations already had a high target density.

{\em GOODS-S:}
As described in \citetalias{bunker+2023b}, we assembled a HST-based catalog in GOODS-S by compiling $z>5.7$ candidates from multiple literature sources that had used a combination of Lyman break dropout criteria and/or photometric redshifts \citep{Bunker2004, YanWindhorst2004, Oesch2010, Oesch2013, Lorenzoni2011, Lorenzoni2013, Yan2010, Ellis2013, McLure2013, Schenker2013, Bouwens2015,Bouwens2021, Finkelstein2015, Harikane2016}. These were supplemented by targets of any redshift from large photometric catalog releases, including \citet{skelton+2014}, \citet{whitaker+2019}, \citet{rafelski+2015}, \citet{guo+2013}, \citet{caldwell+2008}, and \citet{coe+2006}.
Critical to assembling our HST-based catalog was ensuring the astrometry of these literature sources was accurate relative to the GAIA DR2 astrometric frame used in target acquisition. The details of how this was achieved, by relating catalogs to the Complete Hubble Archive for Galaxy Evolution (CHArGE) re-reduction of the HST imaging \citep{Kokorev2022_CHARGE, grizli}\footnote{\url{https://s3.amazonaws.com/grizli-stsci/Mosaics/index.html}}, are given in Appendix A of \citetalias{bunker+2023b}.

To establish photometry and a redshift for target prioritization, we cross-matched targets across each of these catalogs. HST broad-band magnitudes were adopted from the latest available catalog in which the target appeared out of \citet{whitaker+2019}, \citet{skelton+2014}, \citet{rafelski+2015}, and \citet{guo+2013}, or the discovery paper for targets not appearing in any of these catalogs. If the target did not have a reported magnitude, we remeasured the aperture photometry at the given coordinates.
For $z>5.7$ candidates, we adopt the photometric redshifts of the discovery paper. Other targets are assigned a photometric redshift according to the most recent catalog that reports one. We also cross-matched targets with literature spectroscopic redshift catalogs in GOODS-S \citep{vuds_specz, vvds_specz,kurk+2013, stark+2013, kriek+2015, morris+2015, momcheva+2016_3dhst_specz, inami+2017, herenz+2017, vandels_2018, pentericci+2018, maseda+2018}. Where we identify matches, these redshifts supersede any photometric redshifts, provided the quality flagging from those catalogs indicated that the redshift was based on either multiple high S/N emission lines, or a high S/N detection of an asymmetric \Lyalpha emission profile.

{\em GOODS-N:}
The GOODS-N HST-based catalog was assembled in largely the same way. Many of the $z>5.7$ selection papers listed above also extend their samples to GOODS-N \citep{oesch+2014, Bouwens2015, Bouwens2021, Finkelstein2015, Harikane2016}. Again, these were supplemented by targets from large HST photometric releases. \citet{skelton+2014} also covers GOODS-N and was included in our catalog, while we also add in targets from the \citet{barro+2019} catalog.
Astrometry was corrected following exactly the same procedure as in GOODS-S.

Photometry was taken from the most recent catalogs of \citet{barro+2019} and \citet{skelton+2014} if available, otherwise from the discovery paper, or, failing that, from our own remeasured aperture magnitudes.
Photometric redshifts were established in the same way as above, and these were again superseded by spectroscopic redshifts if a positional match was identified with a target in a literature catalog, as described above \citep{reddy+2006, barger+2008, stark+2011, stark+2013, kriek+2015, u+2015, momcheva+2016_3dhst_specz, maseda+2018}.

Additionally, for a small number of our Medium/HST observations in GOODS-N, one of the MSA quadrants partially extended beyond the main GOODS-N CANDELS footprint (see Figure~4 in \citealt{eisenstein+2023a}). 
CANDELS imaging only extended into this area in one band (F814W) of HST/WFC3 imaging \citep{grogin+2011}. Thus, this MSA real estate could not be populated with targets with robust photometric redshifts in this region. We identified eight objects in this area with X-ray counterparts from the Chandra Deep Field North X-ray survey \citep{xue+2016}, and these were included in class 3.0 (see Table~\ref{tab:priorities_Medium_HST}). We then populated the input catalog in this region with targets from more extended imaging, either from Spitzer/IRAC \citep{ashby2013} or ground-based imaging from Subaru \citep{capak+2004}.
The astrometry of these catalogs is less reliable, and thus, all of these targets were placed only in class 8, although $\sim$60 were observed since there was very little competition for shutter real estate in this part of the MSA, and these formed the vast majority of targets from this class that were observed. 

\begin{table*}
\caption{Target prioritisation categories for Medium/HST}  
\centering          
\begin{tabular}{c c c c }   
  \hline\hline
Priority & Redshift & Criteria & Targets observed   \\
\hline                    
  1   & $z>5.7$	& $F160W < 27.5$; V.I. Class$^{1}$ 0	& 68\\
  2.0 & $z>5.7$	& $F160W < 27.5$; V.I. Class 1  & \\
   & 	& $27.5 < F160W < 29$; V.I. Classes 0, 1 & 93 \\
  \\
  3.0 & $1.5 < z < 5.7$ & Rare target (e.g., Quiescent, AGN, ALMA\dots) & 27 \\
  3.5 & $1.5 < z < 5.7$	& $F160W < 23.5$ & 43 \\
  \\
  4.1 & $4.5<z<5.7$	& $F160W < 25.5$	& 14 \\	
  5.1 & $4.5<z<5.7$	& $F160W < 27$ & 87 \\
  6.1 & $4.5<z<5.7$	& $S/N(\Halpha) > 15$ & 44 \\
\\
  4.2 & $3.5<z<4.5$	& $F160W < 25.5$ & 45\\
  5.2 & $3.5<z<4.5$ & $F160W < 27$ & 148 \\
  6.2 & $3.5<z<4.5$ & $S/N(\Halpha) > 15$ & 63 \\
\\
  4.3 & $2.5<z<3.5$	& $F160W < 25.5$ & 122 \\
  5.3 & $2.5<z<3.5$	& $F160W < 27$ & 171 \\
  6.3 & $2.5<z<3.5$	& $S/N(\Halpha) > 15$ & 59 \\
\\
  4.4 & $1.5<z<2.5$ & $F160W < 25.5$ & 176 \\
  5.4 & $1.5<z<2.5$	& $F160W < 27$ & 219 \\
  6.4 & $1.5<z<2.5$	& $S/N(\Halpha) > 15$ & 54 \\
\\
  7   & $z > 1.5$	& Has GAIA2 coords and $F160W > 23.5$ & 265 \\
  7.5 & $z < 1.5$ &	Has GAIA2 coords	$23.5 < F160W< 27$ & 317 \\
  7.6 & $z < 1.5$	& Has GAIA2 coords	$F160W > 27$ & 87 \\
  8   & any $z$ & Anything else with $F160W > 23.5$ & 115 \\
\hline
\label{tab:priorities_Medium_HST}
\end{tabular}
\\
$^{1}${Targets were assigned one of the following visual inspection (V.I.) classes: (0) Most compelling, (1) Plausible $z>5.7$, but less compelling, (2) Real object but likely $z<5.7$, (-1) Reject.}
\end{table*}



{\em Medium/HST target prioritization:}
For each of GOODS-S and GOODS-N, these catalogs were then divided into priority classes for Medium/HST observations following the scheme outlined in Table~\ref{tab:priorities_Medium_HST}.
The highest priority targets in Medium/HST comprised high-redshift galaxies with photometric redshifts $z>5.7$, with the lower redshift cut corresponding to the `$i$-band drop-out' galaxies with the Lyman-break in the HST/ACS F775W filter (see \citealt{Bunker2004}). 
These targets were subdivided based on an F160W magnitude cut, prioritizing the brightest sources. Additionally, we performed a visual inspection of the HST imaging, placing the most robust candidates in our highest class. More marginal targets we retained in Class 2, and unconvincing targets were removed.

The next priority classes comprised targets with low surface density, including very bright ($m_{\rm F160W}<23.5$) $z>2$ objects (to obtain exceptionally high S/N spectra of a small subset), and any targets considered likely to be AGN, quiescent, or Lyman-continuum leakers at $1.5<z<5.7$.

To include more typical star-forming galaxies over this redshift range, we sub-divide our input catalog into four photometric redshift slices ($4.5\leq z<5.7$; $3.5\leq z<4.5$; $2.5\leq z<3.5$ and $1.5\leq z<2.5$), with the higher redshift slices having higher priority given that the surface density of these targets is typically lower.

Each redshift slice is then subdivided into three priority classes, the first two based on the HST F160W magnitude (which is the reddest available band in HST, and the best single-band approximation to a stellar-mass selection). This is then supplemented by a star-formation-rate based selection (using SFRs from the 3DHST catalog of \citealt{skelton+2014}). We converted the SFRs to \Halpha line fluxes, and then to expected $S/N$ for the \Halpha line accounting for typical slit losses of the MSA. Those predicted to have $S/N(\Halpha)>15$ in the NIRSpec prism were included in this priority class.

Class 7 then comprised any other literature source with coordinates that could be robustly tied to the GAIA-DR2 astrometric frame \citepalias[as described in][]{bunker+2023b} with a photometric redshift $z>1.5$, and sources at lower redshift (also with GAIA-DR2) subject to F160W magnitude cuts. Finally, we included anything else that was not at risk of saturating exposures in Class 8.

\subsection{Medium/JWST}
\label{sub:targets_medium_jwst}

For Medium/JWST, our primary target list came from the recent NIRCam images, which extends to longer wavelengths (and in many cases greater depth) than the HST images, and hence revealed more targets with more robust photometric redshifts at high redshift. The target selection criteria are detailed in Table~\ref{tab:priorities_Medium_JWST}

The JADES team produced photometric catalogs (see e.g. \citealt{rieke+2023}) and we performed SED fitting with \beagle and \eazy (see e.g. \citealt{hainline+2024}) to determine photometric redshift probability distributions for each target.
In addition to the SED-fitting-based photometric redshifts, we also used color cuts to identify Lyman-break candidates at $z\gtrsim5.7$.
We note that the photometric catalog was being frequently updated as imaging depth was added and NIRCam data reduction was improved. Target selection was always performed with the latest catalog available to us at the time of designing the observations. For this reason, in general, the input catalogs differ from catalogs associated with later public data releases of the imaging.

Throughout this section, where we refer to magnitudes, these are usually the `CIRC2' apertures from the photometric catalogs, which was selected to approximate the open area of a NIRSpec mirco-shutter. However, for Class $\geq7.0$ where we move to $F444W$-based selections, we switch to using larger apertures, to better approximate a stellar-mass-based selection. For GOODS-S, we used the Kron apertures, while for GOODS-N we used the CIRC4 aperture because the Kron aperture photometry was not available in GOODS-N at the time of target selection.

\begin{table*}
\caption{Target prioritization categories for Medium/JWST}
\centering          
\begin{tabular}{c c c c c }   
  \hline\hline
Priority & Redshift & Criteria (if JWST-based ) & Criteria (if HST-based) & Targets   \\
\hline                    
1 & $z > 8$ &
$m_{\rm UV} < 28.0$ (V.I. Class = 0)$^2$ &
 & 20 \\
2 & $z > 8$ &
$m_{\rm UV} < 28.0$ &
$z>8.5$, $F160W < 28.0$ & 7 \\
3 & $z > 8$ &
$28.0< m_{\rm UV} < 28.5$ &
$z>8.5$, $28.5<F160W < 28.0$ & 11 \\
\\
4 & $5.7 < z < 8$ & $m_{\rm UV}< 26.5$ or & & \\
 &  & L.E.$^{1}$
($F_{\rm line}\geq10^{-17.3}$)
 & & 19 \\
5 & $z > 2$ & $m_{\rm AB}<22$ 
& $2<z<5.7$ $F160W<22$ & 5\\
\\
6.0 \& 6.1 & $5.7 < z < 9$ &
$26.5<m_{\rm UV}<28$ or  & $F160W<28$ & \\
 &  &
L.E.$^{1}$ $10^{-17.8}<F_{\rm line}<10^{-17.3}$, 
 &  & 78 \\  
6.2 & $5.7 < z < 8.5$ &
$28<m_{\rm UV}<28.5$ & $F160W>28$ & 10 \\
\\
7.1 & $4.5<z<5.7$ &
UVJ  \& $F444W < 27$; X-ray sources &
$F160W < 28$ & 3\\
7.2 & $3.5<z<4.5$ &
UVJ  \& $F444W < 27$; X-ray sources &
$F160W < 28$ & 6\\
 7.3 & $2.5<z<3.5$ &
UVJ  \& $F444W < 27$; X-ray sources &
$F160W < 27.5$ & 14 \\
 7.4 & $1.5<z<2.5$ &
UVJ  \& $F444W < 27$; X-ray sources &
$F160W < 27.5$ & 19 \\
7.5 & $4.5<z<5.7$ &
$F444W < 27$ &
$F160W < 28$ & 65 \\
7.6 & $3.5<z<4.5$ &
$F444W < 27$ &
$F160W < 28$ & 108\\
 7.7 & $2.5<z<3.5$ &
$F444W < 27$ &
$F160W < 27.5$ & 177\\
 7.8 & $1.5<z<2.5$ &
$F444W < 26$ &
$F160W < 27.5$ & 155\\
7.9 & $1.5<z<2.5$ & $26<F444W<27$; or  & & \\
 & $z<5.7$ &  L.E.$^{1}$ $10^{-17.9}<F_{\rm line}<10^{-17.3}$  & & 50 \\
\\
\hline
8.0 \& 8.1 & $z>1.5$ & $F444W<28$\,mag or $S/N({\rm H}\alpha)>20$ & $F160W>28.5$ \& has GAIA2 coords & 352  \\
8.2 & $z<1.5$ & $F444W<28$\,mag & $24.5<F160W<29$ \& has GAIA2 coords & 171 \\
8.3 & no redshift cut & $F444W<29$ & $F160W>29$ \& has GAIA2 coords & 99\\
9 & --- & & $F160W>24.5$ 
& 46 \\
\hline
\label{tab:priorities_Medium_JWST}
\end{tabular}
\\
$^{1}${Strong line emitters (L.E., units erg\,cm$^{-2}$\,s$^{-1}$)were selected based on measurements from FRESCO or MUSE, or targets with a F410M excess.}
$^{1}${Visual inspection (V.I.) Class=0 are the most robust candidates.}
\end{table*}

We placed galaxies with photometric redshifts higher than 8 in the top priorities, with those brighter than $m_{AB}=28$\,mag in the rest-frame UV ranked top, followed by those with $28<m_{AB}<28.5$\,mag. 

Priority class 4 covers galaxies with $5.7<z<8$ with a magnitude cut on the filter which best approximates to the rest-frame UV around 1500\,\AA. As with the Deep/HST Class 4 in \citetalias{bunker+2023b} we set a relatively bright magnitude cut such that we would expect the rest-optical emission lines to be well detected to facilitate line ratio diagnostics (with $S/N(\Halpha)>25$ in the prism). For Medium/JWST, this rest-UV cut corresponds to $m_{AB}<26.5$\,mag, following the methodology in Section 2.1 of \citetalias{bunker+2023b}. We supplemented this class with some emission-line selected objects with fluxes $>10^{-17.3}$\,erg\,cm$^{-2}$\,s$^{-1}$ drawn from the FRESCO survey \citetext{\citealp{oesch+2023_fresco}, using a custom data reduction; \citealp{sun+2023}}, from MUSE \citep{inami+2017, herenz+2017} and galaxies exhibiting flux excesses in the F410M NIRCam filter consistent with strong line emission. 

Following this, the next class was a small number of very bright ($m_{AB}<22$\,mag) $1.5<z<5.7$ targets to enable continuum science. We then targeted more galaxies in the range $5.7<z<8$ but fainter than Class 4, split into two rest-UV magnitude bins ($26.5<m_{AB}<28$\,mag and $m_{AB}>28$\,mag). We supplemented this class with emission-line selected objects with fluxes below $<10^{-17.3}$\,erg\,cm$^{-2}$\,s$^{-1}$. In the very first epoch of Medium/JWST observations, taken in January 2023, we had sub-prioritised these line emitters as Class 6.0, slightly ahead of the $26.5<m_{AB}<28$\,mag magnitude cut (Class 6.1). But in all subsequent epochs, these were folded in together as Class 6.1.

For lower-redshift galaxies ($1.5<z<5.7$) we used the same 4 redshift slices as for Medium/HST. We created a class of upweighted sources with low target density comprising candidate passive galaxies \citep[selected through the UVJ color criterion,][and with $F444W<27$\,mag]{williams+2009,leja+2019a}, along with X-ray selected sources \citep{luo+2017}, and these were allocated in descending order of redshift slice. We then turned to the remainder of the galaxies in each redshift bin, where we used the reddest NIRCam wide filter F444W as a proxy for a stellar-mass-selected sample (selecting on $F444W<27$\,mag), placing each redshift slice in turn as before. We note that this differs from our Medium/HST selection (which is F160W based), as we take advantage of the availability of the redder filter to better approximate to a stellar-mass-limited sample.

Any unused MSA real estate was filled with fainter targets as described in Table~\ref{tab:priorities_Medium_JWST}, Classes 8.0--8.3 \& 9.
The exact criteria defining Classes $\geq8.0$ were not fixed across all epochs of the Medium/JWST survey. We note that these lower classes were never constructed with a view toward being able to conduct well-defined sample-based studies, and were rather aimed at potentially capturing a few extra worthwhile spectra with what is otherwise spare MSA real estate.
For our first Medium/JWST pointings in GOODS-S, Class 8.0 was made up of leftover targets for which, based on the UV magnitude, we predicted $S/N(\Halpha)>20$ based on the SFR, and Class 8.1 was remaining candidates with $F444W<28$~mag. However, it turned out that many of the targets in this Class 8.0 were pushing below the noise threshold such that this did not end up being a very successful selection, and in later iterations we did not keep this delineation.
As a result, we advise caution when considering Class $\geq8.0$ objects in the context of sample-based analyses.

\subsection{Ultra-deep 3215}
\label{sub:targets_ultra_deep}

Our primary input catalogs for the ultra-deep NIRSpec MSA observations came from the work of \citet{endsley+2023} at $6\lesssim z<9$ and 
\citet{hainline+2024} at $z>8$, both based on the JADES NIRCam imaging.
These were supplemented with photometric redshifts fits on all galaxies in NIRCam JADES catalog (down to a flux limit) using the \beagle and \eazy codes. These photometric redshifts were combined using a permissive consensus criterion to avoid missing good high redshift candidates.

The ultra-deep spectroscopy in this tier has long exposure times, and, based on our experience with the Deep/HST NIRSpec observation \citepalias{bunker+2023b}, redshifts could be obtained for galaxies as faint as $AB<30$\,mag.
Our highest priority was $z>11$ galaxies with confident photometric redshifts and magnitudes $AB<30$\,mag, followed by those with less confident redshift solutions. The next class was galaxies at $10<z<11$ and $AB<30$\,mag, followed by $8<z<10$ (again split into those with convincing photometric redshifts followed by those which had lower confidence).

\begin{table*}
\caption{Target prioritization categories for 3215 `UltraDeep'}
\centering          
\begin{tabular}{c c c c }   
  \hline\hline
Priority & Redshift & Criteria & Targets   \\
\hline                    
  1.1 & $z>11$ & $m_{AB}<30$ & 4\\
  1.2 & $z>11$ & $m_{AB}<30$ and less reliable phot-$z$& 0\\
    2.1 & $10<z<11$ & $m_{AB}<30$ & 0\\
2.3 & $8<z<10$  & $m_{AB}<30$ & 6\\
  2.4 & $8<z<10$ & $m_{AB}<30$ and less reliable phot-$z$& 2\\
  \\
  3.1 & & rare objects$^{1}$ & 5 \\
  3.2 & & rare objects$^{2}$ & 4\\
  \\
  4.1 & $7<z<8$ & $m_{AB}<30$ from Endsley et al. \\
  &    $5.7<z<8$ &     $m_{AB}<28.5$ from other phot-$z$  & 3\\
  4.2 & $5.7<z<7$ & $m_{AB}<30$ from Endsley et al.  & 8\\
  \\
  5.1 & $4<z<5.7$ & $m_{AB} < 28$  & 15 \\
  5.2 & $4<z<5.7$ & $m_{AB} < 29$  & 23\\
   6.1 & $5.7<z<8$ & $28.5<m_{AB} < 30$  & 14\\
  6.2 & $4<z<5.7$ & $m_{AB} < 30$  & 29\\
  7.1  & $2.5<z<4$ & $25<m_{AB}<28$  & 15\\ 
   7.2  & $2.5<z<4$ & $28<m_{AB}<29$  & 12\\ 
   7.3  & $1.5<z<2.5$ & $25<m_{AB}<28$  & 14\\ 
   7.4  & $1.5<z<2.5$ & $28<m_{AB}<29$  & 15\\ 
  7.5  & $z>1.5$ & $29<m_{AB}<30$  & 17\\ 
  8.1  & $z<1.5$ & $25<m_{AB}<28$ & 21\\ 
 8.2  & $z<1.5$ & $28<m_{AB}<29$ & 7\\ 
 8.3  & $z<1.5$ & $29<m_{AB}<30$ & 5\\ 
  9 & & class 9 objects in Deep/HST & 9\\
  \hline
\label{tab:priorities_3215}
\end{tabular}
\\
$^{1}${Rare objects includes: blue UV slopes, AGN $7<z<12$, MIRI $z>7$, X-ray $z>4$, medium-band $\log(line~flux/erg\,cm^{-2}\,s^{-1})>-18.3$}
\\
$^{2}${Rare objects includes: ALMA, MIRI $z<7$, AGN $4<z<7$, medium-band $\log(line~flux/erg\,cm^{-2}\,s^{-1})<-18.3$}
\end{table*}

The next class comprised galaxies with unusual properties, such as extremely blue rest-UV spectral slopes (2 targets), evidence for quiescence (2 targets), high-redshift AGN (2 targets), an object with a large flux excess in the F410M medium-band filter consistent with strong \OIIIL emission at $z\approx 7$, and FRESCO strong line emitters (2 targets).

We then considered the standard sample of objects, moving down in the redshift range. Class 4 includes targets in the range $5.7<z<8$ from the \citet{endsley+2023} catalog, subdividing these into sub-classes 4.1 and 4.2, based on a redshift cut of 7. These objects went down to our $m_{AB}<30$~mag cut. We supplemented sub-class 4.1 with $5.7<z<8$ targets from the full JADES catalog, for which we impose a brighter magnitude cut of $m_{AB}<28.5$~mag since objects fainter than this had less reliable photometric redshifts. 

We then descend in redshift to $4<z<5.7$ from our JADES photometric catalog, prioritizing $m_{AB}<28$~mag (sub-class 5.1) and then $28<m_{AB}<29$~mag (sub-class 5.2).
Then in class 6, we place leftover objects down to $m_{AB}<30$~mag in these $5.7<z<8$ and $4<z<5.7$ slices (as sub-classes 6.1 and 6.2).

We then move to Class 7 which places objects in redshift slices of $2.5<z<4$ and $1.5<z<2.5$, with each slice subdivided into two classes based on $m_{AB}<28$~mag and $28<m_{AB}<29$~mag.
Class 7.5 then places all remaining $z>1.5$ objects with $m_{AB}<30$~mag.

Class 8 then places $z<1.5$ objects divided into three magnitude slices ($m_{AB}<28$~mag, $28<m_{AB}<29$~mag, $29<m_{AB}<30$~mag). Finally, Class 9 contains filler objects from the HST-based catalogs described above.

\subsection{Target assignment with \empt and visual inspection}\label{s.pointing}

Target placement was performed using the \empt software \citep{bonaventura+2023} and proceeded using the same method as described in \citetalias{bunker+2023b}.
Pointing centers are driven by the highest priority class in each tier; Class~1 for Medium/HST and Medium/JWST, and Class~1.1 for Ultra-deep 3215.
For all candidates with $z>5.7$, we visually inspected the individual images and quality of the photometric fits before running the \empt to ensure they had good redshift fidelity. An inspection of the full input catalog of many tens of thousands of galaxies was not practical, but we did inspect everything that the \empt had allocated shutters to when designing trial MSA configurations. Sources that were badly contaminated by neighboring objects were removed, and the \empt re-run at the same location to assign new targets in place of those rejected (this was typically less than 10 objects per MSA pointing).
In allocating shutters, we require the centroid of the object to fall within an `admittance zone' as described in \citetalias{bunker+2023b}. For the low-dispersion prism, we do not allow the spectra of any target to overlap. As with our Deep/HST observations, we keep the same MSA configuration for the grating spectra as for the prism, which means that the higher-dispersion spectra (which are more extended on the array) do overlap, and we use the prism spectra to avoid confusion in line identifications. The grating spectra of a small number of sources (the highest priority sources, and very bright objects) are protected against overlap by closing the shutters of lower-priority targets in nearby rows. This means that a small number of objects are observed in the prism alone.

\section{NIRSpec/MSA Data Reduction}\label{s.datared}

The data reduction pipeline for this release is the same as in \citetalias{bunker+2023b}. However, the larger dataset enables us to discuss more in detail some of the calibration issues highlighted in DR1.

The pipeline is developed by the ESA NIRSpec Science Operations
Team (SOT) and Guaranteed Time Observations (GTO) NIRSpec teams \citep{alvesdeoliveira+2018,ferruit+2022}. Most of the processing steps are similar to those adopted by the STScI pipeline used to generate the MAST archive products, but the background subtraction, rectification, 1D extraction, and spectra combination steps have been optimized for the targets observed in JADES programs (see details in \citetalias{bunker+2023b}).
In particular, we apply a wavelength correction to compensate for the wavelength bias of non-centered compact sources.
This bias arises for sources that are smaller than the slit width, when they are spatially offset within the shutter along the dispersion axis. The issue is discussed in \cite{ferruit+2022}, and we apply the correction they propose (cf.~their Figure~9).

During the quality assessment of NIRSpec observations, we noted that some shutters failed to open when the MSA was configured at each pointing. These unexpected disobedient shutters might corrupt both the estimate of the background emission and the science spectrum during the data processing workflow. The impact of disobedient shutters is evident in the PRISM/CLEAR observations where the background and target emission are prominent.
Therefore, we initially analyzed the presence of these failed shutters by processing the data without background subtraction and identifying those shutters in which the signal is consistent with no emission. We removed such disobedient shutters from the MSA mask and re-processed the data following the standard procedure. 
Only 3~per cent of the targets are affected by disobedient shutters, reducing the total exposure time dedicated to the selected galaxy. In most cases, we just removed only one of the 3 shutters forming the target slitlet, but for a few targets, the number of disobedient shutters for the slitlet was 2 and even 3.

\begin{figure*}
  \includegraphics[width=\textwidth]{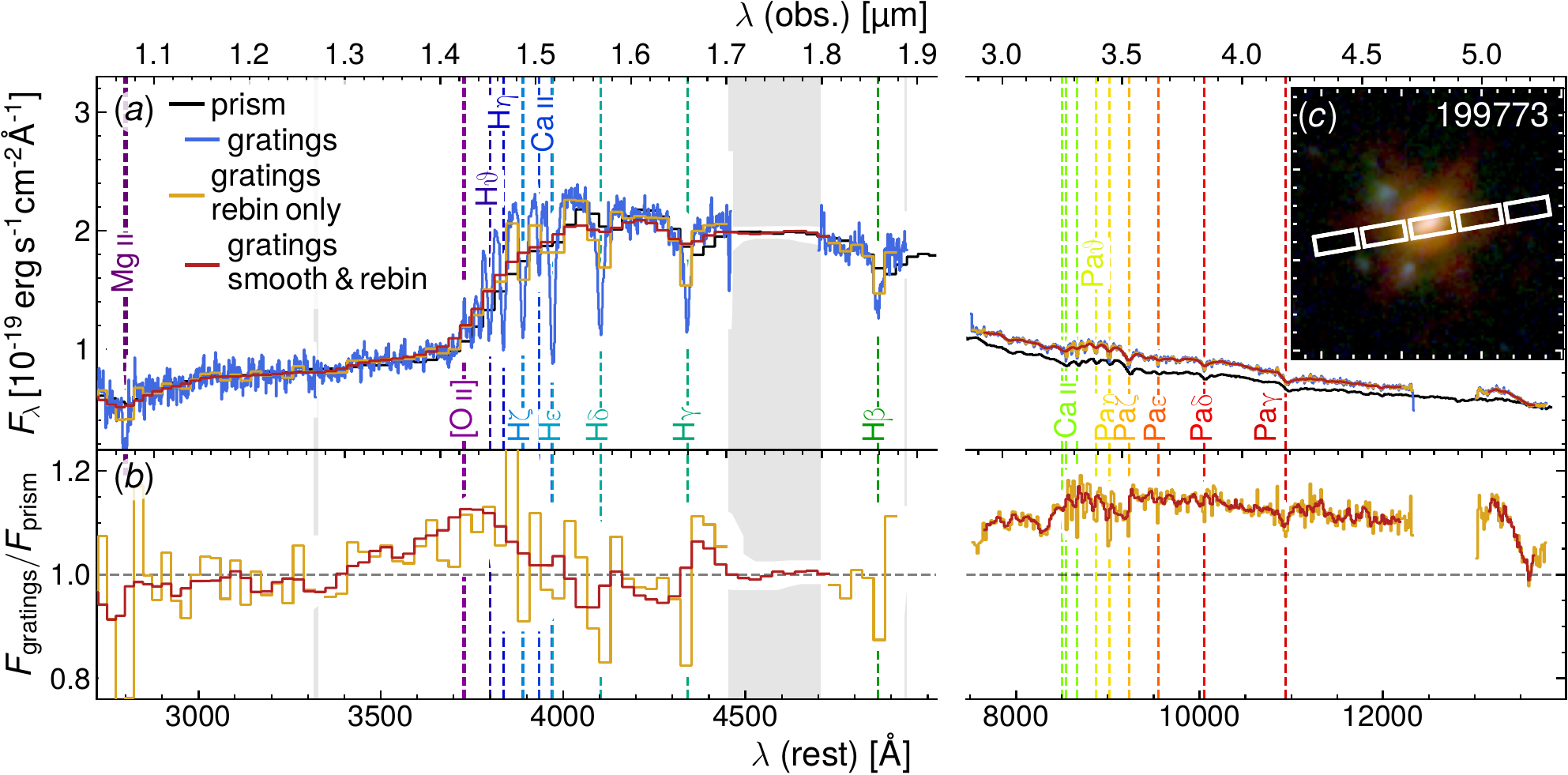}
  \caption{Target ID~199773 from PID~3215, illustrating remaining calibration problems. Panel~(a) shows the prism and gratings data (black and blue, G140M to the left and G395M to the right), and the gratings data after rebinning to the prism grid (yellow) and after matching the nominal resolution of the prism and then rebinning (red; vertical gray areas highlight spectral regions where we interpolated over missing grating data, due to bad pixels or the detector gap; the red line is clipped at the edges of the wavelength range due to the size of the convolution kernel). Panel~(b) shows the ratio between the rebinned and smoothed-then-rebinned grating spectrum and the prism spectrum (same line colors as panel~a).
  The flux calibration mismatch between prism and gratings is wavelength dependent, and is most severe in G395M.
  The galaxy image is shown in panel~(c), with the MSA shutters overlaid. From Z.~Ji~et~al. (in~prep.)}\label{f.obs.probs}
\end{figure*}

In Figure~\ref{f.obs.probs} we illustrate our current flux and wavelength calibration issues using 199773 (panel~c), a massive quiescent galaxy at redshift $z=2.8$, where we detect several stellar absorption lines in the medium-resolution gratings.
In panel~(a) we show the prism spectrum (solid black) and the combined G140M and G395M spectra (blue; no G235M observations are present in PID~3215).
The yellow line is the grating spectrum re-binned to the prism wavelength grid, the red line shows the grating spectrum after matching the resolution of the prism (using a Gaussian kernel) and then rebinning.
In principle, the red and black lines should overlap, but we can see substantial mismatches in both flux level and wavelength; this is illustrated in panel (b), where we show the ratio between the flux densities of the gratings and prism (after rebinning, yellow, and after smoothing and rebinning, red).
At wavelengths $\lambda<1.3$~\mum the flux levels agree to within 5~per cent, but at $1.3-1.5~\mum$ the G140M flux increases reaching a ratio of 10--15~per cent, before going down again from 1.5~\mum.
Being part of PID~3215, this object has no G235M data, but the G395M data show instead a higher flux mismatch of 15~per cent on average with respect to the prism, though with tapering both at the blue and red ends of the grating spectrum (see also Figure~\ref{f.highl.pablo} in Section~\ref{s.highl}, illustrating another source with continuum detected in the medium gratings).

At the same time, we confirm that overall the medium gratings require a lower redshift solution than the prism \citetext{Section~\ref{s.quality} and \citetalias{bunker+2023b}}.


\section{preliminary redshift identification}\label{s.visinsp}

Prior to running the flux measurement software, we measure an initial redshift estimate using a two-step process.
In the first step, we run the spectral modeling software \bagpipes \citep{carnall+2019} on the prism data. Our setup is optimized for time-efficient redshift measurement, by using a parametric star-formation history which may not fit well the stellar continuum.
This step returns both a redshift estimate and a fiducial model spectrum.
A detailed description of this procedure is available in \citetalias{bunker+2023b}; an example \bagpipes model fit is reported in Figure~\ref{f.inzimar} (orange line).
Each galaxy is then visually inspected by at least two team members, who use a rudimentary graphics interface to compare the \bagpipes model to all the available data, including the grating spectra, if available.
To assist in the decision, the interface displays a set of strong spectral features (Section~\ref{s.visz.ss.inzimar}).
In this step, the astronomers can change the redshift and assign a quality flag.

\subsection{Visual redshift determination}\label{s.visz.ss.inzimar}

\begin{figure*}
  \centering
  \includegraphics[width=0.9\textwidth]{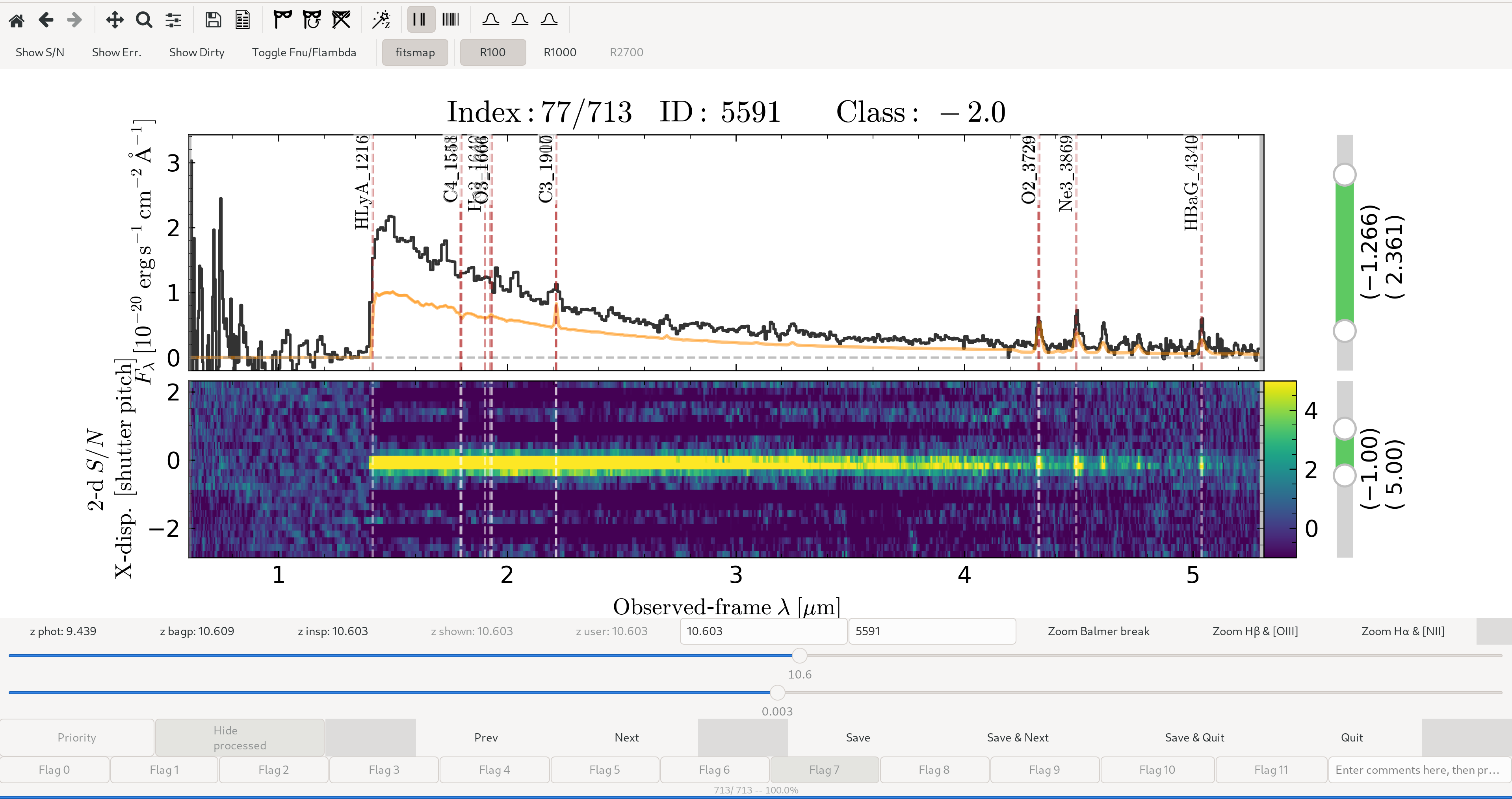}
  \caption{The interface used to visually inspect NIRSpec data, showing GN-z11 (ID~5591 in \mediumjwstgn; \citealp{oesch+2016,bunker+2023a,maiolino+2023a}).
  The console displays simultaneously the 1-d spectrum and 2-d S/N map.
  The user is able to move a set of bright spectral features to be used as reference (vertical dashed lines).
  The bottom row of  console is the set of flags the user can assign.}\label{f.inzimar}
\end{figure*}

The visual inspection is performed using a program which presents the user simultaneously with all the available information in a compact interface.
The console is shown in Figure~\ref{f.inzimar}, open on target ID~5591 \citep[GN-z11;][]{oesch+2016,bunker+2023a,maiolino+2023a} from \mediumjwstgn. The top panel shows the 1-d prism spectrum, with overlaid the \bagpipes model, and a set of reference spectral features (vertical dashed lines). Various buttons enable the user to display the 1-d S/N, the 1-d uncertainty, additional lines and, crucially, data from other dispersers.
The bottom panel shows the 2-d S/N map.
Finally, the console automatically opens \fitsmap \citep{hausen+robertson2022} centered on the current target; \fitsmap gives the user access to the panchromatic \jwst/NIRCam and \hst photometry \citep{rieke_jades_2023}, including the photometric redshifts based on \eazy \citep{brammer+2008,hainline+2024}.

The user is able to change interactively the redshift and judge different solutions. The outcome of this inspection is a user-validated redshift (or no redshift) and a set of flags (Table~\ref{t.zflags}).
Users can optionally enter comments; the only mandatory comment is to specify when there is a serendipitous source in the shutter. Typical comments include prominent or peculiar morphologies and doubt about alternative possibilities.

\begin{table}
\begin{center}
\caption{Flag values and meanings used in the visual inspection. These can be thought of as bit flags, hence in general a target has multiple flags.}\label{t.zflags}
\begin{tabular}{l|l}
    \cline{1-2}
    Value & Description \\ 
    \cline{1-2}
    0$^a$ & Not inspected   \\
    1$^b$ & Impossible to determine  \\
    2           & Tentative       \\
    3           & Peculiar$^c$      \\
    4           & From continuum  \\
    5           & Single prism line \\
    6           & Multiple prism lines \\
    7           & Multiple medium-grating lines \\
    8           & Multiple high-resolution grating lines \\
    9           & Prism data corrupted \\
    10          & Medium-resolution data corrupted \\
    11          & High-resolution data corrupted \\
    \cline{1-2}
\end{tabular}
\end{center}
$^a$ Cannot be combined with other flags.
$^b$ Cannot be combined with other flags, except 9, 10 and 11.
$^c$ Usually a serendipitous source in the shutter. User must enter a comment.
\end{table}

Each galaxy has been inspected by at least two people and by up to four. When the sample is fully inspected, the user sends the resulting catalog of redshifts and flags for merging. The catalogs are compared based on their redshift value.
Redshifts that agree to within a tolerance of a spectral pixel are averaged; targets that have different redshifts or redshift flags are re-inspected and a final decision is taken.
The resulting redshifts are then used as input in various analysis steps, ranging from emission-line measurements (as described in this article), to Bayesian spectral modeling (e.g., \beagle, \prospector).

\begin{figure}
  \includegraphics[width=\columnwidth]{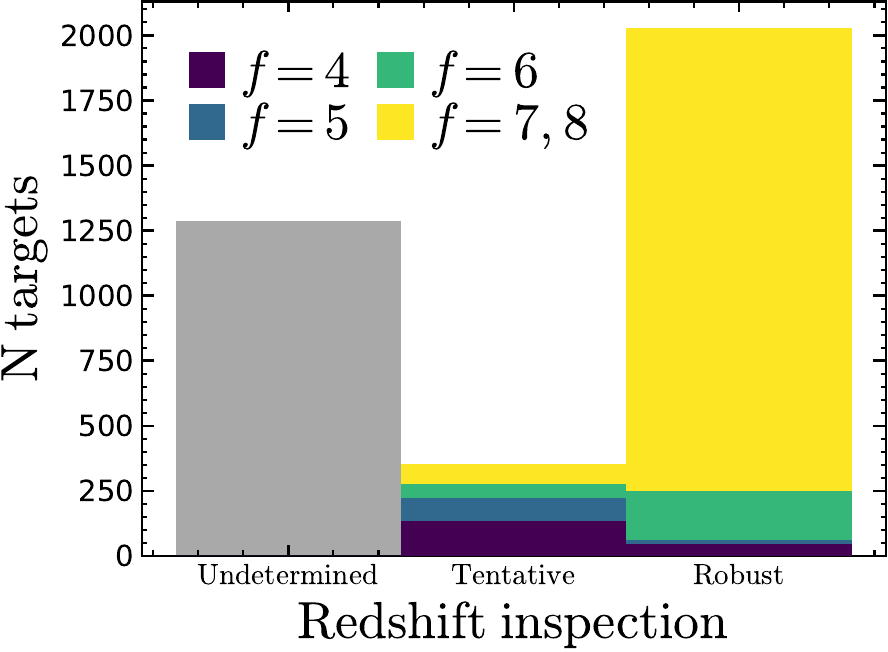}
  \caption{Summary of the visual inspection; the left column shows targets for which a redshift could not be determined (flag 1). Galaxies with uncertain redshifts (center) and with secure redshifts (right) are split by the highest-confidence redshift flag, colored as labeled. Overall, the current success rate of JADES redshifts is 65 per~cent.}\label{f.visz}
\end{figure}
The distribution of visual redshift flags from this procedure is illustrated in Figure~\ref{f.visz}.

\section{prism emission-line fluxes}\label{s.r100}

We use the spectral fitting software \ppxf \citep{cappellari2023}, to model the data as a linear combination of spectral templates.
As input templates, we use a set of simple stellar-population spectra from \fsps \citep[SSP;][]{conroy+2009}.
The spectra were calculated using \textsc{mist} isochrones \citep{choi+2016}, the C3K model atmospheres \citep{conroy+2019}, and a Salpeter initial mass function \citep{salpeter1955}. The spectral resolution is R=10,000 between $0.1<\lambda<3$~\mum; these templates are available from C. Conroy upon reasonable request.
We consider a subset of the templates grid spanning logarithmically ages $0.03\text{--}20$~Gyr and metallicities [Z/H] $-2.5\text{--}0$.
We adjust the age grid to each target, ensuring that the oldest available SSP is consistent with the age of the universe at the redshift of the target \citep[see e.g.,][]{looser+2023}.

In addition to these stellar templates, we use a set of Gaussian templates to represent nebular emission lines.
The gas templates are of three kinds; single Gaussians that represent individual emission lines that are spectrally isolated at any redshift (e.g., \HeIL[5787], \Pabeta), single Gaussians that represent multiple, spectrally blended lines (e.g., $\Halpha+\NIIall$, $\Hgamma+\OIIIL[4363]$), and doublet Gaussians representing doublets with fixed ratios (e.g., \OIIIall, \SIIIall).
A summary of emission-line templates and their redshift range is shown in Table~\ref{t.templates}. Note that the exact set of templates used depends on the source initial redshift; this is because the spectral resolution of the prism is a strong function of wavelength \citep{jakobsen+2022}, causing emission-line groups to be spectrally resolved or unresolved at different redshifts.
Moreover, we include a step function that is meant to capture very strong Balmer jumps \citep[e.g.,][]{cameron+2023c}.
All these templates are bound to have non-negative coefficients in the linear combination.
Finally, we use a 10\textsuperscript{th}-order multiplicative Legendre polynomial to adapt the shape of the continuum to the data; this can be thought of as a combination of physical effects (e.g., dust reddening) and flux calibration (e.g., incorrect slit-loss corrections, for extended objects and for objects with strongly wavelength-dependent morphology).
Before running \ppxf, each input template is smoothed to twice the spectral resolution of the data, the templates are truncated to match the approximate rest-frame wavelength range of the data, and stellar flux blue-ward of \Lyalpha is set to 0.
The templates are additionally convolved with a velocity distribution, modeled as a Gaussian.
We run \ppxf two times for each galaxy; in the first instance, we `tie' the templates in kinematic subsets, constrained to have the same velocity and velocity dispersion. The kinematic groups are: Balmer lines and stellar templates, rest-frame UV lines, rest-frame optical lines and rest-frame NIR lines.
After this first pass, all lines detected to at least  5\textsigma are kept, whereas the others are discarded.
In the second run, we fix the kinematics of the stellar continuum absorption, use only previously detected emission line templates, and remove almost all kinematic groups.
Exceptions to the latter rule are: the blend group formed by \SIIall and the blend $\Halpha+\NIIall$; the group formed by \Hbeta and \OIIIall; the group formed by \Hgamma and \OIIIL[4363]; and the group of \HeIL[10830] and \Pagamma, whose kinematics are always tied together.
We note that \HeIL[10830] is resonant, therefore this emission-line tends to be redshifted relative to the systemic velocity; however, leaving the line kinematics free relative to \Pagamma tended to produce bad fits due to low spectral resolution. Therefore, we opted to keep these lines tied.
These conditions track the setup of \citetalias{bunker+2023b}, and are necessary due to the limited spectral resolution of the prism, particularly in the range 1–2 \mum.
A difference with respect to \citetalias{bunker+2023b} is that we fix the flux ratio between the emission lines of the \OIIIall and \SIIIall doublets.
Other doublets with fixed line ratios are not enforced due to being unresolved
(e.g., \semiOIIIall) or blended with other lines (e.g., \NeIIIall).

After each fit, we post-process the line fluxes as follows. Below redshift $z<2$, we combine \Hbeta and \OIIIall and $\Halpha+\NIIall$ and \SIIall; the line uncertainties are added in quadrature.
Between $2\leq z < 5.3$, we combine the flux from the \OIIIall doublet.
Unlike for DR1, the flux of \Lyalpha is never provided, due to the difficulty of modeling the source continuum in the vicinity of this line. We refer the reader to G.~Jones (in~prep.) for \Lyalpha emitters.

The resulting best-fit spectra were visually inspected for artifacts and bad fits.
The most common of these are low equivalent width emission lines near the Balmer break, emission lines due to contaminants, and outliers, especially in shorts-affected observations.
The low equivalent width emission lines near the Balmer break arise when the shape of the break is not fit correctly, and the algorithm may use \OIIIall, \NeIIIall and \Hdelta to add to the continuum. Contaminants and artifacts may escape the sigma-clipping in \ppxf when they fall close to strong emission lines in the intended target.
All these instances were masked in the data table, and are flagged with a dedicated flag \verb|PRISM_flux_flag|.

\begin{table*}
\begin{center}
\caption{List of the emission lines fit in the prism spectra. All wavelengths are in vacuum.}\label{t.templates}
\begin{tabular}{ll|lcll}
    \hline
    & Line(s) & $\lambda$ [\AA] & $z$ range & Column name & Notes \\ 
    \hline
    & \CIVall                 & 1549.48         &      ---     & \verb|C4_1549|          & \\
    & \HeIIL+\semiOIIIall     & 1650.00         &      ---     & \verb|Blnd_He2_O3_1650| & \\
    & \CIIIall                & 1907.71         &      ---     & \verb|C3_1907| & \\
    & \MgIIall                & 2799.94         &      ---     & \verb|Mg2_2796| & \\
    & \OIIall                 & 3728.49         &      ---     & \verb|O2_3727| & \\
    & \NeIIIall               & 3869.86,3968.59 & $0 <z < 5.3$ & \verb|Ne3_3869|,\verb|Ne3_3968| & \\
    & \NeIIIL[3869]           & 3869.86         & $z \geq 5.3$ & \verb|Ne3_3869| & \\
    & \NeIIIL[3968]+\Hepsilon & 3968.59         & $z \geq 5.3$ & \verb|Ne3_3968|      & \\
    & \Hdelta                 & 3728.49         &      ---     & \verb|HD_4102| & \\
    & \Hgamma+\OIIIL[4363]    & 4341.65         & $0 <z < 5.3$ & \verb|Blnd_HG_O3 | & \\
\rdelim\{{2}{*}[] & \Hgamma                 & 4341.65         & $z \geq 5.3$ & \verb|HG_4341|  & \\
    & \OIIIL[4363]            & 4363.44         & $z \geq 5.3$ & \verb|O3_4363|  & \\
\rdelim\{{2}{*}[] & \Hbeta    & 4862.64         & $0 < z < 2 $ & \multirow{2}{*}{\texttt{Blnd\_HB\_O35007d}} & \\
    & \OIIIall                & 4960.30,5008.24 & $0 < z < 2 $ & & \\
\rdelim\{{2}{*}[] & \Hbeta    & 4862.64         & $2 \leq z < 5.3$ & \verb|HB_4861| & \\
    & \OIIIall                & 4960.30,5008.24 & $2 \leq z < 5.3$ & \verb|O3_5007d| & \\
\rdelim\{{2}{*}[] & \Hbeta    & 4862.64         & $z \geq 5.3$ & \verb|HB_4861| & \\
    & \OIIIall                & 4960.30,5008.24 & $z \geq 5.3$ & \verb|O3_4959|,\verb|O3_5007| & \\
    & \HeIL[5875]             & 5877.25         &      ---     & \verb|He1_5875| & \\
    & \OIall                  & 6302.05,6363.67 &      ---     & \verb|O1_6300| & \\
\rdelim\{{2}{*}[] & \Halpha+\NIIall & 6564.52   & $0 < z < 2 $ &  \multirow{2}{*}{\texttt{Blnd\_HA\_N2\_S2}} & \\
    & \SIIall                 & 6725.00         & $0 < z < 2 $ & & \\
\rdelim\{{2}{*}[] & \Halpha+\NIIall & 6564.52   & $ z \geq 2 $ & \verb|HA_6563| & \\
    & \SIIall                 & 6725.00         & $ z \geq 2 $ & \verb|S2_6725| & \\
    & \HeIL[7065]             & 7067.14         &      ---     & \verb|He1_7065| & \\
    & \SIIIall                & 9071.10,9533.20 &      ---     & \verb|S3_9069|,\verb|S3_9532| & \\
    & \Padelta                & 10052.12        &      ---     & \verb|PaD_10049| & \\
\rdelim\{{2}{*}[] & \HeIL[10829]            & 10832.06        &      ---     & \verb|He1_10829| & \\
    & \Pagamma                & 10940.98        &      ---     & \verb|PaG_10938| & \\
    & \Pabeta                 & 12821.43        &      ---     & \verb|PaB_12818| & \\
    & \Paalpha                & 18755.80        &      ---     & \verb|PaA_18751| & \\
    \hline
\end{tabular}
\end{center}
The set of templates used to fit any given galaxy depends on its initial redshift guess; this is because the spectral resolution of the prism is a strong function of wavelength \citep{jakobsen+2022}, causing emission-line groups to be spectrally resolved or unresolved at different redshifts. Empty redshift ranges indicate the template is used at all redshifts.
Rows connected by curly braces indicate emission-line pairs/groups that have tied velocity and velocity dispersion.

\end{table*}
\begin{table}
\begin{center}
\caption{Structure of the prism flux table. The full list of emission lines is reported in Table~\ref{t.templates}; all fluxes are in units of \fluxcgs[-18].}\label{t.prism}
\begin{tabular}{ll}
    \hline
    Column name               & Description \\
    \hline
    \texttt{NIRSpec\_ID}      & ID of the target in eMPT$^\ddag$ \\
    \texttt{TIER}             & Name of subset$^\ddag$ \\
    \texttt{PID}              & Program ID \\
    \texttt{RA\_TARG}         & Target right ascension [degrees] \\
    \texttt{Dec\_TARG}        & Target declination [degrees] \\
    \texttt{Field}            & Name of field (GS or GN) \\
    \texttt{NIRCam\_ID}       & ID of matched NIRCam source$^\dagger$\\
    \texttt{RA\_NIRCam}       & NIRCam right ascension [degree] \\
    \texttt{Dec\_NIRCam}      & NIRCam declination [degree] \\
    \texttt{x\_offset}        & Intra-shutter target offset [arcsec] \\
    \texttt{y\_offset}        & Intra-shutter target offset [arcsec] \\
    \texttt{ObsDate}          & Date of observations \\
    \texttt{Priority}         & Target Priority \\
    \texttt{assigned\_Prism}  & \texttt{True} if has prism observations \\
    \dots                     & \dots \\
    \texttt{assigned\_G395H}  & \texttt{True} if has G395H observations \\
    \texttt{nDither\_Pr}      & Number of dithers for prism \\
    \texttt{nDither\_Gr}      & Number of dithers for gratings \\
    \texttt{nInt\_Prism}      & Number of integrations for prism \\
    \dots                     & \dots \\
    \texttt{nInt\_G395H}     & Number of integrations for G395H \\
    \texttt{tExp\_PRISM}      & Exposure time for prism [s] \\
    \dots                     & \dots \\
    \texttt{tExp\_G395H}      & Exposure time for F290LP/G395H [s] \\
    \texttt{DR\_flag}         & \texttt{True} if problem in reduction or shorts \\
    \texttt{PRISM\_flux\_flag}  & \texttt{True} if at least one line flagged \\
    \texttt{z\_Spec}          & Redshift (both prism and gratings)      \\
    \texttt{z\_Spec\_flag}    & Redshift flag (both prism and gratings) \\
    \texttt{z\_PRISM}         & Prism-based redshift                    \\
    \texttt{C4\_1549\_flux}   & \CIVall flux             \\
    \texttt{C4\_1549\_err}    & \CIVall flux uncertainty \\
    \dots                     & \dots                                   \\
    \texttt{PaA\_18751\_flux} & \Paalpha flux                   \\
    \texttt{PaA\_18751\_err}  & \Paalpha flux uncertainty \\
    \hline
\end{tabular}
\end{center}
$^\ddag$ \texttt{NIRSpec\_ID} are not unique in the table, but the combination of 
\texttt{NIRSpec\_ID} and \texttt{TIER} is unique.
$^\dagger$ \texttt{NIRCam\_ID}s are unique, but whether they match the \texttt{NIRSpec\_ID}s depends on target selection (\hst vs \jwst selection), as well as on whether the NIRCam catalog was revised after the NIRSpec observation (which may result in sources being lost to blending and to crossing the non-detection threshold).
\end{table}

\section{medium-resolution gratings emission-line fluxes}\label{s.r1000}

We fitted the medium-resolution spectra using  \texttt{QubeSpec}'s\footnote{\url{https://github.com/honzascholtz/Qubespec}} fitting module. Each emission line was fitted using a single Gaussian component and the continuum was fitted as a power law. This simplistic approach is sufficient for describing a narrow range of the continuum around an emission line of interest ($\pm100 ~\AA$), because usually the continuum is poorly detected. The majority of the emission lines are fitted in isolation except for a group of emission lines that are close to each other. We show the full list of emission lines fitted in this work and the groups fitted together in Table \ref{grating.eml}.

\begin{table*}
\begin{center}
\caption{List of the emission lines fit in the medium-resolution grating spectra. All wavelengths are in vacuum. Rows connected by curly braces indicate emission lines that were fitted using the same redshift and FWHM during the same fit because they are sufficiently close in wavelength that the continuum can be modeled simultaneously. }\label{grating.eml}
\begin{tabular}{ll|ll}
    \hline
    & Line(s)                   & $\lambda$ [\AA]  & Column name \\ 
    \hline
    \rdelim\{{3}{*}[]& \CIVall  & 1549.48          & \verb|C4_1549|  \\
    & \HeIIL                    & 1640.00          & \verb|He2_1640| \\
    & \semiOIIIall              & 1663.00          & \verb|O3_1663|  \\
    & \CIIIall                  & 1907.71          & \verb|C3_1907|  \\
    \rdelim\{{2}{*}[] & \OIIall & 3728.49          & \verb|O2_3727|  \\
    & \NeIIIL[3869]             & 3869.86          & \verb|Ne3_3869| \\
    & \Hdelta                   & 3728.49          & \verb|HD_4102|  \\
\rdelim\{{2}{*}[] & \Hgamma     & 4341.65          & \verb|HG_4341|  \\
    & \OIIIL[4363]              & 4363.44          & \verb|O3_4363|   \\
\rdelim\{{2}{*}[] & \Hbeta      & 4862.64          & \verb|HB_4861| \\
    & \OIIIall                  & 4960.30,5008.24  & \verb|O3_5007| \\
    & \HeIL[5875]               & 5877.25          & \verb|He1_5875| \\
    & \OIall                    & 6302.05          & \verb|O1_6300| \\
\rdelim\{{3}{*}[] & \Halpha     & 6564.52          &  \verb|HA_6563| \\
    &   \NIIall                 & 6585.27, 6549.86 & \verb|N2_6584|   \\
    & \SIIall                   & 6718.29, 6732.67 & \verb|S2_6718|, \verb|S2_6732| \\
    & \HeIL[7065]               & 7067.14          & \verb|He1_7065| \\
    & \SIIIall                  & 9071.10,9533.20  & \verb|S3_9069|,\verb|S3_9532| \\
    & \Padelta                  & 10052.12         & \verb|PaD_10049| \\
    & \HeIL[10829]              & 10832.06         & \verb|He1_10829| \\
    & \Pagamma                  & 10940.98         & \verb|PaG_10938| \\
    & \Pabeta                   & 12821.43         & \verb|PaB_12818| \\
    & \Paalpha                  & 18755.80         & \verb|PaA_18751| \\
    \hline
\end{tabular}
\end{center}
\end{table*}

To estimate the model parameters we use \texttt{QubeSpec}, a Bayesian modeling code implemented with the Markov-Chain Monte-Carlo (MCMC) integrator \texttt{emcee} \citep{foreman-mackey+2013}. To measure the emission-line fluxes, we need to set prior probabilities for each of the variables. The peaks of the Gaussian profiles and the continuum normalization are given a log-uniform prior, while the FWHMs are set to a uniform distribution spanning from the minimum spectral resolution of the NIRSpec/MSA ($\sim$200 \kms) up to a maximum of 800~\kms. The prior on the redshift was a truncated normal distribution centered on the redshift from the visual inspection and with a standard deviation of 300~\kms and with a maximum allowed deviation of 1,000~\kms. 

For emission lines that lie in the overlap of the gratings, we fit both sets of the data and report the properties of the fit with the highest SNR.
We do not attempt to stack these spectral overlaps due to different line spread function and potential flux calibration offsets between the gratings. This will be further investigated in a future data release. 

We fit only a single Gaussian per emission line in the medium-resolution grating. We note that there are some objects with detected outflows or broad line regions. These fits will be further investigate in I.~Juod\v{z}balis et~al. (in~prep.) and S.~Carniani et~al. (in~prep.).

After the initial fitting run, we visually inspect every model for any incorrect fits or spurious line detection that are caused by unflagged outliers. These flagged fits are then refitted and re-inspected. The fluxes are calculated using the MCMC chains (after discarding the burn-in chains) and the final reported values and their uncertainties are the median value and standard deviation from the chains. We only report detections at SNR$>5$; for non-detections we report the 1-\textsigma uncertainties for the user to define their own upper limits. The final redshift from the medium-resolution spectra is the redshift inferred from the best detected emission line.

The structure of the gratings emission line catalog is presented in Table \ref{grating.eml_structure} in units of $\times 10^{-18}$ erg s$^{-1}$ cm$^{-2}$. The names of the individual emission lines column are the same as reported in Table \ref{grating.eml}. For the emission line doublets with fixed line ratios (such as \NIIall and \OIIIall) we only report the flux of the stronger emission line. Alongside the fluxes and their uncertainties, the initial rows are the same as for the prism table.

\begin{table}
\begin{center}
\caption{Structure of the gratings flux table. The initial rows are the same as for the prism
(between \texttt{NIRSpec\_ID} and \texttt{z\_PRISM}; cf.~Table~\ref{t.prism}); all fluxes are in units of \fluxcgs[-18].
}\label{grating.eml_structure}
\begin{tabular}{ll}
    \hline
    Column name               & Description \\
    \hline
    \texttt{NIRSpec\_ID}      & ID of the target in eMPT$^\ddag$ \\
    \texttt{TIER}             & Name of subset$^\ddag$ \\
    \dots                     & \dots                  \\
    \texttt{tExp\_G395H}      & Exposure time for F290LP/G395H [s] \\
    \texttt{z\_Spec}          & Redshift (both prism and gratings)      \\
    \texttt{z\_Spec\_flag}    & Redshift flag (both prism and gratings) \\
    \texttt{z\_PRISM}         & Prism-based redshift                    \\
    \texttt{C4\_1549\_flux}   & \CIVall flux             \\
    \texttt{C4\_1549\_err}    & \CIVall flux uncertainty \\
    \dots                     & \dots                                   \\
    \texttt{PaA\_18751\_flux} & \Paalpha flux                   \\
    \texttt{PaA\_18751\_err}  & \Paalpha flux uncertainty \\
    \hline
\end{tabular}
\end{center}
$^\ddag$ \texttt{NIRSpec\_ID} are not unique in the table, but the combination of 
\texttt{NIRSpec\_ID} and \texttt{TIER} is unique.
\end{table}

\section{Quality assessment}\label{s.quality}

\subsection{Redshift combination and comparison: prism vs medium gratings}

\begin{figure}
  \includegraphics[width=\columnwidth]{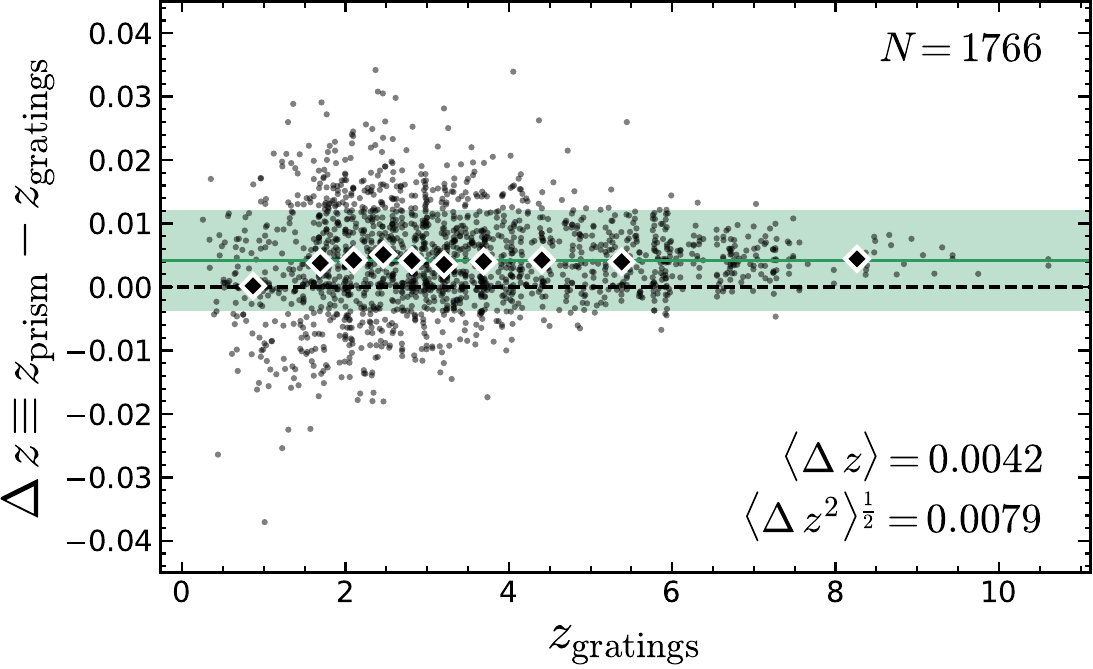}
  \caption{Comparison of redshifts between the prism and the medium-resolution gratings. We find a redshift-independent offset $\Delta\,z=0.0042$ (blue line), consistent with \citetalias{bunker+2023b}; the shaded region is the standard deviation.
  The increased dispersion at low redshifts is expected from the strong dependence of the prism spectral resolution with wavelength. The black diamonds are the moving median; there is some evidence for a reduced bias around $z<1$.
  }\label{f.redshift.zcomp}
\end{figure}

In Figure~\ref{f.redshift.zcomp} we compare the redshift measurements from the prism and from the medium gratings, where both are available.
Having defined $\Delta z \equiv z_\mathrm{prism} - z_\mathrm{gratings}$, we find a mean of $0.0042\pm0.0002$ and a standard deviation of 0.079, consistent with the findings of \citetalias{bunker+2023b}.
This statistically significant offset points to a residual wavelength calibration problem in the prism or medium gratings.
We find the offset to be redshift independent; a line fit with the robust least trimmed squares algorithm
\citep[using the python implementation {\tt \href{https://pypi.org/project/ltsfit/}{ltsfit}};][]{rousseeuw+driessen2006,cappellari+2013a} is consistent with a flat slope of $0.0002\pm0.0001$.
The increased scatter at low redshift is a consequence of the rapid increase of the prism spectral resolution with wavelength; multiplying $\Delta z$ by $R/(1+z)$ the scatter becomes approximately uniform with $z$ (where $R$ is the prism spectral resolution of the \OIIIL line at redshift $z$, and $z$ is the redshift of each source).

Whenever we have a strong line detection (5~\textsigma) in the medium gratings, we adopt the redshift of this line as the object redshift (flag A).
Using a single emission line is warranted because medium-resolution fits are made only for galaxies with a visual-inspection flag of 7 (Section~\ref{s.visinsp}). We checked that there are no cases where the grating and prism spectra disagree by more than $\Delta\,z = 0.05$, so we can rule out any mis-identified lines. Large offsets ($|\Delta\,z-\langle \Delta\,z\rangle|>0.015$ (Figure~\ref{f.redshift.zcomp}) were visually inspected, and are mostly due to uncertainties in the \Hbeta-\OIIIall blend and to low signal-to-noise data.
If no lines have been detected in the medium gratings, we use the prism redshift, requiring at least two emission lines for a secure redshift (flag B), or the combination of a line and/or a strong continuum break (for a less secure or less precise redshift, flag C).
An even lower class is reserved for redshifts identified as tentative in the visual inspection; in this case, we report the visual-inspection redshift (flag D). All other redshifts are assigned -1 (flag E).
To summarize, the final redshift flags are:
\begin{compactitem}
  \item[A] Redshift from at least one emission line in the medium-resolution grating.
  \item[B] Redshift from two or more prism emission lines.
  \item[C] Redshift from the continuum, or from the continuum and a single prism emission line.
  \item[D] Tentative, from visual inspection.
  \item[E] No redshift.
\end{compactitem}
Note that the first three flags are the same as in \citetalias{bunker+2023b}.

\begin{figure}
  \includegraphics[width=\columnwidth]{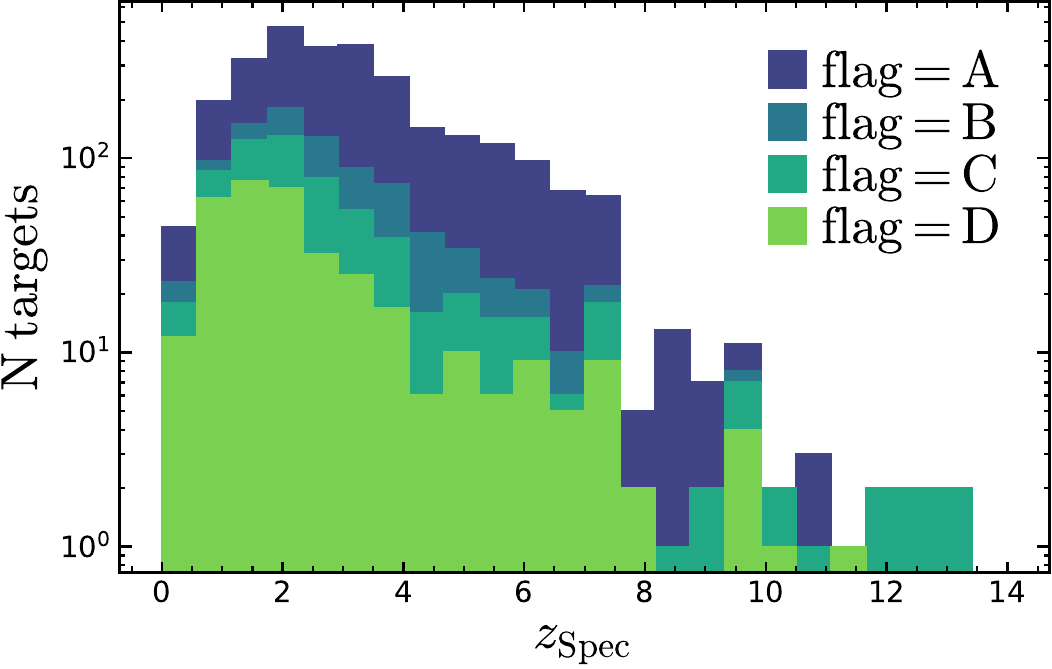}
  \caption{Redshift distribution of the sample, color-coded by the final redshift flag (Section~\ref{s.quality}).
  We note a drop in redshift distribution at $z_\mathrm{Spec} \sim 7.5$; this reflects a similar drop in the
  distribution of photometric redshifts of the targets selected for observation.
  }\label{f.finalz}
\end{figure}

\begin{figure}
  \includegraphics[width=\columnwidth]{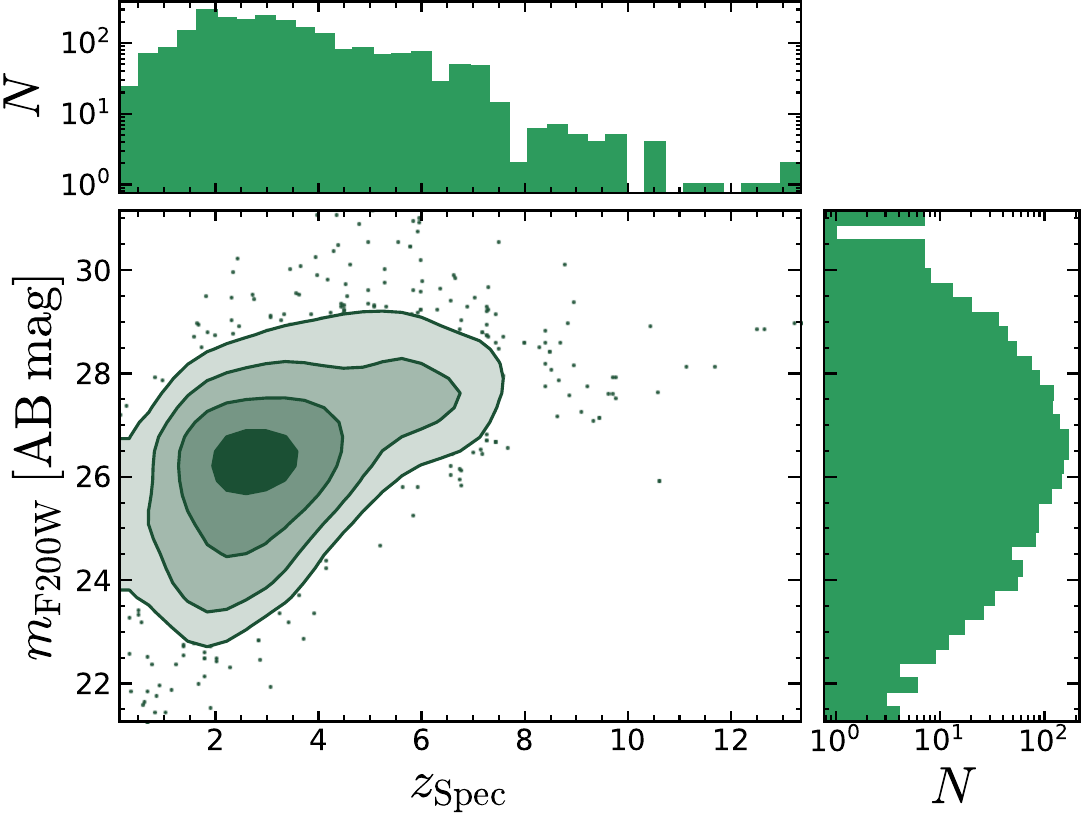}
  \caption{Redshift vs magnitude distribution of the sample; NIRSpec deep spectroscopy can measure redshifts for targets fainter than 30~mag.
  }\label{f.zmag}
\end{figure}

The combined redshift distribution of the sample is shown in Figure~\ref{f.finalz}, color-coded by flag. There is a drop in the distribution at $z_\mathrm{Spec} \sim 7.5$, which reflects at least in part a similar dearth of targets in the distribution of photometric redshifts of the targets selected for observation.
The overall distribution of the spectroscopic sample vs magnitude is displayed in Figure~\ref{f.zmag} (we show only targets with flag A--C, and with a secure match in NIRCam). The effect of the cosmic evolution of the luminosity function is clearly visible. There is a lack of galaxies fainter than 29~mag at redshifts lower than $z_\mathrm{Spec}\lesssim 2$ and higher than $z_\mathrm{Spec}\gtrsim 9.5$; this is caused by both sample selection and sensitivity as follows.
At low redshifts, NIRCam photometry becomes less able to clearly distinguish line excesses, because the spacing of strong emission lines reduces as $1+z$; the lower sensitivity of NIRSpec at wavelengths $\lambda < 1~\mum$ compounds the problem.
At redshifts higher than $z_\mathrm{Spec}\approx 9.5$, instead, the strongest emission lines (\OIIIall) are redshifted out of the NIRSpec coverage, so redshifts measurements rest solely on the \Lyalpha drop and on less prominent lines -- both of which are harder to detect in faint targets.

\begin{figure}
  \includegraphics[width=\columnwidth]{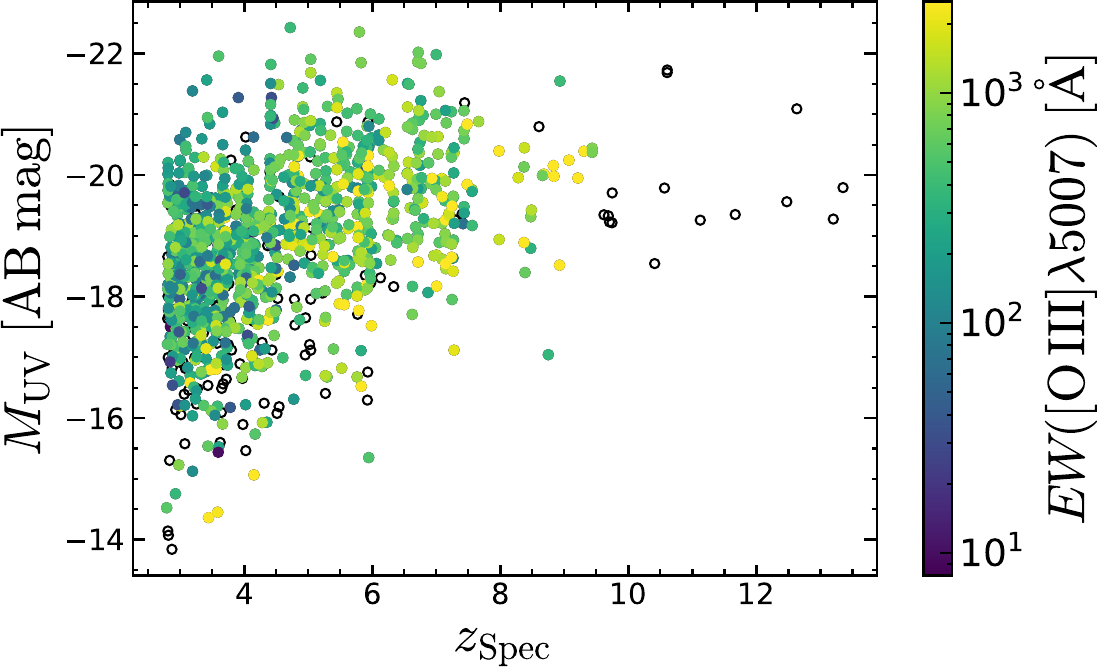}
  \caption{Redshift vs UV magnitude distribution of the sample, color-coded by the equivalent width of \OIIIL.
  Magnitudes were calculated directly from the prism spectra, using aperture corrections estimated by comparing the
  prism magnitude to the 0.35-arcsec NIRCam magnitude in the filter nearest to rest-frame 1,500~\AA.
  \OIIIL falls outside of the NIRSpec wavelength range at $z\gtrsim9.5$.
  }\label{f.zmuv}
\end{figure}

In Figure~\ref{f.zmuv} we show $M_\mathrm{UV}$ vs redshift, for the sample where magnitudes could be measured directly from the NIRSpec data; to this end, we used a nominal top-hat filter between rest-frame 1,400 and 1,600~\AA. The resulting magnitudes were corrected for aperture effects upscaling by the ratio between the observed and synthetic magnitude in the NIRCam band nearest in wavelength. The color coding is the equivalent width of \Halpha+\NIIall, measured directly on the prism data.
There are clear trends of EW with both $M_\mathrm{UV}$ at fixed redshift and with redshift at fixed $M_\mathrm{UV}$; the first trend arises from the sub-linear slope of the star-forming main sequence, where galaxies have lower specific star-formation rate with increasing stellar mass. The second trend follows the decreasing normalization of the star-forming sequence with increasing cosmic time.

\subsection{Flux comparison: prism vs medium gratings}

\begin{figure*}
  \includegraphics[width=\textwidth]{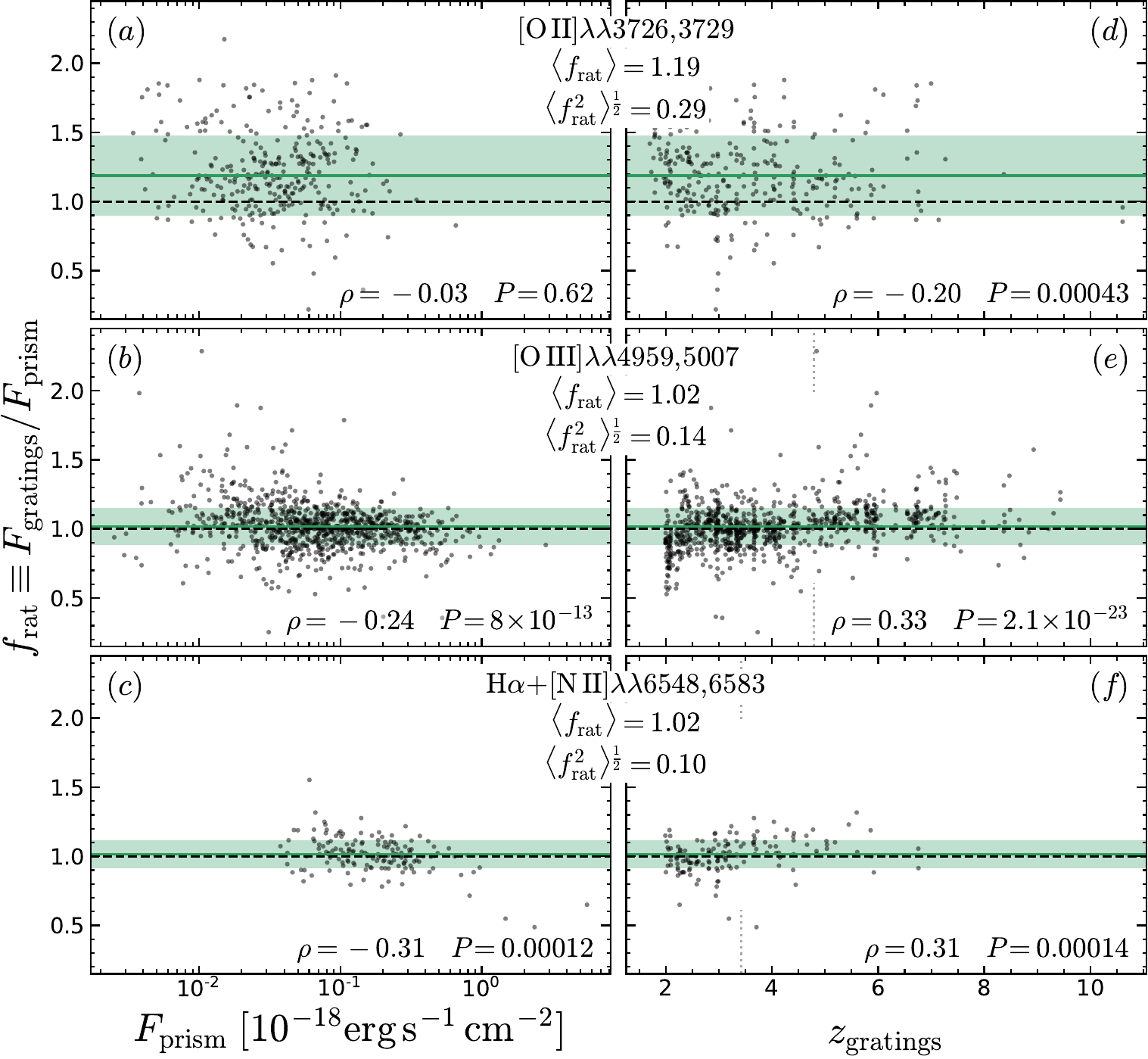}
  \caption{
  Comparison of the emission-line fluxes measured from the prism and from the medium-resolution gratings, as a function of prism flux (left column) and redshift (right column). The central inset between the columns are the median and standard deviation for each emission-line complex (also displayed as green horizontal lines and shaded regions), while the bottom right corner of each panel reports the Spearman rank correlation coefficient and associated p-value. \OIIall (panels~a and~d) shows the regime where the gratings-to-prism flux ratio $f_\mathrm{rat}$ is dominated by systematics in the continuum model; \OIIIall (panels~b and~e) illustrates with high significance the strong correlation between $f_\mathrm{rat}$ and redshift, which indicates the flux-calibration mismatch between prism and gratings is most severe in G395M. The dotted vertical lines in panels~(e) and~(f) show the redshift where \OIIIL and \Halpha are observed at 2.9~\mum -- the bluest wavelength captured by the G395M/F290LP disperser/filter combination.
  Finally, \Halpha+\NIIall shows that the total fluxes measured in the prism (where we do not separate \Halpha from \NIIall) match well the fluxes measured in the gratings (where the emission-line complex is well resolved); the trends with flux and redshift are consistent with what we see for \OIIIall.
  }\label{f.redshift.fcomp}
\end{figure*}

In Figure~\ref{f.redshift.fcomp} we compare the flux measurements from the medium gratings to the corresponding measurements from the prism; the top, middle and bottom rows show respectively \OIIall, \OIIIall and \Halpha+\NIIall.
For \OIIIall, we consider only galaxies at $z>2$, wherein the prism catalog the doublet is clearly separated from \Hbeta; for \Halpha+\NIIall, we take the grating measurements of \Halpha and \NIIL and add them, upscaling \NIIL by 1.34 to take into account \NIIL[6548].
For each set of emission lines, we consider the ratio $f_\mathrm{rat} \equiv F_\mathrm{gratings}/F_\mathrm{prism}$, and study this value as a function of $F_\mathrm{prism}$ (left column) and redshift; due to the requirement to have both prism and medium-grating measurements, redshift is always $z_\mathrm{gratings}$.

For \OIIall, we find that the grating fluxes are 17~per cent higher than the prism values, with large scatter (25 per cent) and a significant decreasing trend with redshift.
We interpret this discrepancy as due to how the continuum is modeled in the prism; in particular, using an incorrect value of the spectral resolution can significantly affect the recovered \OIIall flux in the low-resolution regimes found at low redshift.

For \OIIIall, we find good agreement (median ratio 1.01) but a large scatter (albeit smaller than for \OIIall, 14~per cent).
We believe the better agreement is due to the fact that the galaxy continuum around 5,000~\AA is relatively featureless, compared to the region of \OIIall.
We find a decreasing trend with $F_\mathrm{prism}$ (panel~b) and an increasing trend with $z_\mathrm{grating}$ (panel~d).
A partial-correlation analysis confirms that the redshift correlation is the main one, and that the flux correlation arises from the anti-correlation between flux and redshift.
The fact that the ratio increases with redshift is in agreement with the findings of Figure~\ref{f.obs.probs}, where it seems that the flux discrepancy between the prism and medium-resolution gratings is smallest in G140M and highest in G395M.

Finally, panels~c and~f show \Halpha+\NIIall; the results here are consistent with what found for \OIIIall.
The significant correlation between $f_\mathrm{rat}$ and flux is driven by the outliers at $F_\mathrm{prism}>60$~\fluxcgs[-18]; removing these points also removes the correlation.

Figure~\ref{f.redshift.fcomp} shows jumps in $f_\mathrm{rat}$ at certain values of $z_\mathrm{grating}$; for example, in Figure~\ref{f.redshift.fcomp}e these jumps happen at $z_\mathrm{grating}\approx 2.5$ and 4.7, which correspond to when the observed wavelength of \OIIIall moves respectively from G140M into G235M, and from G235M into G395M.
To estimate the average flux calibration offset in our data, we divide the sample in three redshift bins, determined by when \OIIIall (our brightest line on average) is observed with G140M, G235M or G395M (respectively, $0<z_\mathrm{grating}<2.5$, $2.5<z_\mathrm{grating}<4.7$ and $z_\mathrm{grating}>4.7$).
In these bins, we find an \textit{average} value of $f_\mathrm{rat}$ of $0.90\pm0.03$, $1.00\pm0.01$, and $1.10\pm0.02$. However, the scatter is large, and galaxies with a clearly detected continuum (e.g., Figs.~\ref{f.obs.probs} and~\ref{f.highl.pablo}) show that the flux-calibration mismatch is wavelength dependent.
We did not find statistical evidence for a dependence of $f_\mathrm{rat}$ on the target location on the MSA.

To further investigate this discrepancy between the medium gratings and prism we compared the fluxes of emission lines that are observed in two different gratings. This occurs for lines in the region of the spectrum probed by two configurations simultaneously; i.e. $1.6 \lesssim \lambda \lesssim 1.8~\mum$ for G140M/F070LP and G235M/F170LP, and $2.9 \lesssim \lambda \lesssim 3.1~\mum$ for G235M/F170LP and G395M/F290LP. We find that the average flux ratios are $f_\mathrm{G140M/G235M} = 1.07\pm0.01$ and $f_\mathrm{G395M/G235M} = 1.01\pm0.01$.
These results seem to contradict the findings from comparing the emission-line fluxes from the gratings to the prism (where G140M is lower than the prism, G235M is consistent, and G395M is higher than the prism; Figure~\ref{f.redshift.fcomp}). However, this analysis focuses on a specific region of the wavelength range, where two gratings overlap, whereas the emission-line comparison spans the entire wavelength range of NIRSpec. As shown in Figure~\ref{f.obs.probs}, the flux-calibration bias between the prism and G395M seems to be wavelength-dependent.

\subsection{Accuracy of the wavelength calibration}\label{s.wavecal}

\begin{figure*}
  \includegraphics[width=\textwidth]{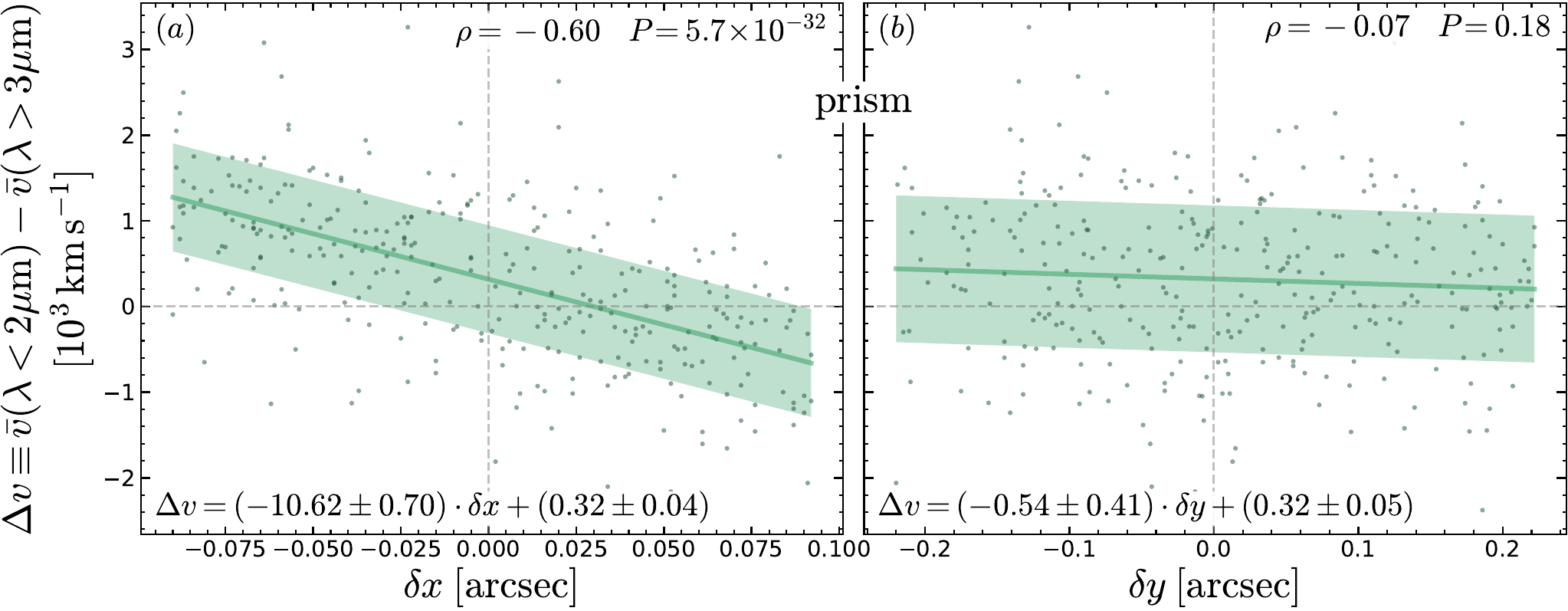}
  \includegraphics[width=\textwidth]{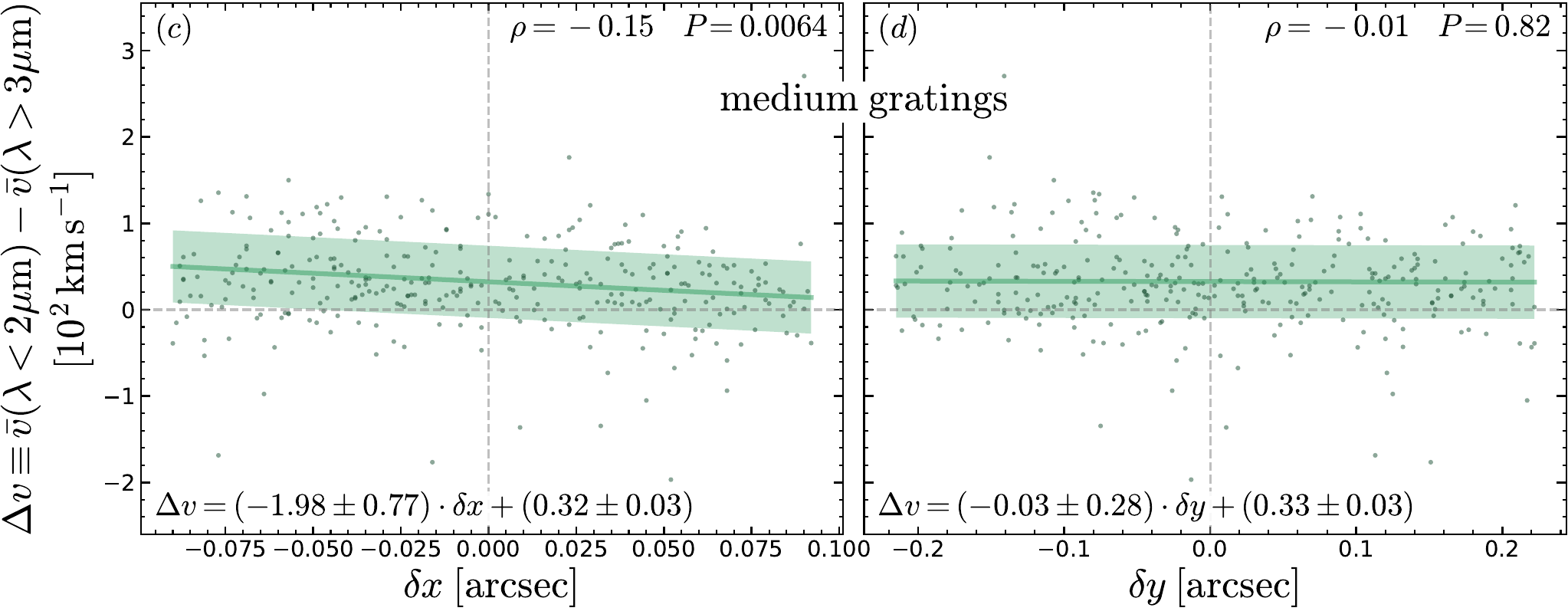}
  \caption{Wavelength calibration bias as a function of intra-shutter source position; for both the prism (top) and medium gratings (bottom). $\delta x$ is the spatial offset of the target with respect to the center of the micro shutter, measured along the dispersion direction. For an unbiased wavelength solution, we would expect $\Delta v = 0$ at $\delta x=0$ and $\delta y=0$, and no correlation.
  The observed correlation with $\delta x$ means that the correction we adopted for the intra-shutter source position is insufficient \citetext{see Section~\ref{s.datared} and \citealp{ferruit+2022}}; $\delta x$ increases toward bluer wavelengths. Neglecting intra-shutter source position entirely would result in an even larger bias than what reported here.
  }\label{f.quality.shutters}
\end{figure*}

To assess the accuracy of the wavelength calibration, we use the metric $\Delta v$; for each target, $\Delta v \equiv v(1<\lambda<2~\mum) - v(\lambda >3~\mum)$,
where $v(1<\lambda<2~\mum)$ is the mean velocity of emission lines with observed wavelengths between 1 and 2~\mum, and $v(\lambda >3~\mum)$ is the mean velocity of emission lines with observed wavelengths between 3~\mum and the maximum wavelength.
For this test, we consider only galaxies with more than three independent emission lines having $S/N>7$. In Figure~\ref{f.quality.shutters} we show $\Delta~v$ as a function of the intra-shutter source positions, $\delta x$ (closely aligned along the dispersion direction) and $\delta y$; the gray dots are individual galaxies, the green line with bands is a robust least-squares fit \citep{cappellari+2013a}.
For an unbiased solution, both the zero-point and slope of the best-fit line should be zero. In contrast, we find  an average zero-point offset of 300~\kms (for the prism) and 30~\kms (for the gratings); these values correspond to 0.1--1 pixels (prism) to 0.1--0.3 pixels (medium gratings). These offsets indicate an overall bias of the wavelength solution.
In addition to this zero-point offset, there is also a clear negative correlation between $\Delta v$ and $\delta x$ for both prism and gratings. The correlation for the prism is both stronger and more statistically significant, reaching an excursion of 0.5--3~pixels.
Before interpreting these correlations, we remark that the coordinates of the MSA are opposite to the pixel coordinates of the detector on the focal plane assembly \citetext{see \citealp{ferruit+2022}, their Figure~4; and \citealp{jakobsen+2022}, their Figure~4}; this means that positive $\delta x$ are offset toward bluer wavelength values.
With this in mind, the anti-correlation we find means that the correction to the wavelength solution due to intra-shutter offsets of compact sources \citep[Section~\ref{s.datared} and][]{ferruit+2022} is insufficient.
A solution to these remaining calibration issues is beyond the time constraints of this data release, and will be presented in a future work. In Appendix~\ref{a.wavecal} we further show that on average the wavelength bias of the prism depends on the global position of the source in the MSA.

\section{Using the NIRSpec data products}\label{s.limitations}

In this section, we provide a concise summary on how to use the data products, and a list of the limitations of the current data release. The prospective user of the data provided in this release is encouraged to consider these limitations carefully.
\begin{compactitem}
  \item \textbf{Data reduction problems and short circuits.} Spectra with data reduction problems are flagged with \verb|DR_flag|=\verb|True|;
  if a redshift was given, it is guaranteed to be accurate from visual inspection. However, for any purpose other than redshift, these data should be visually inspected to assess whether they are suitable. Under this flag we also collect observations affected by MSA short circuits, which may present abnormally bright background, including steep background gradients across the detector. In all cases, this results in lower signal-to-noise spectra than one would expect given the source magnitudes and integration time.
  In the most severe cases, shorts cause incorrect background subtraction and no useful observations (Section~\ref{sec:observations}, Appendix~\ref{a.shorts}).
  \item \textbf{Aperture correction.} Aperture corrections assume the target has point-source geometry; for extended sources, this implies both the total flux and the color of the spectrum are inaccurate. For calculating emission-line ratios over long wavelength separations (e.g., \Halpha/\Hbeta, \OIIIL/\OIIall, \Paalpha/\Halpha), we recommend using aperture corrections derived from the photometry.
  Sources more extended than one shutter should be considered with particular care, or even excluded (see Background subtraction).
  \item \textbf{Background subtraction.} The background subtraction strategy is optimized for compact sources; while shutters affected by contaminants are pre-identified and not considered in the subtraction, shutters affected by the same source cause self subtraction. Depending on the source size and spatial gradients, this may bias the shape of the spectrum and the total flux.
  \item \textbf{Noise spectrum.} The noise spectrum is based on variance-conserving resampling, to mitigate the effect of correlated noise \citep{dorner+2012}. A full analysis of correlated noise in NIRSpec will be presented in a future work (P.~Jakobsen, in~prep.)
  \item \textbf{Wavelength calibration.} There is a discrepancy between the wavelength calibration of the prism and gratings (causing a typical $\Delta z = 0.0042$; Figure~\ref{f.redshift.zcomp}). In addition, we find an overall offset in the wavelength calibration of both the prism (mean value 300~\kms) and for the gratings (mean value 30~\kms). We apply a correction for the wavelength offset due to the intra-shutter position of each source, but there is still a residual bias.
  After correcting empirically for this bias, we show that the prism wavelength offset depends on the spatial location in the MSA (Section~\ref{s.wavecal} and Appendix~\ref{a.wavecal}); for the medium gratings the residual wavelength dependence on the intra-shutter position is milder, and we find no detectable trend with spatial location on the MSA.
  \item \textbf{Flux calibration.} The relative flux calibration between the prism and gratings is accurate to within 15~per cent, and depends on the wavelength (e.g., figures~\ref{f.obs.probs}, \ref{f.highl.pablo}, \ref{f.redshift.fcomp}). The user is encouraged to consider this problem when measuring flux ratios, particularly when comparing between different dispersers and wavelengths.
\end{compactitem}

\section{highlights}\label{s.highl}

\begin{figure*}
  \includegraphics[width=\textwidth]{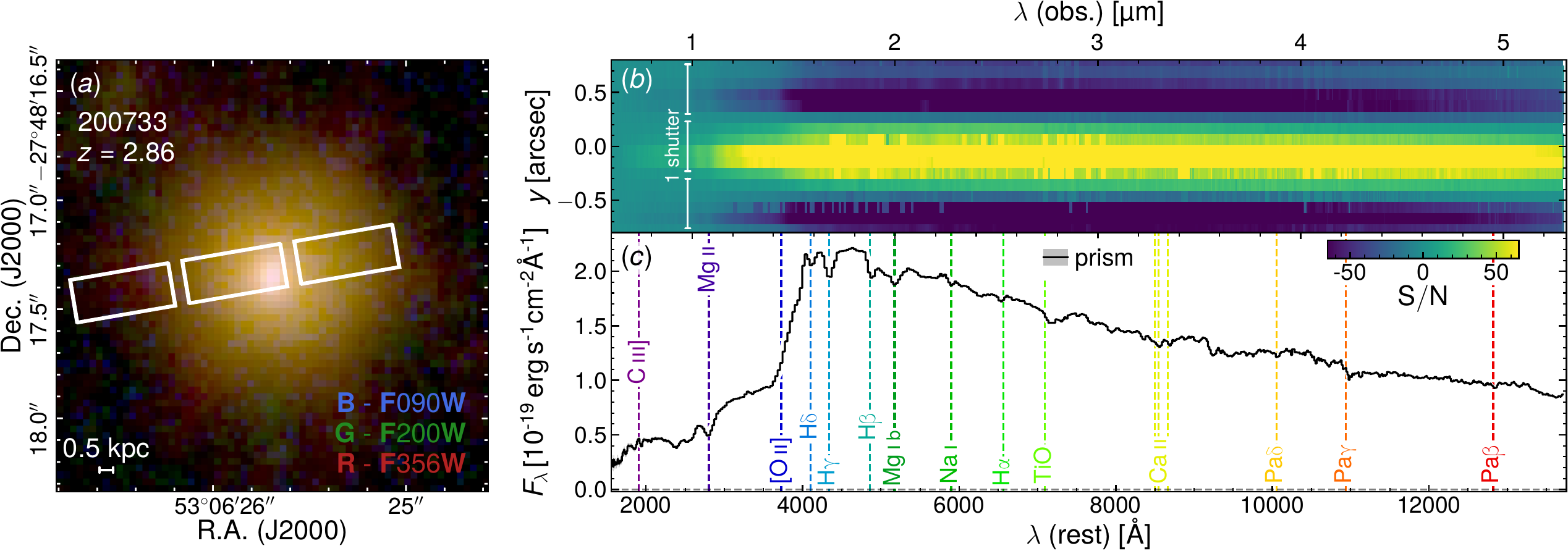}
  \caption{A low-redshift quiescent galaxy, displaying a smooth, peaked light distribution and an evolved stellar population. Panel~(a) shows a false-color NIRCam image, with the adopted filters in the bottom right corner. Panel~(b) is the NIRSpec/MSA 2-d signal-to-noise map of the prism spectrum; three shutters are indicated. Here and in all other 2-d maps (Figs.~\ref{f.highl.agn}b--\ref{f.highl.gsz12}b), negative signal-to-noise is caused by the nod-and-subtract strategy for removing the background. Panel~(c) is the 1-d, 5-pixel box-car extracted spectrum. Besides the strong 4000-\AA break, several stellar and ISM absorption features are detected. Notice that the aperture correction applied in this data reduction is optimized for point-like sources; extended galaxies like 200733 would require both a different aperture correction and a different background-subtraction strategy, to avoid self subtraction.
  From Z.~Ji et~al. (in~prep.).
  }\label{f.highl.qulow}
\end{figure*}

\begin{figure*}
  \includegraphics[width=\textwidth]{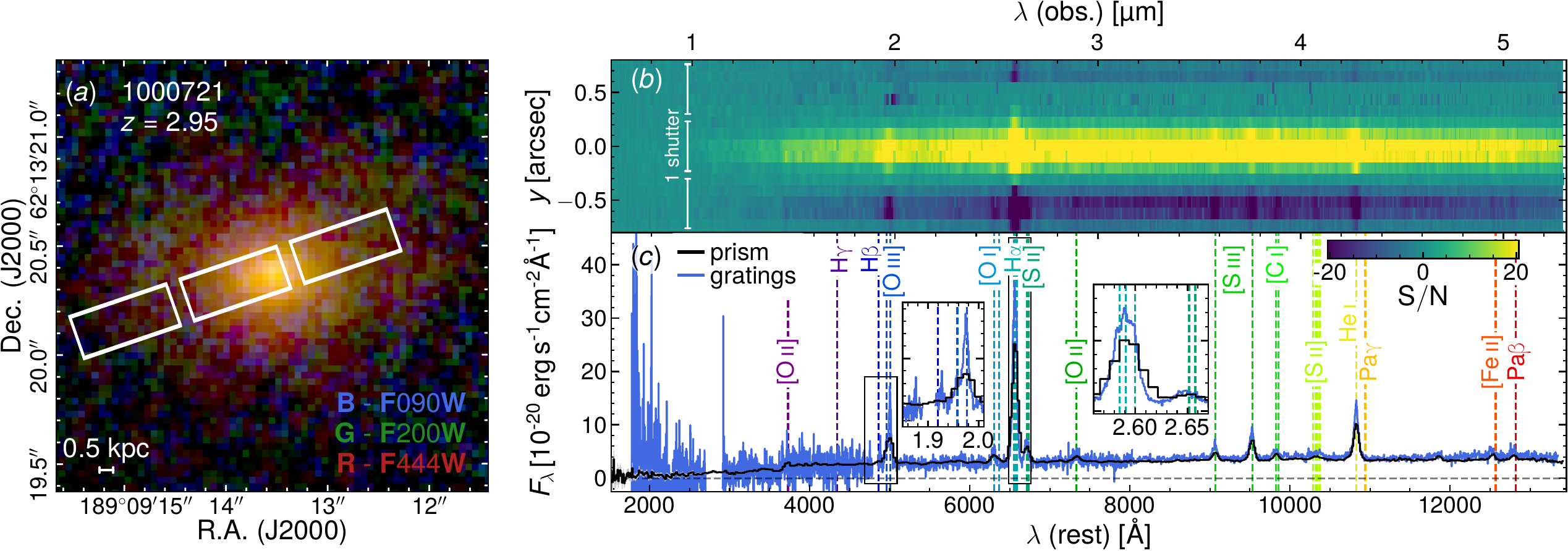}
  \caption{Example of a dust-reddened AGN host, with clear outflows (as seen in the broad component of forbidden \OIIIall, \SIIall and~\SIIIall), stellar continuum (there is evidence of a Balmer break) and high temperature (from the detection of several auroral lines, \OIIAuall and \SIIAuall).
  The blue line in panel~c is the (spliced) medium-resolution grating spectra (gaps are due to the gaps between the NIRSpec detectors). The black rectangles show details of the \Hbeta--\OIIIall and \Halpha--\NIIall--\SIIall emission-line regions.
  All other symbols are the same as Figure~\ref{f.highl.qulow}.
  }\label{f.highl.agn}
\end{figure*}

\begin{figure*}
  \includegraphics[width=\textwidth]{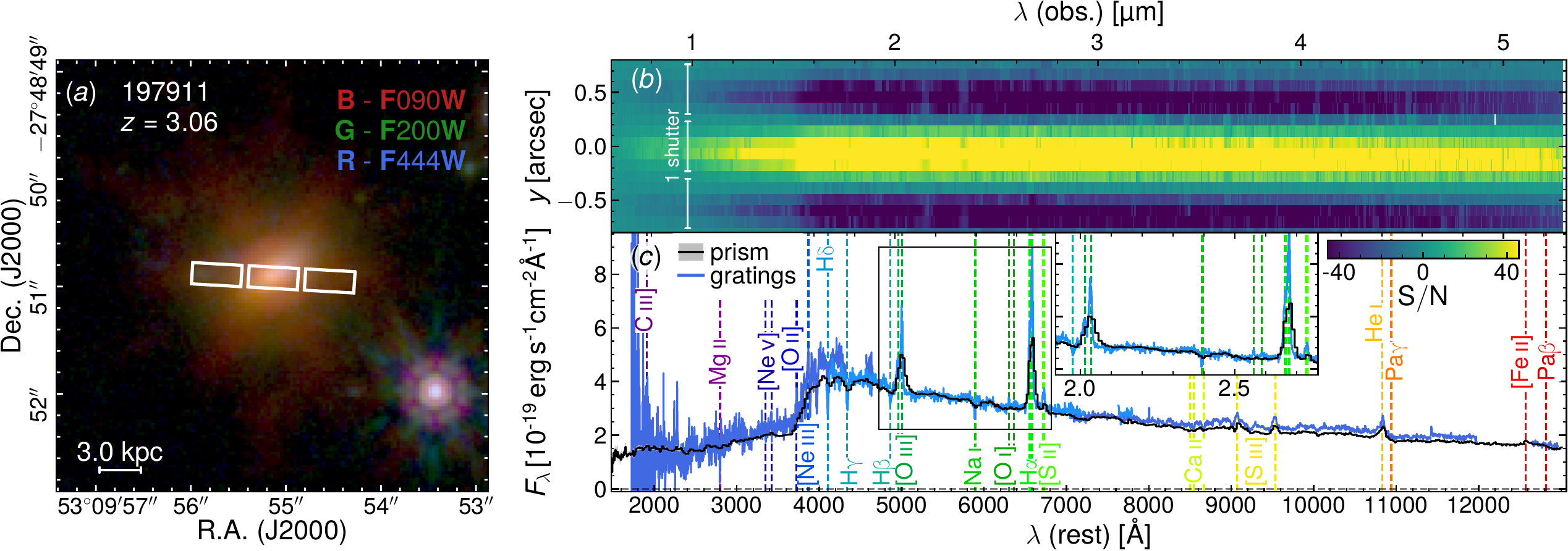}
  \caption{`Pablo's Galaxy', a UVJ-quiescent galaxy at $z=3$ hosting an X-ray and MIR AGN and multi-phase outflows. This system displays a rich set of stellar and ISM absorption lines \citep[\MgIIall and \NaIall, the latter tracing a fast neutral-gas outflow;][]{deugenio+2023c}. Emission lines trace both low-ionization gas \OIIall, \OIall, \NIIall, \SIIall\ -- possibly due to shocked or stripped gas, as well as higher-ionization species (\NeVall).
  Note the flux-calibration offset between G140M (blue) and G235M (light blue: 10~per cent) and G235M and G395M (blue; 7~per cent). Like 200733 (Figure~\ref{f.highl.qulow}), Pablo's galaxy is fairly extended, meaning the standard aperture correction and background subtraction are not optimal.
  From J.~Scholtz et al. (in~prep).
  All symbols are the same as Figure~\ref{f.highl.qulow}.
  }\label{f.highl.pablo}
\end{figure*}

\begin{figure*}
  \includegraphics[width=\textwidth]{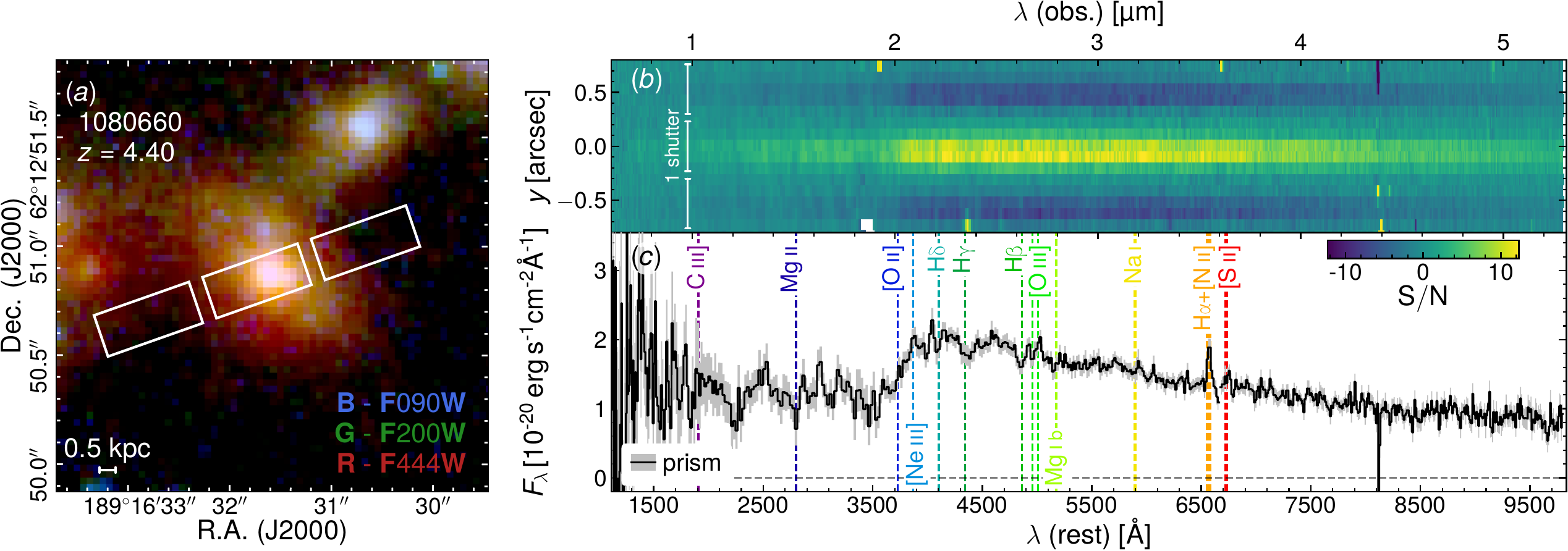}
  \caption{Example of a quiescent galaxy at $z=4.4$, ID~1080660. This object displays a complex morphology with both a blue core (white in panel~a) and a redder extended component (in green).
  The prominent Balmer break and flat rest-UV spectrum indicate this galaxy is an early quiescent system.
  We see clear emission from the $\Halpha + \NIIall$ blended complex, from \OIIIall, and from \MgIIall,
  indicating AGN activity and, possibly, ongoing outflows, as seen in quiescent galaxies at lower redshifts.
  All symbols and panels are the same as Figure~\ref{f.highl.qulow}.
  }\label{f.highl.quiescent}
\end{figure*}

\begin{figure*}
  \includegraphics[width=\textwidth]{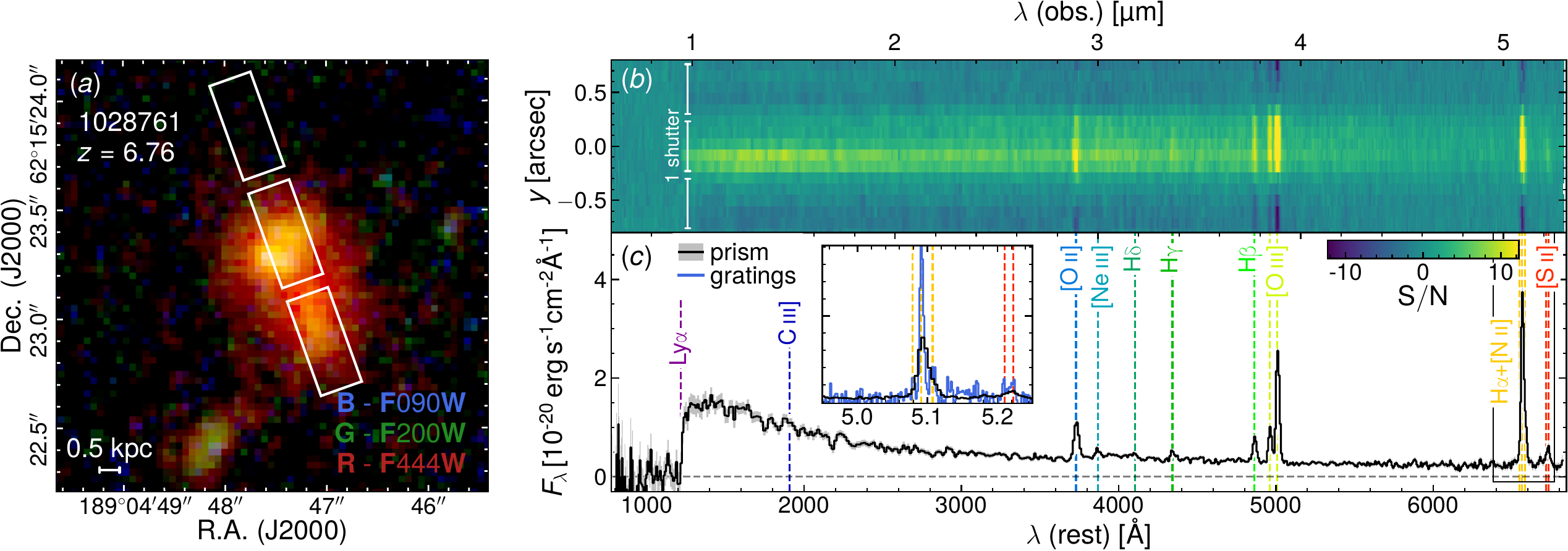}
  \caption{Example of a high-redshift galaxy with high metallicity and/or signatures of shocked gas, ID~1028761.
  The NIRSpec 2-d signal-to-noise map (panel~b) clearly reveals spatially extended emission; a different background subtraction strategy is clearly required for this class of targets;
  in cases like this, using the provided emission-line fluxes may result in unphysical line ratios, or harder-to-identify bias.
  From A.~Cameron et~al. (in~prep.).
  All symbols and panels are the same as Figure~\ref{f.highl.qulow}.
  }\label{f.highl.n2s2}
\end{figure*}

\begin{figure*}
  \includegraphics[width=\textwidth]{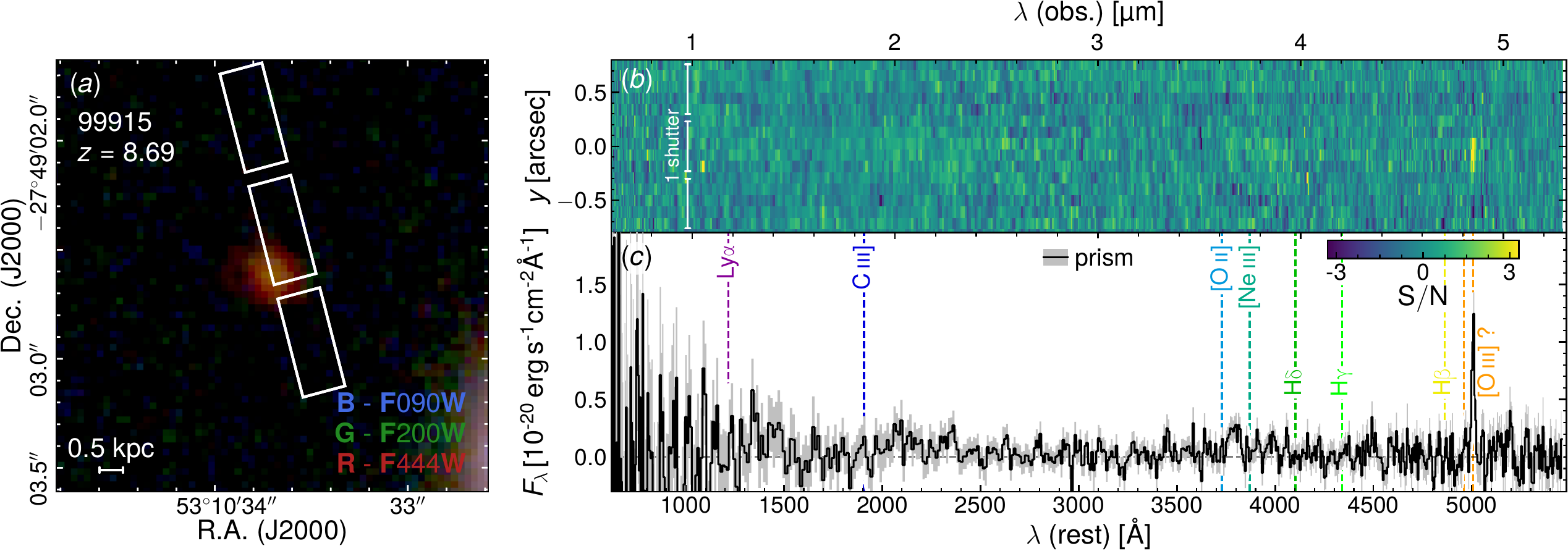}
  \caption{Example of a high-redshift galaxy with uncertain redshift, ID~36424.
  The clear emission line at 4.8~\mum could be identified as \Halpha, but this would be inconsistent with the photometric break between F090W and F115W.
  A tentative identification of the line as \OIIIL is presented here; while consistent with the photometric break, this solution is itself problematic due to the missing \OIIIL[4959]. We can rule out an artifact, because the line is also detected in our G395M spectrum and in the NIRCam grism \citetext{from FRESCO; \citealp{oesch+2023}; see also F.~Sun et al., in~prep.}.
  All symbols and panels are the same as Figure~\ref{f.highl.qulow}.
  }\label{f.highl.uncertain}
\end{figure*}

\begin{figure*}
  \includegraphics[width=\textwidth]{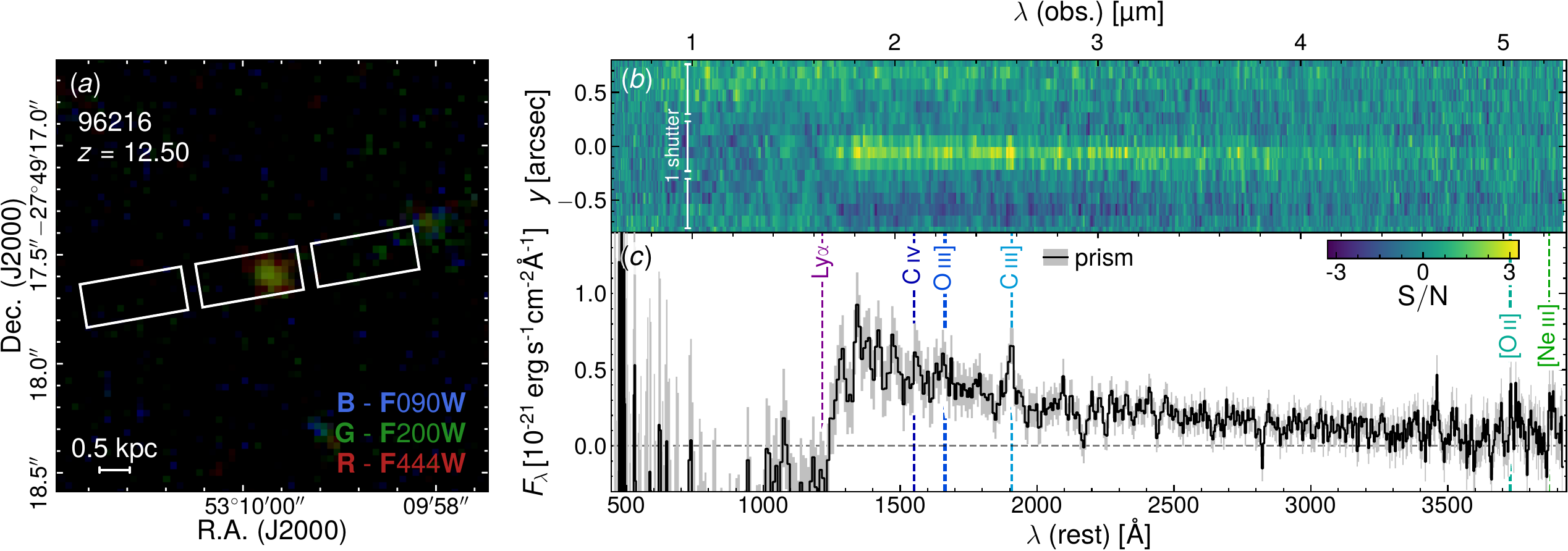}
  \caption{One of the highest-redshift sources in this data release, 96216 \citetext{JADES-GS-z12-0, \citealp{curtis-lake+2023,robertson+2023}; 2773, \citetalias{bunker+2023b}}.
  This system shows the highest-redshift detection of metals to date \citep{deugenio+2023d}.
  All symbols and panels are the same as Figure~\ref{f.highl.qulow}.
  }\label{f.highl.gsz12}
\end{figure*}

The diversity of the JADES DR3 spectroscopic sample is illustrated by comparing some highlights (Figs.~\ref{f.highl.qulow}--\ref{f.highl.gsz12}). In Figure~\ref{f.highl.qulow} we show 200733, an example of a low-redshift quiescent galaxy at $z=2.86$. NIRCam photometry (panel~a) indicates a smooth, peaked light profile with an extended halo, suggesting a high ($n\gtrsim2$) S\'ersic index and, therefore, a dynamically evolved system. The 1-d spectrum (panel~c) displays the signatures of an evolved stellar population, with a distinct 4000-\AA break -- typical of old ($>1.5$~Gyr), metal-rich stellar populations, accompanied by a possible, fainter Balmer break ($\approx 3750~\AA$), indicating a younger (0.5--1~Gyr old) stellar population.
The galaxy displays a number of stellar absorption features with high equivalent width; the Balmer series, \permittedEL[Mg][i][\textlambda\textlambda][5167--][5184], and the `calcium triplet', \permittedEL[Ca][ii][\textlambda\textlambda][8498--][8662].

At slightly higher redshift ($z=2.95$) we find 1000721 (Figure~\ref{f.highl.agn}), an AGN-host galaxy with type-1 AGN and high-velocity outflows, a secure signature of an active supermassive black hole. The host galaxy is clearly seen in NIRCam; the lack of flux in F090W (cf.~Figure~\ref{f.highl.qulow}a) indicates high dust reddening, which is indeed seen in the NIRSpec data (panel~c).
Emission from the \OIIIall and \Halpha+\NIIall complexes is spatially extended (panel~b), indicating a resolved disc or narrow-line region.
In the medium-resolution grating spectra (blue line in panel~c) the presence of an outflow can be clearly seen in \OIIIall and \SIIall, while \Halpha shows evidence of both ionized gas outflows and a broad line region. The prism spectrum (black line) reveals a Balmer break, indicating that the continuum emission is dominated by stars; a number of auroral lines is readily detected in rest-frame $r-$ and $Y-$bands.

`Pablo's Galaxy' (197911, GS-10578; Fig~\ref{f.highl.pablo}) is a marvelous massive galaxy at $z=3.06$, identified as quiescent via the UVJ color-color diagram \citep{williams+2009}.
This extraordinary galaxy displays stellar rotation \citep{deugenio+2023c}, an X-ray and MIR-detected type-2 AGN \citep{circosta+2019}, fast ionized-gas outflows in \OIIIL, and neutral-gas outflows with high mass loading \citep[\NaIall absorption in panel~c;][]{deugenio+2023c}.
Medium-resolution observations spanning the entire NIRSpec wavelength enable the study of \MgIIall emission and absorption, high-ionization species (\NeVall), electron densities (\OIIall and \SIIall), and stellar \textalpha-elements abundance (\permittedEL[Mg][i][\textlambda\textlambda][5167--][5184], \permittedEL[Ca][ii][\textlambda\textlambda][8498--][8662]).
Stringent upper limits on \Pabeta disfavor a dust-obscured starburst.

1080660 is an example of a higher-redshift quiescent galaxy at $z=4.4$ (Figure~\ref{f.highl.quiescent}), among the highest-redshift quiescent galaxies known \citep[cf.][]{carnall+2023b,nanayakkara+2022}.
NIRCam (panel~a) shows two interlopers (north west; photometric redshift 2.5) and a possible dusty companion to the east (1080661, with photometric redshift 3.8, but this value is highly uncertain due to the dusty nature of this target). 1080660 itself has an evolved morphology, consisting of a bright central core and more extended emission along the north-east--south-west direction; the extended emission appears significantly redder than the core (green vs white in the false-colour image of panel~a), suggesting a possible central starburst, as seen in some local post-starburst galaxies \citep{deugenio+2020} and, recently, in NIRCam imaging \citep{wright+2023}. 
The spectrum (panel~c) exhibits a clear Balmer break, \Hdelta, \Hgamma and \Hbeta absorption, and \OIIIall and \Halpha+\NIIall emission of relatively low equivalent width. There is no evidence for spatially extended nebular emission in the 2-d (panel~c).

Figure~\ref{f.highl.n2s2} shows 1028761, a merger between a relatively un-obscured galaxy and a dusty galaxy at $z=6.76$. This system displays high values of the emission-line ratios \OIIall/\OIIIL, \NIIL/\Halpha and \SIIall/\Halpha, characteristic of high-metallicity gas or shock-dominated emission.

In Figure~\ref{f.highl.uncertain} we show 99915, which displays a strong single emission line at 4.8~\mum. The line is also seen in the G395M grating spectrum, and in the NIRCam grism spectrum (F.~Sun et al., in~prep.), which rules out an artifact. The galaxy also shows a photometric drop between the NIRCam F090W and F150W filters, which at face value rules out a solution where the emission line is \Halpha.
Identifying the line as either \Hbeta or \OIIIL would match the \Lyalpha drop seen in NIRCam, but is not without problems.
If the line was \Hbeta, we would expect to observe \Hgamma at about half the \Hbeta flux; similarly, the tentative solution at $z=8.69$, which identifies the observed line with \OIIIL, would require observing \OIIIL[4959] at about one third of the line flux.
None of these accompanying lines (the putative \Hgamma nor \OIIIL[4959]) are seen either in prism, grating, or NIRCam grism, leaving this object as a tentative redshift determination.

Finally, in Figure~\ref{f.highl.gsz12} we show JADES-GS-z12-0 \citep{curtis-lake+2023,robertson+2023}, which at the time of this data release is the highest-redshift detection of a metal emission line \citep{deugenio+2023d}. The data included in this release consists separately of observations from PID~3215 and PID~1210.
In this data release, we use the redshift from \cite{deugenio+2023d}, based off clearly detected \CIIIall emission. This value is lower than the redshift reported in our previous articles \citetext{\citealp{curtis-lake+2023}, \citealp{robertson+2023}, \citetalias{bunker+2023b}}; the latter was measured from the wavelength of the \Lyalpha drop, assuming only IGM absorption, and the discrepancy with the \CIIIall redshift is explained by DLA \citep[Damped \Lyalpha absorption; e.g.,][]{wolfe+2005}.
Increasing evidence is building up that a substantial fraction of $z>10$ galaxies may have DLA absorption \citep[e.g.,][]{heintz+2023b}, which may bias photometric and \Lyalpha-drop redshifts to higher values \citetext{e.g., \citealp{deugenio+2023d}; see especially \citealp{hainline+2024b}.}

\section{Conclusions}\label{s.conclusions}

In this work, we presented new and updated JADES NIRCam and NIRSpec observations obtained up until October 2023 in the two GOODS fields.
The spectra include both medium-depth and deep observations up to redshift $z \sim 13$, reaching the deepest unlensed spectroscopic observations to date (up to 45~hours on source).
The sample size and data quality of the spectra are a testament to the success of the NIRSpec/MSA instrument.
The high success rate of the redshift identification validates the selection criteria, including the quality of NIRCam data, the accuracy of the photometric analysis, and the precision of the photometric redshift determination with \eazy and \beagle.

We release fully reduced and calibrated images and spectra, and present catalogs of photometry, photometric redshifts, spectroscopic redshifts, and emission-line fluxes.
Insight from this large sample enabled us to pin down some remaining challenges in the data reduction: a mismatch in the redshift and flux-calibration between different dispersers, background subtraction and slit-loss corrections appropriate for extended sources, and residual wavelength calibration issues.
Future calibration programs will certainly address these problems.
An additional challenge is deriving an accurate selection function for correctly weighting each galaxy, which we will provide in the next data release.

In the future, significantly larger samples of the general galaxy population in the redshift and mass range probed by JADES would require a substantial investment of \jwst time, or a revised observing strategy.
Larger samples of specific classes of objects will still be crucial for understanding rare types, where JADES has only hinted at the potential (e.g., high-redshift quiescent galaxies, extremely reddened galaxies, little red dots, $z>8$ \Lyalpha emitters).

In the meantime, the current sample is the largest extragalactic sample with low- and medium-resolution spectroscopy spanning 0.6--5.3~\mum; the depth of the medium and deep spectra, and the synergy with medium- and wide-band imaging, enables for the first time a statistical study combining morphology and rest-frame optical spectroscopy of galaxies between the peak of the star-formation rate density and the first few hundred Myr after the Big Bang.




\bigskip

FDE, JS, RM, TJL, JW, WMB, XJ, IJ and CS acknowledge support by the Science and Technology Facilities Council (STFC), by the ERC through Advanced Grant 695671 ``QUENCH'', and by the UKRI Frontier Research grant RISEandFALL. RM also acknowledges funding from a research professorship from the Royal Society.
AJC, AJB, JC, AS and GCJ acknowledge funding from the ``FirstGalaxies'' Advanced Grant from the European Research Council (ERC) under the European Union's Horizon 2020 research and innovation programme (Grant agreement No. 789056)
SC, EP and GV acknowledge support by European Union’s HE ERC Starting Grant No. 101040227 - WINGS.
ECL acknowledges support of an STFC Webb Fellowship (ST/W001438/1).
SA, BRP and MP acknowledge support from Grant PID2021-127718NB-I00 funded by the Spanish Ministry of Science and Innovation/State Agency of Research (MICIN/AEI/ 10.13039/501100011033).
MP also acknowledges support from the Programa Atracci\'on de Talento de la Comunidad de Madrid via grant 2018-T2/TIC-11715.
CNAW, BER, BDJ, DJE, PAC, EE, MJR and FS acknowledge JWST/NIRCam contract to the University of Arizona NAS5-02015.
BER also acknowledges support from the JWST Program 3215. DJE is supported as a Simons Investigator.
The Cosmic Dawn Center (DAWN) is funded by the Danish National Research Foundation under grant DNRF140.
ST acknowledges support by the Royal Society Research Grant G125142.
H{\"U} gratefully acknowledges support by the Isaac Newton Trust and by the Kavli Foundation through a Newton-Kavli Junior Fellowship.
This research is supported in part by the Australian Research Council Centre of Excellence for All Sky Astrophysics in 3 Dimensions (ASTRO 3D), through project number CE170100013.
ALD thanks the University of Cambridge Harding Distinguished Postgraduate Scholars Programme and Technology Facilities Council (STFC) Center for Doctoral Training (CDT) in Data intensive science at the University of Cambridge (STFC grant number 2742605) for a PhD studentship.	
Funding for this research was provided by the Johns Hopkins University, Institute for Data Intensive Engineering and Science (IDIES).
PGP-G acknowledges support from grant PID2022-139567NB-I00 funded by Spanish Ministerio de Ciencia e Innovaci\'on MCIN/AEI/10.13039/501100011033, FEDER, UE.
DP acknowledges support by the Huo Family Foundation through a P.C. Ho PhD Studentship.
MSS acknowledges support by the Science and Technology Facilities Council (STFC) grant ST/V506709/1.
RS acknowledges support from a STFC Ernest Rutherford Fellowship (ST/S004831/1).
NCV acknowledges support from the Charles and Julia Henry Fund through the Henry Fellowship.
The research of CCW is supported by NOIRLab, which is managed by the Association of Universities for Research in Astronomy (AURA) under a cooperative agreement with the National Science Foundation.
This work was performed using resources provided by the Cambridge Service for Data Driven Discovery (CSD3) operated by the University of Cambridge Research Computing Service (www.csd3.cam.ac.uk), provided by Dell EMC and Intel using Tier-2 funding from the Engineering and Physical Sciences Research Council (capital grant EP/T022159/1), and DiRAC funding from the Science and Technology Facilities Council (www.dirac.ac.uk)
The authors acknowledge use of the lux supercomputer at UC Santa Cruz, funded by NSF MRI grant AST 1828315.
This study made use of the Prospero high performance computing facility at Liverpool John Moores University.


%

\vspace{5mm}
\facilities{\jwst(NIRCam), \jwst(NIRSpec/MSA), \hst(ACS), \hst(WFC3)}
\dataset[10.17909/8tdj-8n28]{http://dx.doi.org/10.17909/8tdj-8n28}.


\software{
{\tt \href{https://pypi.org/project/astropy/}{astropy}} \citep{astropyco+2013},
{\tt \href{https://pypi.org/project/corner/}{corner}} \citep{foreman-mackey2016}, {\tt \href{https://sites.google.com/cfa.harvard.edu/saoimageds9}{ds9}} \citep{joye+mandel2003},
{\tt \href{https://pypi.org/project/emcee/}{emcee}} \citep{foreman-mackey+2013},
{\tt \href{https://github.com/ryanhausen/fitsmap}{fitsmap}} \citep{hausen+robertson2022},
{\tt \href{https://github.com/cconroy20/fsps}{fsps}} \citep{conroy+2009,conroy_gunn_2010}, 
{\tt \href{https://pypi.org/project/ltsfit/}{ltsfit}} \citep{cappellari+2013a},
{\tt \href{https://pypi.org/project/matplotlib/}{matplotlib}} \citep{hunter2007},
{\tt \href{https://pypi.org/project/numpy/}{numpy}} \citep{harris+2020},
{\tt \href{https://pypi.org/project/ppxf/}{ppxf}} \citep{cappellari2017, cappellari2023},
{\tt \href{https://pypi.org/project/scipy/}{scipy}} \citep{jones+2001}, {\tt \href{https://github.com/AstroJacobLi/smplotlib}{smplotlib}} \citep{smplotlib}, and
{\tt \href{https://www.star.bris.ac.uk/~mbt/topcat/}{topcat}}\citep{taylor2005}.}



\appendix

\section{Observations affected by MSA short circuits}\label{a.shorts}

\begin{figure*}
  \includegraphics[width=\textwidth]{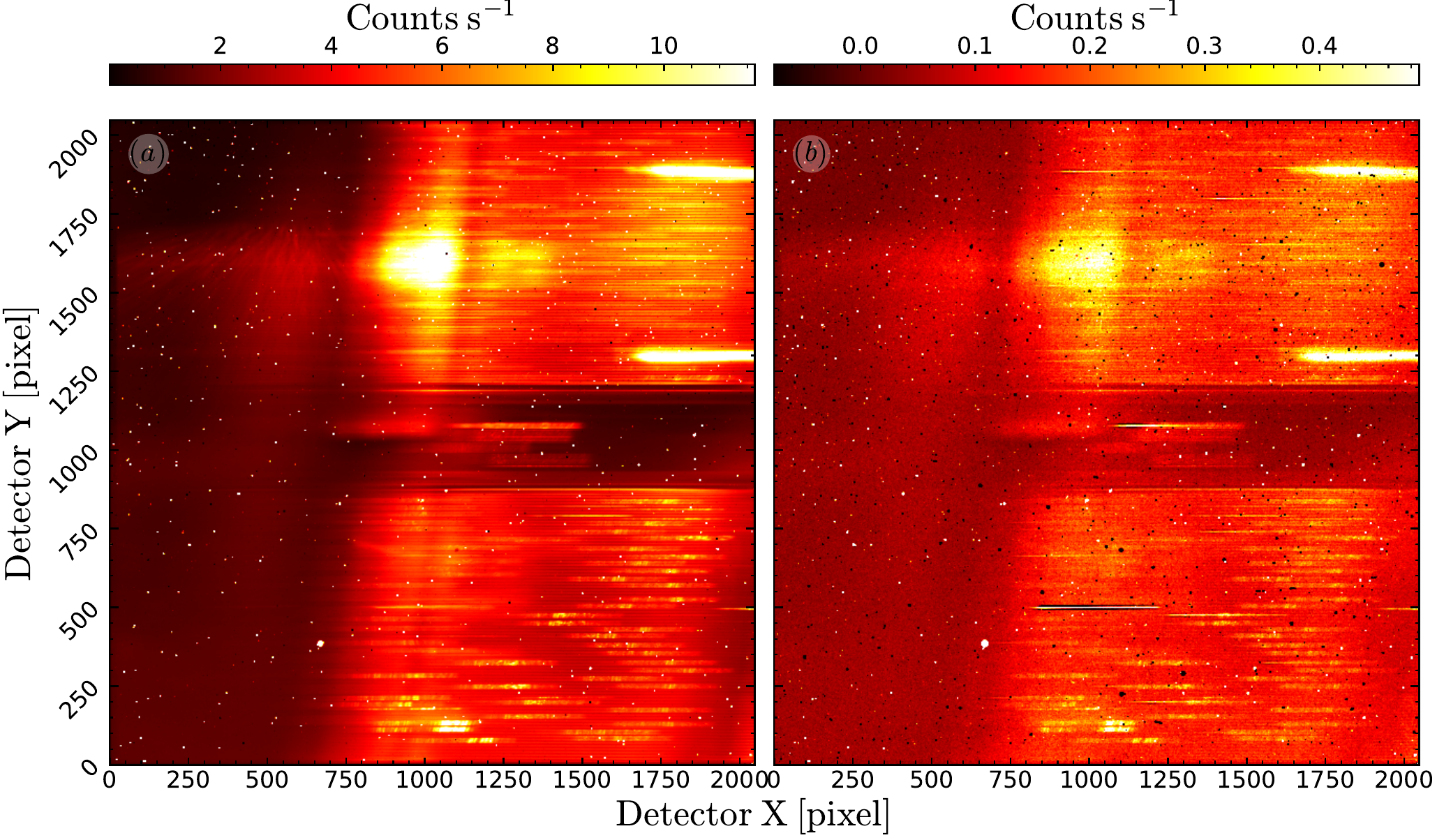}
  \caption{Imaging count-rate maps from NIRSpec detector NRS1 carried out during one of the visit of program PID~1180. Left and right panels illustrate the effect of the short circuits on the count-rate maps before (left) and after (right) background subtraction.}\label{f.shortscrt}
\end{figure*}

Some observations in this data release were affected by MSA short circuits, or `shorts'. For most programs, these were only a minority, and were excluded from the data reduction (e.g., PID~3215, Section~\ref{s.3215}).
However, for PID~1180 shorts affected two thirds of the initial observations.
The brightness of the shorts emission can vary drastically between occurrences: the brightest can render the entire integration unusable (Figure~\ref{f.shortscrt}), while the faintest may contaminate only a few sources near the affected region of the field of view.

As we argued in Section~\ref{s.1180}, these shorts-contaminated data are still useful to measure redshifts, and are included in this data release (Figure~\ref{f.shorts}).

There are three major ways in which shorts affect the quality of the data. First, shorts increase the background level, reducing the signal-to-noise of the observations.
Second, the shorts' background is tied to a given exposure, and, therefore can vary between different sky and nod positions; this causes our background subtraction to fail, because our strategy assumes the same background between different nods.
Third, the combination of increased and varying background affects subsequent steps of the pipeline, which can cause excessive or insufficient outlier removal.
Users are encouraged to treat these observations with caution, particularly for measurements using the spectral continuum.
The flag \verb|DR_flag| in the published tables identifies all spectra where there is a problem in the data reduction, or where the observations were affected by shorts (including when the actual contamination is low).

\begin{figure*}
\includegraphics[width=\textwidth]{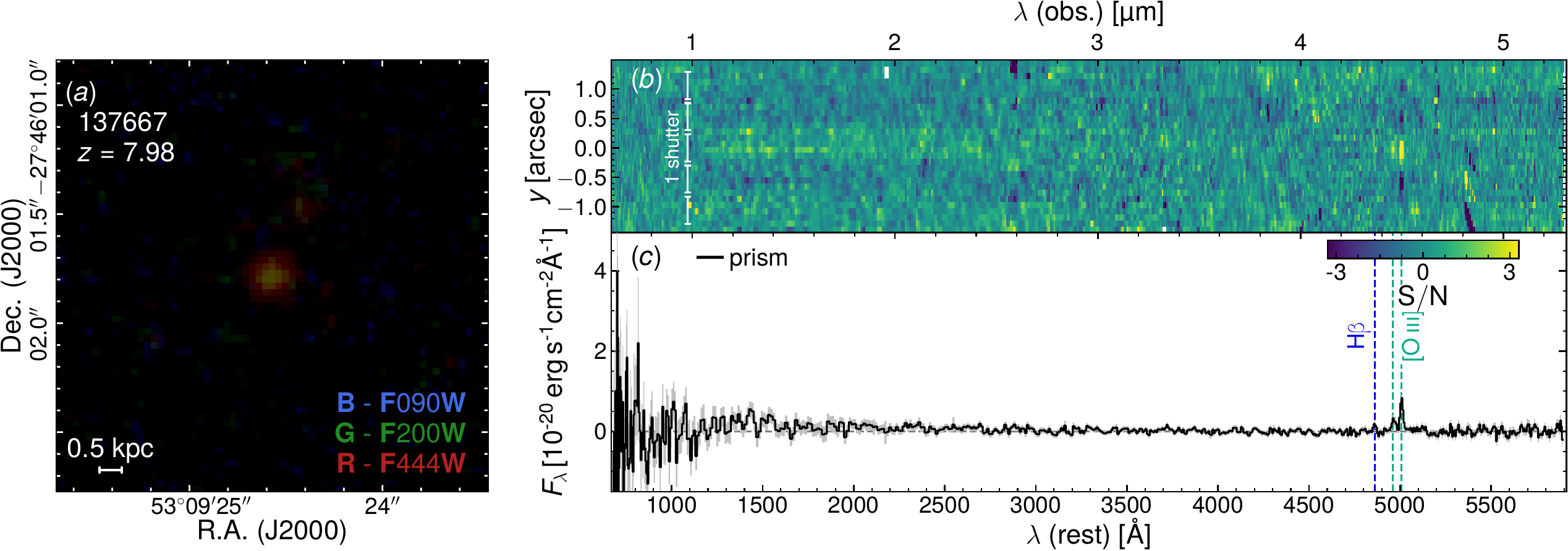}
\includegraphics[width=\textwidth]{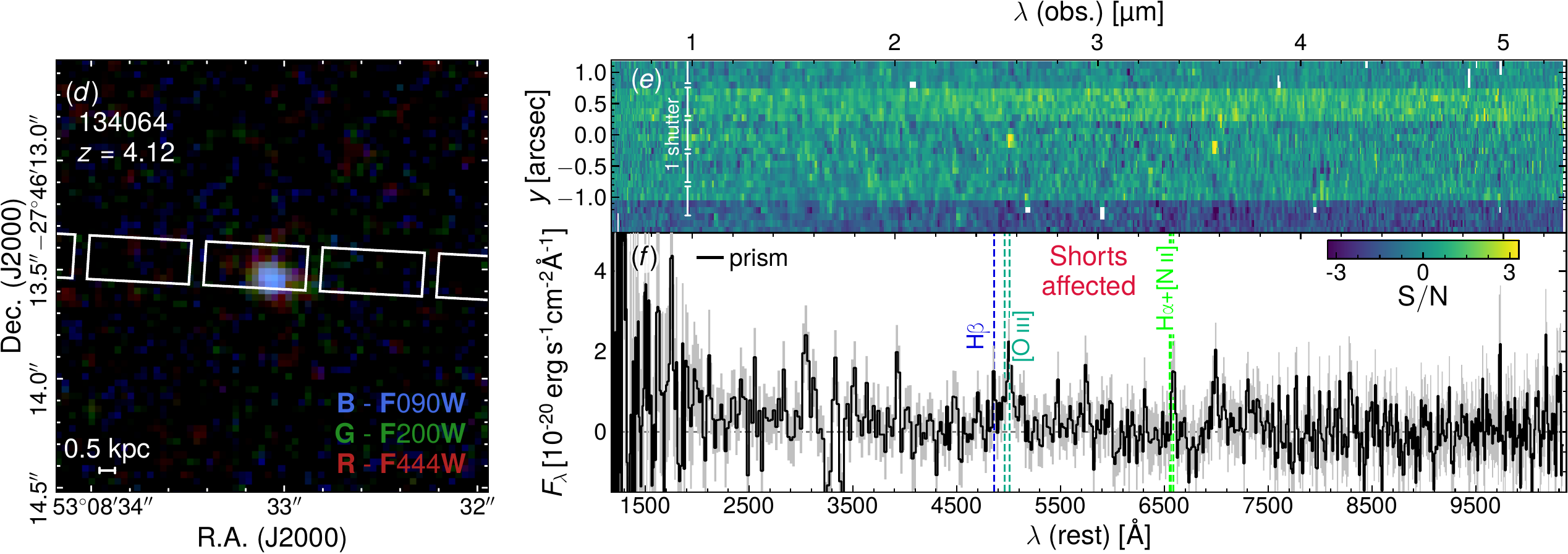}
\caption{A selection of targets from program 1180, comparing a successful observation (panels~a--c) and an observation affected by `shorts' (panels~d--l).
The successful observation shows the expected signal-to-noise both in the 2-d map (panel~b) and in the extracted 1-d spectrum (panel~c).
In contrast, panels~(e), (h) and~(k) show clear problems. Panel~(e) shows different background levels between different integrations (captured at different nod positions, resulting in horizontal striping). Panels~(h) and~(k) have similar background to panel~(b), but the shorts affect the data reduction pipeline in other ways.
For 148429, the shorts are likely responsible for the excess of outliers in the region 1--2~\mum; nevertheless, the data can still be used to measure the galaxy redshift.
For 212327 (whose redshift is known from the medium-resolution grating), the shorts caused a mis-alignment of the already extended source; the resulting data is unusable.
}\label{f.shorts}
\end{figure*}
\addtocounter{figure}{-1}

\begin{figure*}
\includegraphics[width=\textwidth]{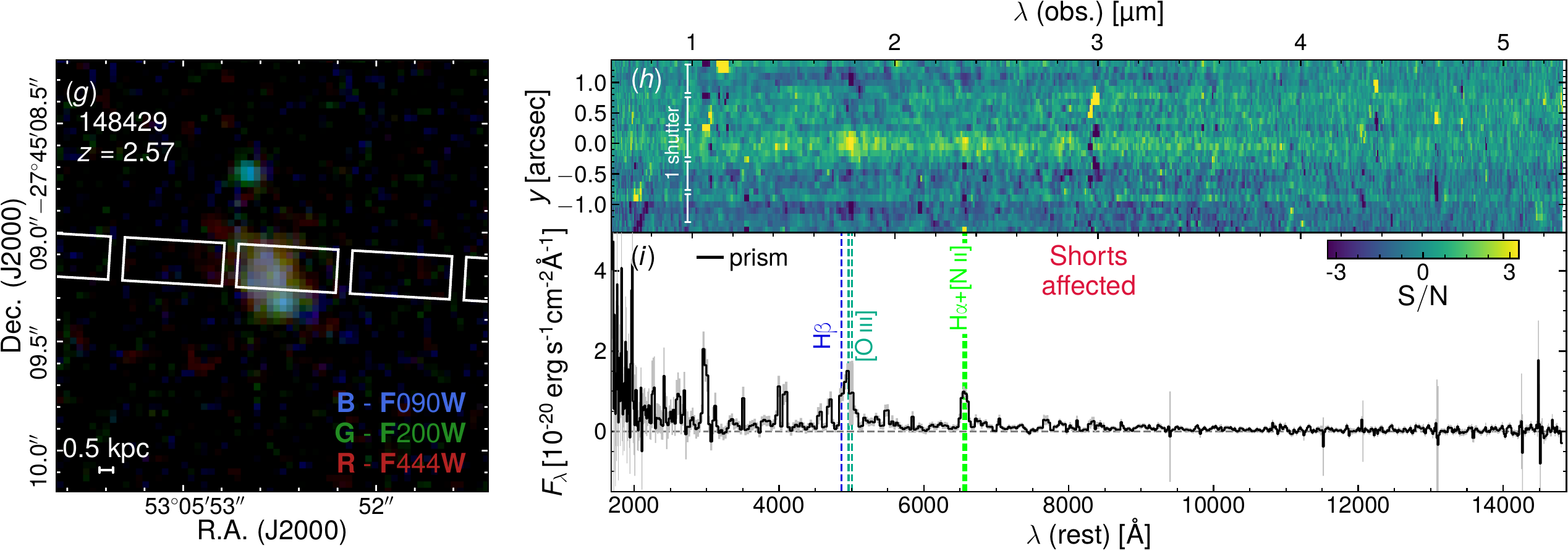}
\includegraphics[width=\textwidth]{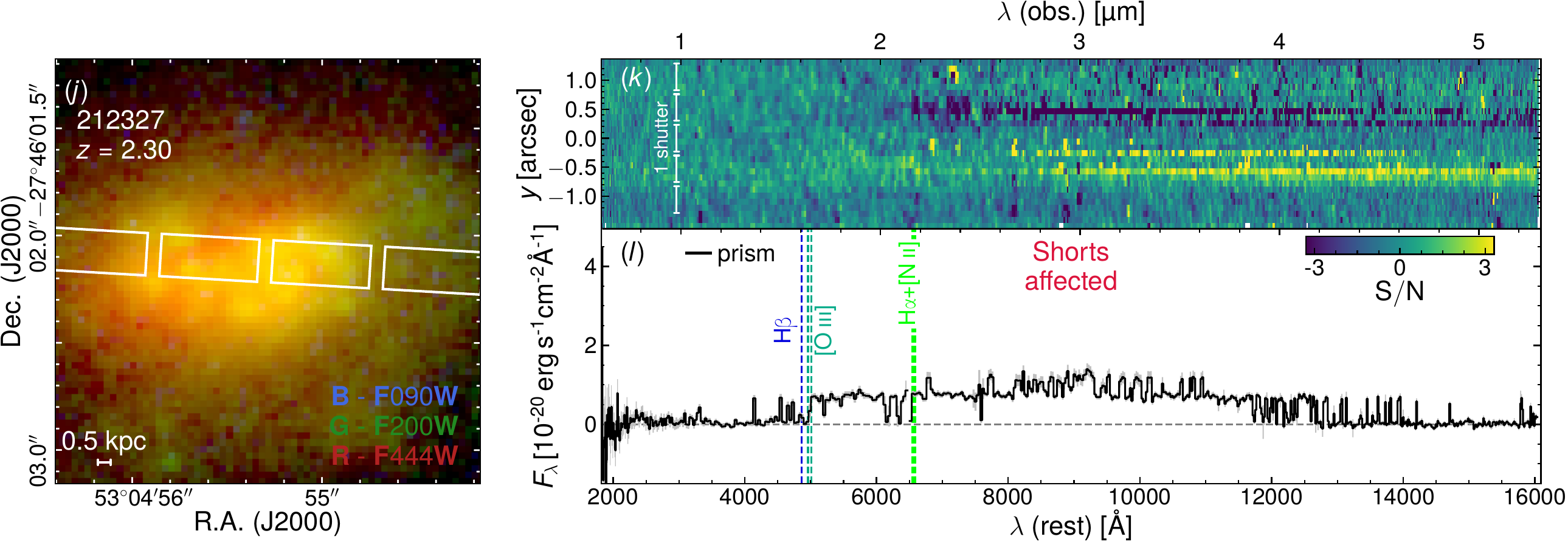}
\caption{(continued).}
\end{figure*}

\section{Comparison with DR1}\label{s.dr1comp}

\begin{figure}
  \includegraphics[width=\columnwidth]{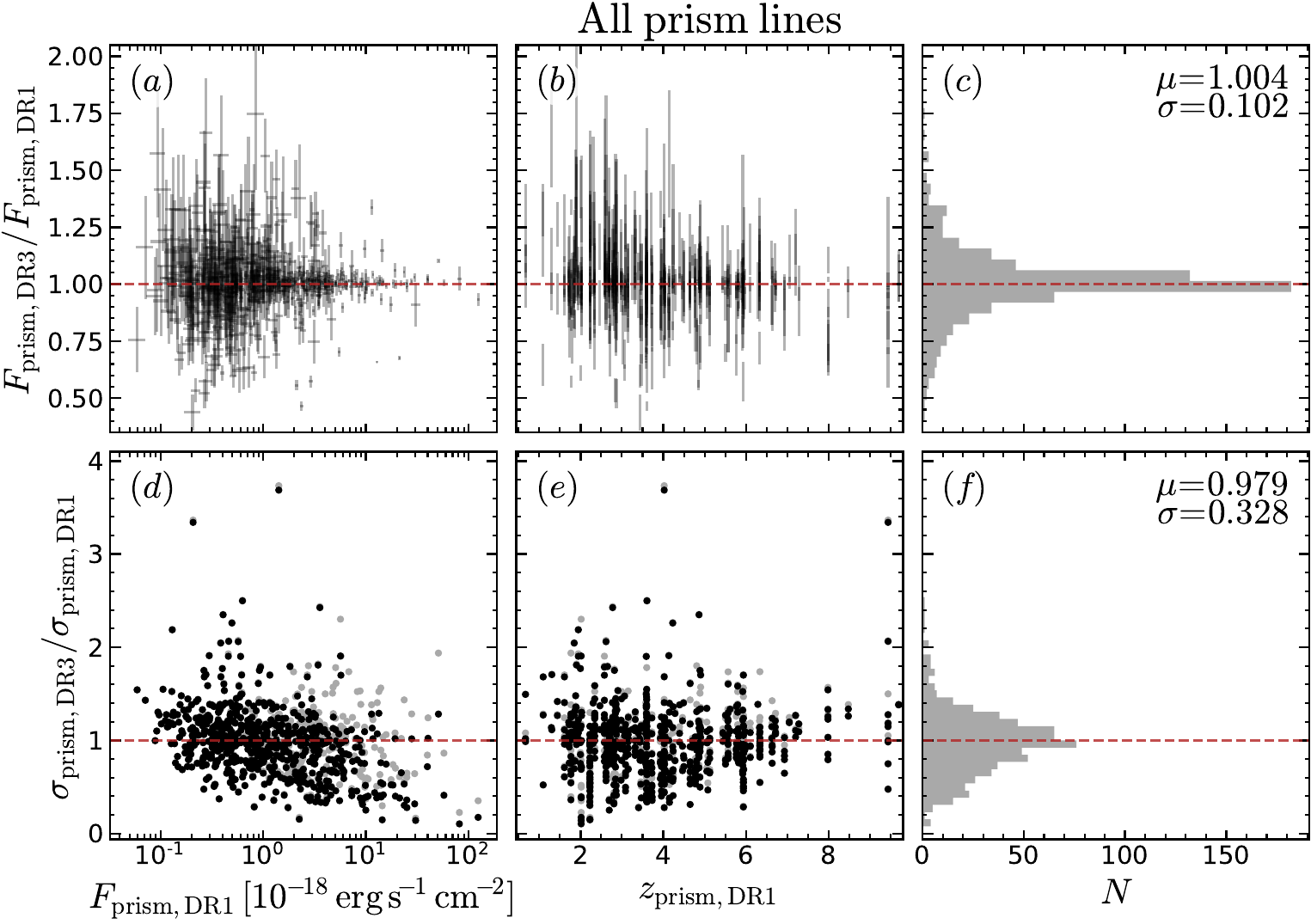}
  \caption{Comparison of the prism emission-line fluxes and their uncertainties between DR1 and DR3; the top row compares the fluxes, the bottom row compares the uncertainties.
  We find a statistically significant correlation between the ratio of uncertainties and flux (panel~\subreflett{f.r100.fluxcomp}{d}), indicating that our \ppxf implementation tends to underestimate the line noise in the high-S/N regime, relative to DR1.
  The gray points in the bottom panel illustrate the effect of adding 1~per cent relative uncertainties to the DR3 flux measurements.
  }\label{f.r100.fluxcomp}
\end{figure}

In this section, we compare the measurements obtained by applying the algorithms used in this data release (DR3), to the data previously released in DR1 \citepalias{bunker+2023b}.
We remark that NIRSpec DR1 and DR3 use the same data reduction, hence the spectra are exactly the same as in \citetalias{bunker+2023b}.
In Figure~\ref{f.r100.fluxcomp} we compare the flux (top rows) and uncertainties (bottom rows) between DR3 and DR1. We find excellent agreement for the overall flux measurements, even though some bright lines ($F>$\fluxcgs[-18][], panel~a) display highly significant differences.

Unlike for flux, the measurement uncertainties from DR3 and DR1 are different, with the present values 3~per cent smaller. We find a statistically significant trend with flux, suggesting the mismatch in the uncertainties is due to systematics at the bright end of the sample; adding 1~per cent systematic uncertainty to the flux measurements removes this correlation (gray points in panel~d).

\begin{figure}
  \includegraphics[width=\columnwidth]{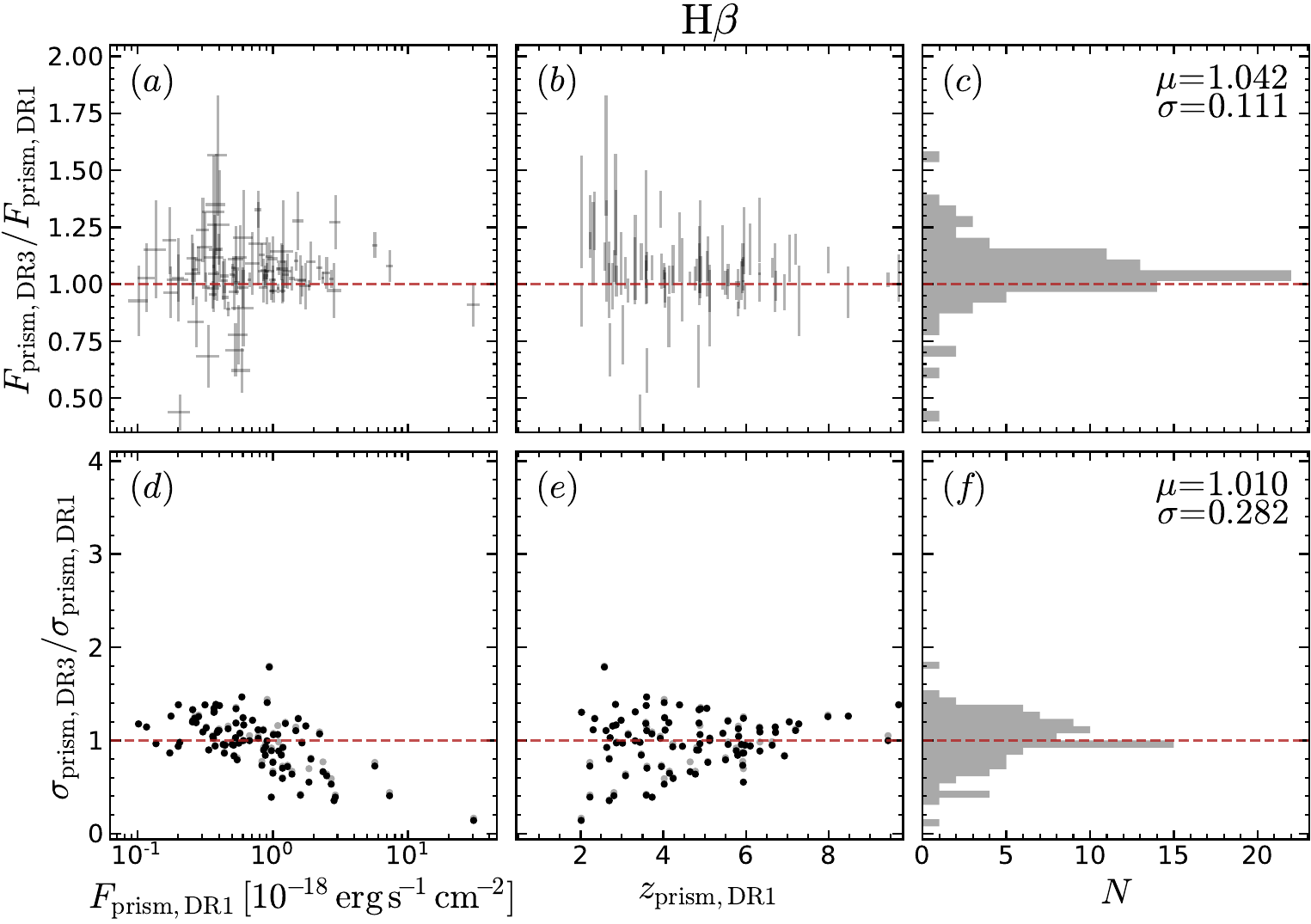}
  \caption{Comparison of the \Hbeta flux measured from the prism between DR1 and DR3; the symbols are the same as in Figure~\ref{f.r100.fluxcomp}. Below redshift $z=2$,
  \Hbeta is blended with \OIIIall and is not reported in this figure.
  The \Hbeta flux ratio between DR3 and DR1 is 1.04, while for all emission lines
  the ratio is 1.003.
  }\label{f.r100.hbfluxcomp}
\end{figure}

The agreement presented in Figure~\ref{f.r100.fluxcomp} varies from line to line; in Figure~\ref{f.r100.hbfluxcomp}, as an example, we show \Hbeta. Here the DR3 fluxes are 4~per cent higher, with the discrepancy correlating with flux and decreasing with redshift, as expected from the continuum correction applied by \ppxf.

A similar result is found for the medium-resolution gratings, where again we find excellent agreement in the overall flux (within 1~per cent), but smaller uncertainties (by 16~per cent).

\begin{figure}
  \includegraphics[width=\columnwidth]{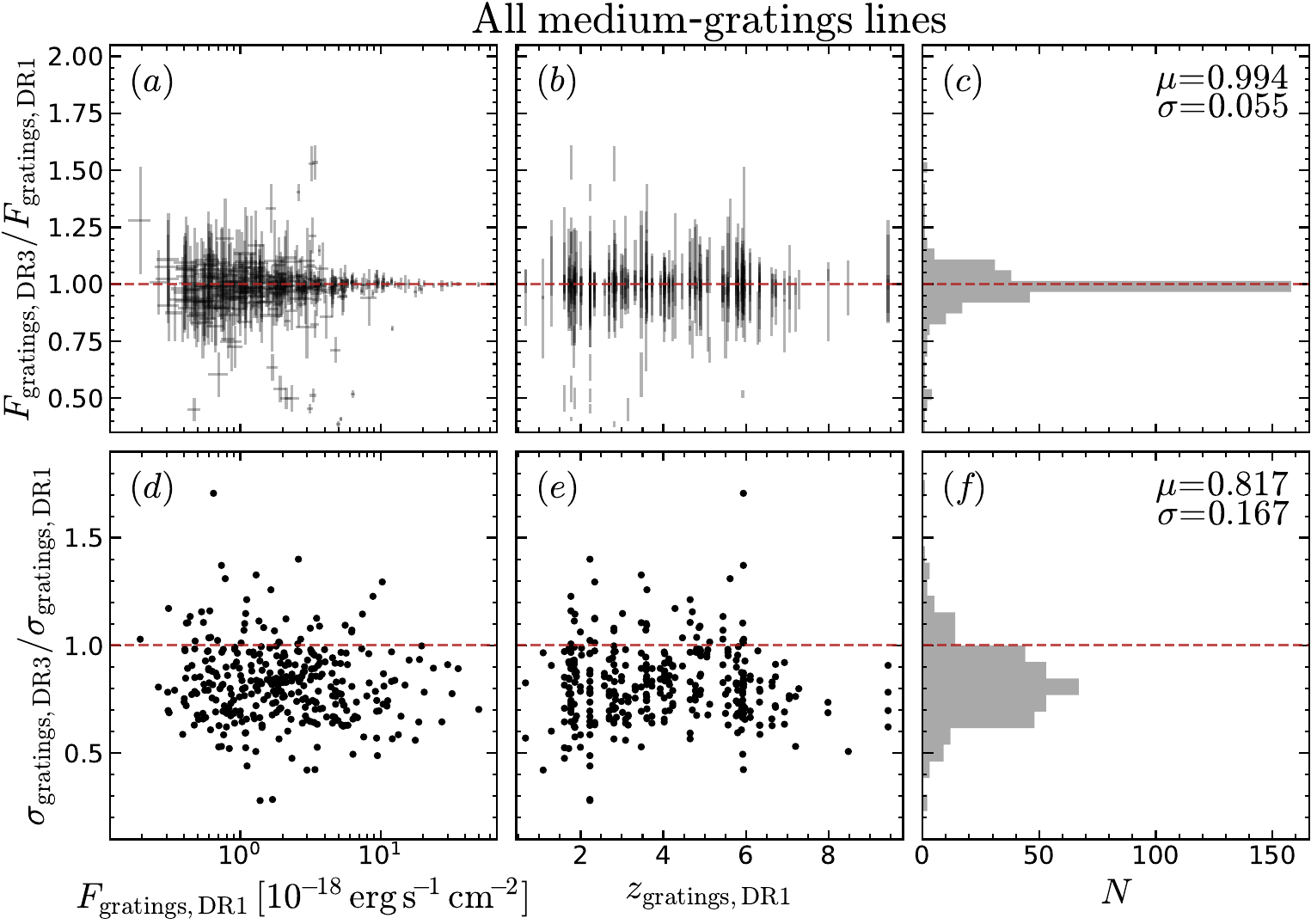}
  \caption{Comparison of the medium-resolution emission-line fluxes and their uncertainties between DR1 and DR3; the top row compares the fluxes, the bottom row compares the uncertainties. Fluxes are in excellent agreement, while we find the uncertainties to be smaller by 20~per cent, on average.
  }\label{f.r1000.fluxcomp}
\end{figure}
\section{Field dependence of the prism wavelength calibration bias}\label{a.wavecal}

\begin{figure*}
\includegraphics[width=\textwidth]{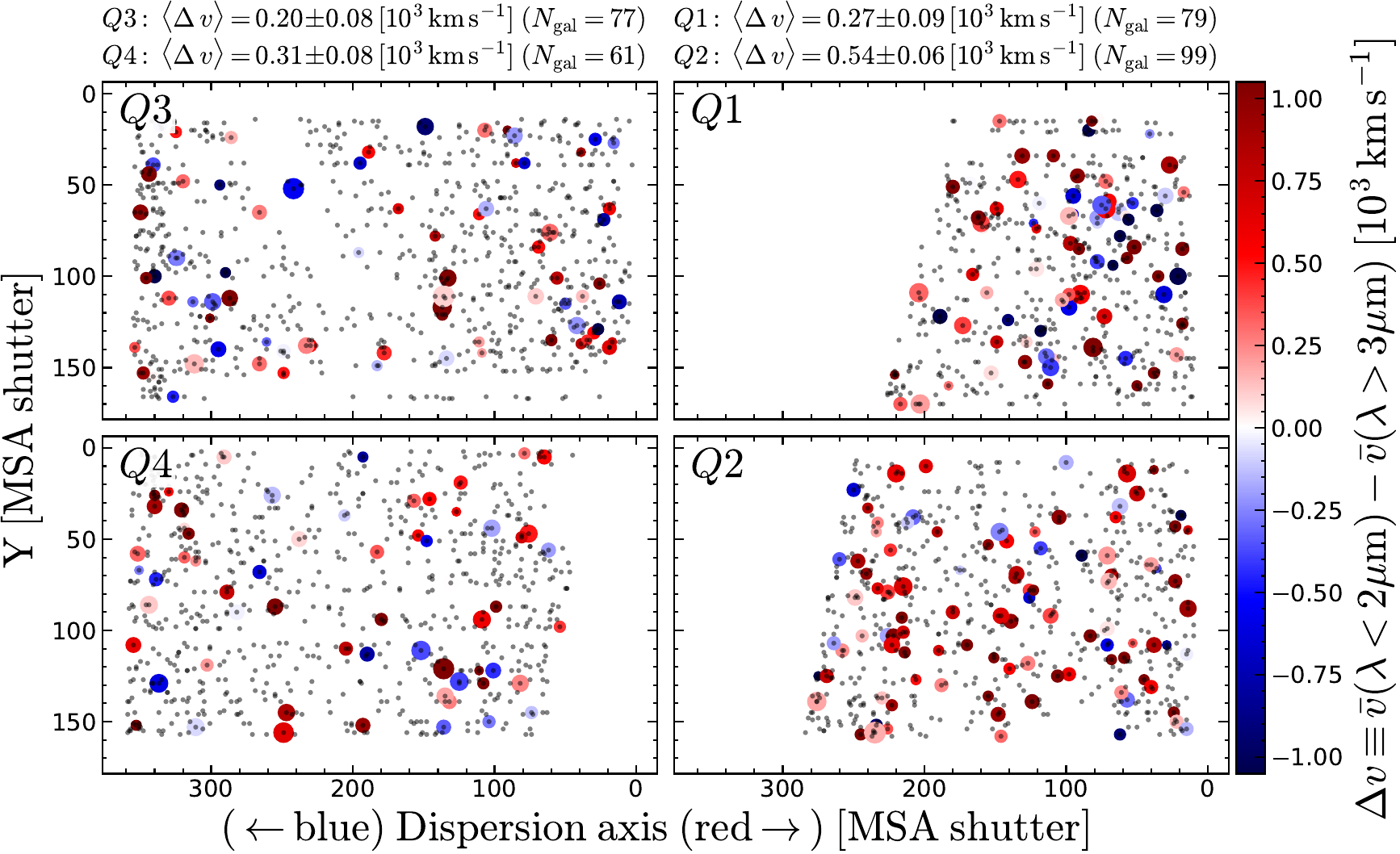}
\caption{Wavelength solution bias $\Delta v$ as a function of location on the MSA; the four rectangles are the four MSA quadrants Q1--Q4, with the location of the sample galaxies marked by small gray dots. The large dots represent galaxies with more than four emission lines detected at $S/N>5$, for which we could calculate the velocity offset between lines observed at wavelengths $1-2$~\mum and lines observed at $3-5.3~\mum$. $\Delta v$ has been corrected empirically for residual intra-shutter wavelength bias (Section~\ref{s.quality}. $\langle \Delta v \rangle$ is the average of $\Delta v$ inside each of the quadrants, illustrating a trend of increasing bias diagonally from Q3 to Q2.
}\label{f.quadrants}
\end{figure*}

In this appendix, we show that the bias in the wavelength calibration of the prism spectra has a residual dependence on the source position within the MSA.
We calculate $\Delta v$ for each galaxy (with the same definitions and cuts as in Section~\ref{s.quality}), then average $\Delta v$ inside each of the four MSA quadrants.
This reveals a bias only in quadrant 2 (hereafter: Q2), with a 4-\textsigma detection.
We then use the best-fit linear relation of Figure~\ref{f.quality.shutters} to remove the wavelength bias trend with $\delta x$, and repeat the test. The results are shown in Figure~\ref{f.quadrants}; there is a clearly detected zero-point offset in $\langle \Delta v \rangle$, as expected from Figure~\ref{f.quality.shutters}.
The offset is smallest in Q3 (2.5-\textsigma significance) and highest in Q2 (8~\textsigma), with intermediate values in Q1 and Q4 (3- to 4-\textsigma significance).
The diagonal direction from Q3 to Q2 is also the direction where the field-dependent spectral resolution increases, suggesting that the observed bias could be due to insufficient correction of the wavelength bias due to intra-shutter source position.

\bibliography{apjs_jadesdr3.bbl}
\bibliographystyle{config/aasjournal}

\end{document}